\documentclass[letterpaper]{ar-1col}

\usepackage{natbib}
\usepackage{graphics,graphicx}
\usepackage{wrapfig}
\usepackage{amsmath}
\usepackage{color}
\usepackage{lscape}
\usepackage{float}

\definecolor{boxtext}{cmyk}{1,.45,0,.18}
\definecolor{boxback}{cmyk}{.15,.07,0,.03}

\begin{document}

\input epsf.tex 
\input epsf.def
\input psfig.sty

\jname{Annu. Rev. Astron. and Astrophys.}
\jyear{2014}
\jvol{}

\title{On the Cool Side: Modeling the Atmospheres of Brown Dwarfs and Giant Planets}

\markboth{Marley \& Robinson}{Brown Dwarf and Giant Planet Atmospheres}

\author{M.~S. Marley$^1$, and T.~D. Robinson$^1$
\affil{$^1$
NASA Ames Research Center, MS 245-3, Moffett Field, CA 94035, USA}}


\begin{abstract}
The atmosphere of a brown dwarf or extrasolar giant planet controls
the spectrum of radiation emitted by the object and regulates its
cooling over time. While the study of these atmospheres has been
informed by decades of experience modeling stellar and planetary
atmospheres, the distinctive characteristics of these objects present 
unique challenges to forward modeling. In particular, complex 
chemistry arising from molecule-rich atmospheres, molecular opacity 
line lists (sometimes running to 10 billion absorption lines or more) 
multiple cloud-forming condensates, and disequilibrium chemical 
processes all combine to create a challenging task for any modeling 
effort. This review describes the process of incorporating these 
complexities into one-dimensional radiative-convective equilibrium 
models of sub-stellar objects. We discuss the underlying mathematics 
as well as the techniques used to model the physics, chemistry, 
radiative transfer, and other processes relevant to understanding 
these atmospheres. The review focuses on the process of the creation 
of atmosphere models and briefly presents some comparisons of model 
predictions to data. Current challenges in the field and some comments 
on the future conclude the review.
\end{abstract}

\maketitle

\section{Introduction}
\label{sec:intro}

The atmosphere of a brown dwarf or giant planet, despite constituting a 
negligible fraction of the total mass, plays a crucial role in controlling 
the evolution and appearance of the object.  By connecting the deep, 
convective interior with the thermal radiation pouring out from the 
object, the atmosphere regulates how quickly the interior can cool over 
time.  The atmosphere also imprints the varied signatures of gases, 
condensates, gravity, and the temperature profile onto emitted thermal 
radiation, thereby controlling the spectral signature of the object. 
Thus, understanding the spectrum and evolution over time of a giant 
planet or brown dwarf requires a working knowledge of the atmosphere. 
Especially for freely floating brown dwarfs, almost everything we know 
about an object depends upon our ability to understand its atmosphere.

The atmospheres of brown dwarfs and giant planets, however, are complex. 
Because these bodies are relatively cool (by stellar standards), chemical 
equilibrium favors the formation of molecules which often have opacities 
that vary strongly with wavelength. In addition, at these temperatures 
condensates can form, adding the seemingly intractable complexity of cloud 
physics to the problem. Unlike stellar atmospheres, whose `photosphere' 
(or the region over which thermal optical depth is near unity) is typically 
well defined, the strongly wavelength-dependent opacity in brown dwarf and 
giant planet atmospheres leads to a photosphere that varies with 
wavelength, and whose physical location can vary by several pressure scale 
heights.

The aim of this review is to discuss the techniques, and challenges, 
related to the construction of model atmospheres for brown dwarfs and young 
giant planets. For specificity, we will consider those processes which 
influence the atmospheres of the L, T, and Y dwarf spectral types as 
well as the directly-imaged planets.  We will explore influences on the 
thermal structure, composition, and energy transport in these objects. While 
our focus is primarily on self-luminous objects, we will also briefly 
consider the problem of absorption of incident light, relevant to worlds 
orbiting a primary star. 

In this review we will consider the problem of constructing a self-consistent, 
one-dimensional atmospheric structure model that converts a given internal  
heat flux from the deep, convective interior to radiation that departs the 
top of the atmosphere. This type of model is usually termed a 
`radiative-convective' model, and aims to represent, as an average, the 
complex three-dimensional structure of the atmosphere. We will explore how 
gravity, atmospheric composition, gaseous and cloud opacity, and incident 
flux all influence the thermal profile and spectra of emitted radiation. We 
will demonstrate the utility such models have for the interpretation of 
observations, and also consider their limitations for studying complex 
problems, including time variability and atmospheric dynamics.

A number of previous reviews have covered topics related to 
radiative-convective modeling of planetary and sub-stellar atmospheres.  Of 
historical interest are reviews by \citet{pecker1965}, who discusses 
modeling of stellar atmospheres, and \cite{vardya1970}, who examines 
properties and modeling of low-mass stars.  Several decades after these 
reviews, \citet{allardetal1997} revisited models and spectroscopic properties 
of very low mass stars.  These authors also discussed early work in the 
modeling of brown dwarfs, including grain formation and opacity. Of course, 
techniques and data for the modeling of condensates, gas and grain opacities, 
and radiative transfer have progressed substantially in the intervening years.  

As the atmosphere controls the cooling rate of the interior, an understanding 
of brown dwarf or giant exoplanet evolution crucially depends upon the 
construction of realistic, non-gray atmosphere models for describing the 
surface boundary condition \citep{saumonetal1994,chabrier&baraffe1997}.  The 
evolution of very low mass stars and brown dwarfs is reviewed by 
\citet{burrowsetal2001}, while \citet{saumon&marley2008} present more recent 
calculations. \citet{burrowsetal2001} review the data and techniques for 
computing atmospheric chemistry, abundances, and opacities, compare several 
cloud models, and review exoplanet and brown dwarf evolution and spectral 
properties. \citet{kirkpatrick2005} more completely discusses L and T dwarf 
colors and spectra and the classification of such objects. 

The theory of giant planets, emphasizing interior structure, evolution, 
and what we have learned from studying Jupiter and Saturn, is explored 
by \citet{hubbardetal2002}.  Very recently, \citet{helling&casewell2014} 
have reviewed the current state of brown dwarf observations, and discuss 
related modeling with an emphasis on clouds, including a comparison of a 
number of cloud models.  Finally, \citet{catling2014} reviews the physics 
of planetary atmospheres, with an emphasis on Solar System worlds, and 
the equations and processes that govern atmospheric composition, 
chemistry, thermal structure, radiative transfer, and circulation.

One reason that the study of cool atmospheres is especially interesting 
is that it lies at the frontier of two fields: astronomy and planetary 
science.  
Indeed among the early theoretical examinations of the first 
indisputable brown dwarf, Gliese~229B, one set of models 
\citep{marleyetal1996} had a heritage traceable to studies of the atmospheres 
of Titan and Earth, while two others \citep{tsujietal1996,allardetal1996} 
relied on modified stellar atmospheres codes.   This convergence of 
theory from both hot stellar atmospheres and cold planetary atmospheres 
towards what had once been a non-man's land of atmospheric structure theory 
(in the realm of effective temperatures of around 500--2500~K) has enhanced 
the field and provided valuable checks and balances to the theoretical 
development. Now mostly explored, this review aims to provide a guide for 
exploration of this new and interesting territory.

We begin this review with an overview of the physics that govern 
radiative-convective modeling of brown dwarfs and giant planets 
(Section~\ref{sec:overview}).  Following this, we discuss, in turn, the 
processes that are central to constructing a proper one-dimensional 
physical model: radiative transfer (Section~\ref{sec:radiation}), 
convection (Section~\ref{sec:convection}), atmospheric chemistry 
(Section~\ref{sec:chemistry}), gas opacities 
(Section~\ref{sec:opacity}), and cloud formation and condensate 
opacities (Section~\ref{sec:condense}).  Of course, each of these 
topics could easily merit their own dedicated review, and the references 
to classic papers and textbooks cited in this review provide excellent 
opportunities for more detailed follow-up reading.  We, then, outline 
how these physics are used to actually derive a thermal profile, and 
present some relevant data-model comparisons.  Finally, we conclude 
with a discussion of some current issues, and prospects for the future.
 
\section{Physics Overview}
\label{sec:overview}

Our focus specifically on atmospheres of brown dwarfs and giant planets 
simplifies our discussion to the physics of predominantly hydrogen-helium 
objects, which have been relatively well studied. This review takes the 
perspective that the atmosphere is a semi-infinite column of gas in 
hydrostatic equilibrium. The goal is to understand how the gravity, internal 
heat, energy transport mechanisms, composition, and cloud structures of such 
an atmosphere influence the thermal profile and, consequently, the properties 
of its emitted radiation. 

\begin{figure}
  \centering
  \includegraphics[scale=0.55]{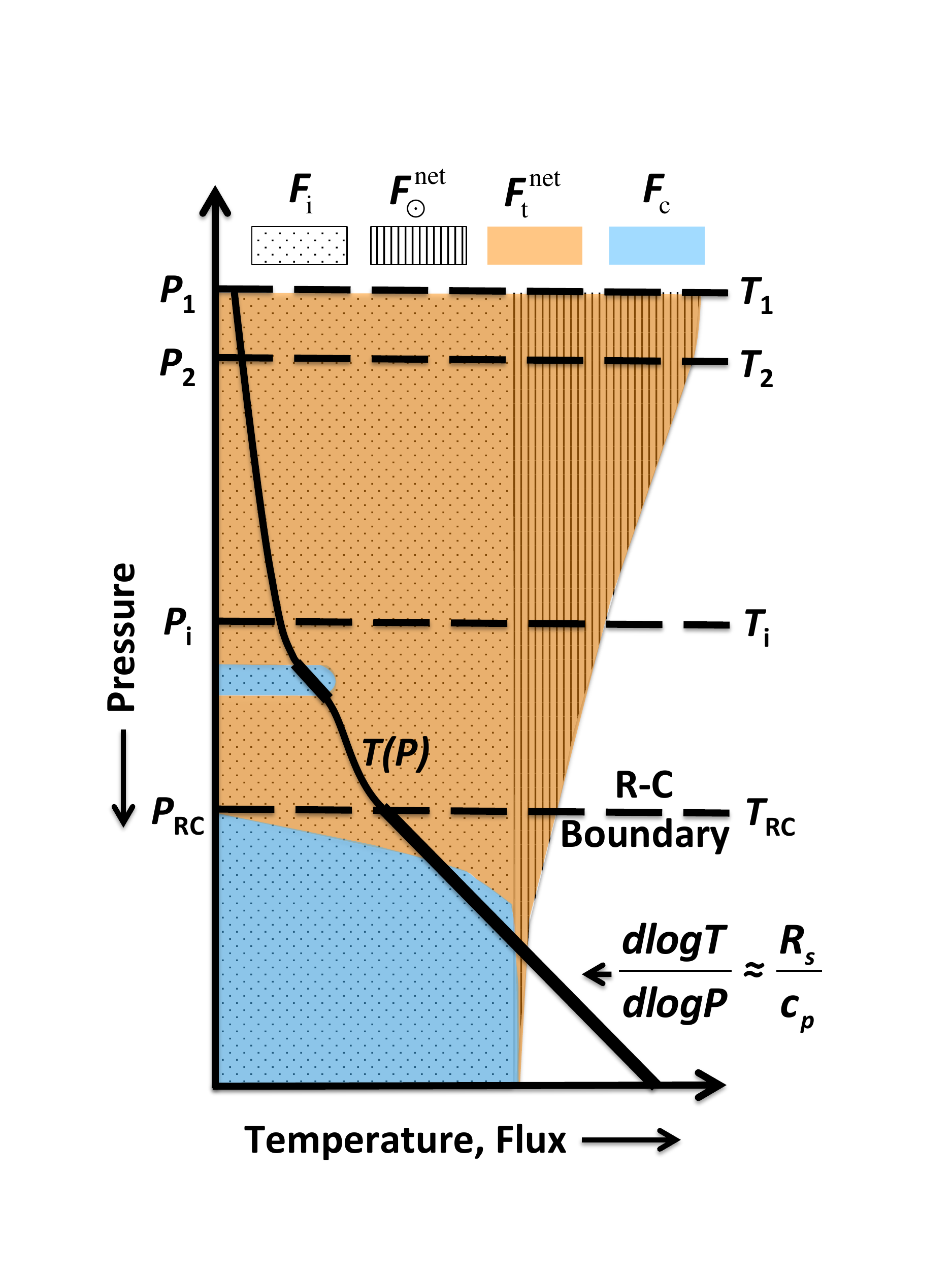}
  \caption{Schematic depiction of a thermal structure model.  The 
  vertical axis is pressure, increasing downwards, which is the 
  independent variable, and the horizontal axis shows, relatively, 
  temperature and energy flux.  Model levels are shown (horizontal 
  dashed lines), and the solid line is the thermal structure 
  (i.e., temperature) profile, where bolded lengths indicate a 
  convective region.  Level pressures and temperatures are 
  indicated with associated sub-scripted symbols, and `RC' indicates 
  the radiative-convective boundary.  In equilibrium, net thermal 
  flux ($F^{\rm{net}}_{\rm{t}}$, orange) and the convective flux 
  ($F_{\rm{c}}$, blue) must sum to the internal heat flux 
  ($F_{\rm{i}}$, dotted) and, for an irradiated object, the net 
  absorbed stellar flux ($F^{\rm{net}}_{\odot}$, striped).  Note 
  that the internal heat flux 
  is constant throughout the atmosphere, whereas the schematic profile 
  of net absorbed stellar flux decreases with increasing pressure, and 
  eventually reaches zero in the deep atmosphere.  At depth, convection 
  carries the vast majority of the summed internal and stellar fluxes, 
  but is a smaller component in detached convective regions (upper blue 
  region).}
  \label{fig:cartoon}
\end{figure}

Figure~\ref{fig:cartoon} shows a schematic of a one-dimensional model 
atmosphere.  The vertical coordinate is pressure, $P$, which is defined 
on a grid of model levels.  In hydrostatic equilibrium, where the 
gravitational force acting on any given atmospheric slab is balanced by 
the vertical pressure gradient force, the fluid atmosphere obeys
\begin{equation}
  \frac{dP}{dz} = -g \rho \ ,
  \label{eqn:hydrostat}
\end{equation}
where $z$ is altitude, $g$ is the acceleration due to gravity, and $\rho$ 
is the atmospheric mass density.  After inserting the ideal gas law and 
re-arranging, this expression becomes
\begin{equation}
  \frac{dP}{P} = -\frac{gm}{k_BT}dz = -\frac{dz}{H} \ ,
\end{equation}
where $m$ is the mean molecular mass in the atmosphere, $k_B$ is 
Boltzmann's constant ($k_B = 1.381 \times 10^{-16}$ erg K$^{-1}$), and 
$T$ is temperature.  Here we have also defined the atmospheric pressure 
scale height, $H=k_{B}T/mg$, which, for an isothermal layer of the 
atmosphere, is the e-folding distance for pressure, such that the 
pressure-altitude relation is
\begin{equation}
  P(z)  = P(z_0) e^{-\left(z - z_0 \right)/H} \ ,
  \label{eqn:hydrostat_solve}
\end{equation}
where $P(z_0)$ and $z_0$ are the pressure and altitude at the base of 
the layer, respectively.  Thus, the physical distance between two 
adjacent model pressure levels shown in Figure~\ref{fig:cartoon} can 
be determined using the layer pressure scale height and 
Equation~\ref{eqn:hydrostat_solve}.

\begin{textbox}
\section{COLUMN DENSITIES}
  A useful quantity derived using the equation 
  of hydrostatic equilibrium (Equation~\ref{eqn:hydrostat}) is the 
  column mass, which is the integrated mass per unit area above a 
  given atmospheric level, given by  
  $\mathcal{M} = \int_{z}^{\infty} \rho dz = P(z)/g $.  The column molecular 
  number density, $\mathcal{N}$, can be similarly defined, and, using the 
  integral definition, the equation of hydrostatic equilibrium, and 
  assuming an isothermal atmosphere, the column number density is 
  related to the number density profile simply by $\mathcal{N} = n(z)H$.  
  This quantity is especially helpful when estimating optical 
  depths---given an absorption cross section per molecule, 
  $\sigma_{\rm{a}}$, the optical depth is roughly 
  $\tau(z) = \sigma_{\rm{a}} \mathcal{N}(z)$.
\end{textbox}

Pressure- or altitude-dependent atmospheric properties, such as 
temperature (shown in Figure~\ref{fig:cartoon}), chemical composition, or 
wavelength-dependent thermal flux, are determined either at the model 
levels or for the model layers (i.e., at the level mid-points).  A key model 
input parameter is the internal heat flux, $F_{\rm i}$, which, for a 
non-irradiated world, will set the effective temperature via 
$\sigma T_{\rm eff}^{4}=F_{\rm i}$, where $\sigma$ is the Stefan-Boltzmann 
constant ($5.67 \times 10^{-5}$~erg~cm$^{-2}$~s$^{-1}$~K$^{-4}$).  
In steady state, this 
energy flux is constant with pressure throughout the atmosphere, and is 
represented by the dotted region in Figure~\ref{fig:cartoon}.

At great depths in the interior of a brown dwarf or gas giant, the electron 
density is high and thermal photons cannot propagate far, so energy transport 
is dominated by convection.  Reviews by \citet{stevenson&salpeter1976} and 
\citet{hubbard&smoluchowski1973} discuss this point.  At large ages, very 
massive brown dwarfs can develop small conductive cores \citep{lunineetal1986,
chabrieretal2000}, but this is far below the atmosphere. Because convection in 
these dense atmospheres is very efficient, the gradient in the deep thermal 
temperature profiles, $\nabla = d\log T/d\log P$,  is expected to closely follow 
convective adiabats (i.e, thermal profiles of constant entropy), with 
$\nabla = \nabla_{\rm ad}$ \citep{baraffeetal2002}.  These assumptions 
can break down if there are composition gradients that impede convection 
or deep windows in molecular opacity \citep[e.g.,][]{guillotetal1994,
leconte&chabrier2012}, which are eventualities that we ignore here. 

Convection delivers thermal energy to/through the base of the atmosphere 
(represented by the blue shaded region at large pressures in 
Figure~\ref{fig:cartoon}), and thermal radiative transport (represented 
by the orange shaded region in Figure~\ref{fig:cartoon}) begins to 
become more important as the atmosphere thins with increasing radius 
(or decreasing pressure). If there are wavelength regions that are both 
low opacity and which overlap with the local Planck function, then energy 
can be radiated away through these opacity `windows'. At some point, as 
progressively more energy is carried away from the increasingly tenuous 
atmosphere by radiation, the temperature profile no longer changes as 
steeply with altitude, indicating that convection has ceased.  Above this 
level, referred to as the `radiative-convective boundary' (or R-C boundary), 
energy is carried by radiation, and the atmospheric thermal profile is 
governed by radiative equilibrium.

\begin{figure}
  \centering
  \includegraphics[scale=0.65]{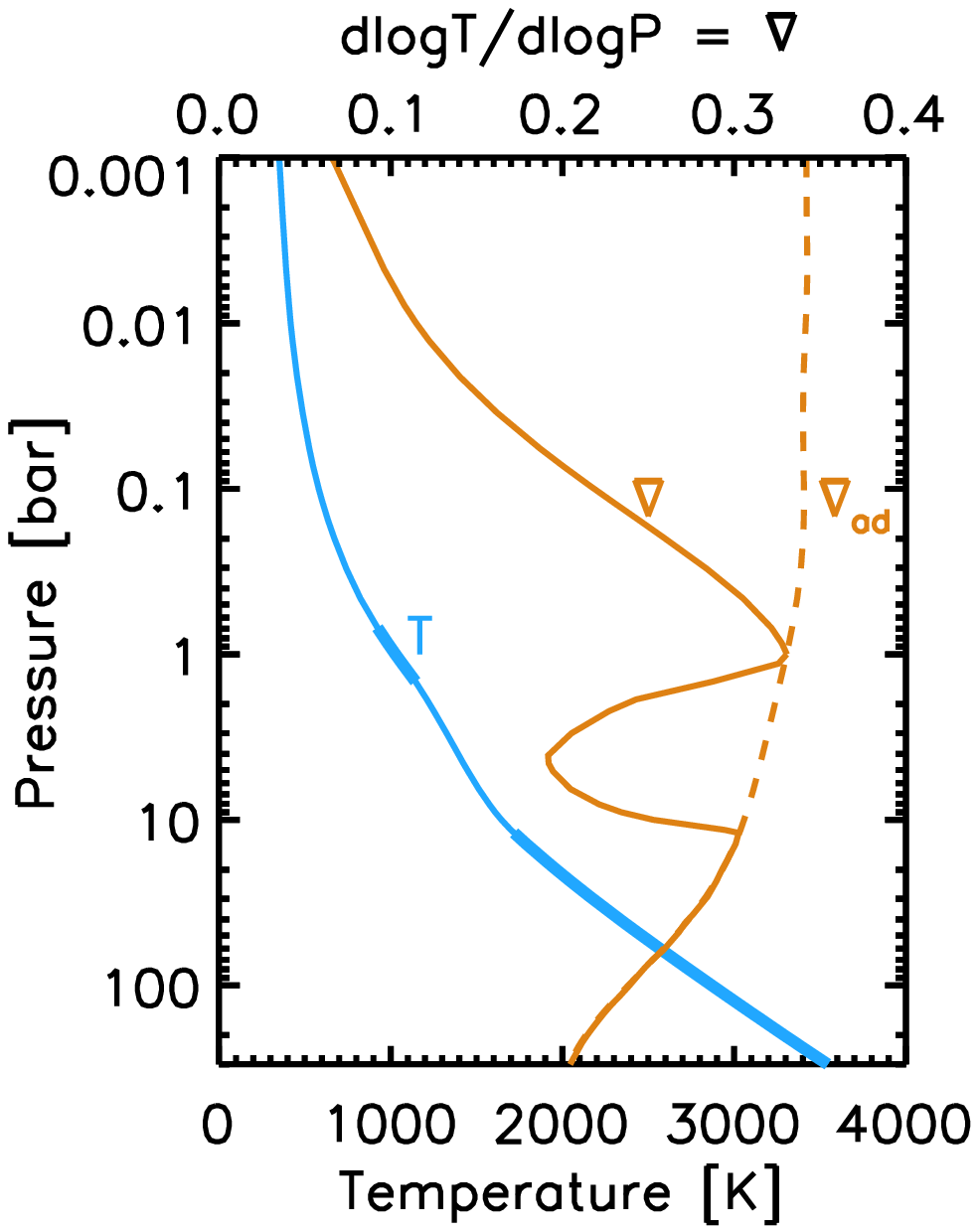}
  \vspace{0.5cm}
  \caption{{\normalsize (\textbf{Two pages.}) Illustration of some of 
  the influences on a radiative-convective model, as well as model output. 
  The first panel shows a computed thermal profile (blue), for a cloud-free 
  mid-T dwarf, showing temperature as a function of pressure in 
  the atmosphere. Orange curves compare the computed atmospheric 
  temperature gradient ($\nabla$, solid) with the local adiabatic gradient 
  ($\nabla_{\rm{ad}}$, dashed). The model has two convection zones 
  (thickened blue), and in both regions $\nabla=\nabla_{\rm ad}$. The 
  second panel helps illustrate why these two convective zones 
  form---each sub-plot shows, for the indicated pressure level, spectra 
  of the local Planck function (blue), the scaled net thermal flux 
  (orange), and the column absorptivity (dark gray) (i.e, $1-e^{-\tau}$, 
  where $\tau$ is the optical depth between the top-of-atmosphere and 
  the indicated pressure).  The absorptivity for $\tau=1$ is shown as 
  a horizontal line in light gray.  In the deepest level 
  ($P=100$~bar), convection is carrying the internal heat flux, and the 
  net thermal flux is small.  However, near $P=10$~bar, windows in the 
  opacity spectrum align with the local Planck function, thus allowing 
  thermal radiation to carry the internal flux, and the atmosphere forms 
  a deep radiative layer.  Further up in the atmosphere, at $P=1$~bar, 
  the local Planck function moves into a region of strong water vapor and 
  methane opacity, thereby re-invigorating convection, which then carries 
  some part of the internal heat flux.  Finally, by $P=0.5$~bar, 
  the atmosphere is in strict radiative equilibrium, which remains towards 
  all smaller pressures (or larger altitudes).}}
  \label{fig:dblconvect}
\end{figure}

In some cases, as the temperature falls with increasing altitude and the 
peak of the Planck function moves to ever longer wavelengths, this peak may 
again overlap with a wavelength region of strong opacity. This can again 
impede thermal radiative energy transport and re-invigorate convection over 
a small vertical region, called a `detached convective zone' (see the 
smaller shaded blue region at lower pressures in 
Figure~\ref{fig:cartoon}).  Figure~\ref{fig:dblconvect} better illustrates 
the physics of such a situation, and shows model profiles of temperature, 
temperature gradient, and the 
adiabatic gradient of  for a cloud-free 
late T dwarf. For this model case a  detached convective region is apparent
near 1~bar.  For four different pressure levels, spectra of the local 
Planck function, net thermal radiative flux, and absorptivity of the 
overlying column of gas are also provided.  The detached convective region forms 
when the peak of the local Planck function overlays strong water vapor and 
methane absorption bands at 2.7 and 3.3~$\mu$m, respectively.

As altitude increases, the atmosphere becomes optically thin to thermal 
radiation at most wavelengths. However, some strong molecular bands may 
remain optically thick to fairly high altitudes and will continue to radiate, 
even while the atmosphere is in general optically 
thin.  For this reason the equilibrium radiative profile is not the same as 
that expected for a gray atmosphere (i.e., an atmosphere where the opacity 
is treated as a wavelength-independent quantity), which reaches a 
constant-with-altitude `skin temperature'  
\citep[for a more detailed discussion, see][p.~298]{pierrehumbert2010}.

If the atmosphere is irradiated (i.e., receiving energy from a parent star), 
then thermal radiative transport and convection must carry the internal 
heat flux as well as the net absorbed stellar flux.  In 
Figure~\ref{fig:cartoon}, this is shown conceptually by the vertically-striped 
flux profile on the right side. For Jupiter, the internal and absorbed 
incident fluxes are about the same \citep{haneletal1981}. In the case of Uranus 
the internal flux is vanishingly small \citep{pearletal1990}, while for the 
so-called `hot Jupiters' the incident flux dominates completely over the 
internal contribution.

For a given atmosphere, the internal heat flux and profile of net absorbed 
stellar flux, along with the opacity structure, control the thermal structure.  
The equilibrium temperature profile obeys
\begin{equation}
  F_{\rm t}^{\rm net}(P) + F_{\rm c}(P) + F^{\rm net}_{\odot}(P) - F_{\rm i} = 0 \ ,
\label{eqn:equilibrium}
\end{equation}
where $F_{\rm t}^{\rm net}$ is the net thermal flux (defined below in 
Equation~\ref{eqn:netflx}), $F_{\rm c}$ is the convective flux, 
$F^{\rm net}_{\odot}$ is the net stellar flux, and, as before, $F_{\rm i}$ 
is the internal heat flux.  Methods for computing these flux profiles, and 
for determining the thermal structure that satisfies 
Equation~\ref{eqn:equilibrium}, are the focus of this review.

\section{Radiation}
\label{sec:radiation}

A central problem in understanding the equilibrium thermal structure of a 
sub-stellar atmosphere is to understand energy transport by radiation through 
the atmosphere.  For irradiated bodies, the absorption of stellar flux 
throughout the atmosphere and/or at the surface drives the climate physics that 
determine thermal structure.  All worlds lose energy by emitting thermal 
radiation, and thermal radiative transport of energy occurs throughout all 
atmospheres. In this section we discuss various aspects of this problem, 
focusing on expressions for the radiative energy fluxes.

\subsection{Radiative Transfer}

Heating and cooling by radiative transfer occurs due to a gradient in the net 
radiative energy flux,
\begin{equation}
q_{r} = \frac{g}{c_P}\frac{dF^{\text{net}}_{r}}{dP} \ ,
\label{eqn:heatrate}
\end{equation}
where $q_{r}$ is the radiative heating rate (in, for example, K s$^{-1}$), 
$c_P$ is the local atmospheric heat capacity, and $F^{\text{net}}_{r}$ is 
the net radiative flux density (typically referred to as just the `radiative 
flux').  The net radiative flux is partitioned between an upwelling stream 
(i.e., towards lower pressures) and a downwelling stream (i.e., towards 
higher pressures), with
\begin{equation}
F^{\text{net}}_{r} = F^{+}_{r} - F^{-}_{r} \ ,
\label{eqn:netflx}
\end{equation}
where a ``+" indicates upwelling and a ``-" indicates downwelling.  Note that 
the total radiative heating rate is determined from the bolometric net flux, with 
\begin{equation}
F^{\text{net}}_{r} = \int_{0}^{\infty} F^{\text{net}}_{r,\nu} d\nu \ ,
\end{equation}
where $F^{\text{net}}_{r,\nu}$ is the spectrally resolved net flux.  Thus, 
in the absence of biases in the calculations, uncertainties in the spectrally 
resolved fluxes at a particular frequency or wavelength can cancel those at 
other frequencies.  A common issue for thermal radiative transport models, 
though, is that, in very opaque regions of the atmosphere, the upwelling and 
downwelling fluxes approach the same value, so that the net thermal flux is 
calculated as the difference between two typically large and similar numbers, 
requiring high accuracy in the computation of $F^{+}_{r}$ and $F^{-}_{r}$.

Most radiative transfer models, in one way or another, solve the 1-D, 
plane-parallel radiative transfer equation (RTE),
\begin{equation}
\mu \frac{dI_{\nu}}{d\tau_{\nu}} = 
          I_{\nu}\left(\tau_{\nu},\mu,\phi\right) 
          - S_{\nu}\left(\tau_{\nu},\mu,\phi\right) \ ,
\end{equation}
where $I_{\nu}$ is the spectral radiance, $\tau_{\nu}$ is the 
frequency-dependent extinction optical depth (which increases towards higher 
pressures), $\mu$ is the cosine of the zenith angle, $\phi$ is the azimuth 
angle, and $S_{\nu}$ is the `source function'.  Optical depth can be 
determined from the absorption coefficient (see Section~\ref{sec:opacity}), 
$k_{\nu}$ (in units of cm$^{2}$~g$^{-1}$, or equivalent), via the 
differential relation 
\begin{equation}
  d\tau_{\nu} = -k_{\nu} \rho_{a} dz \ ,
\end{equation}
where $\rho_{a}$ is the mass density of the absorber (and absorbers 
simply combine linearly in $k_{\nu} \rho_a$).  The source function, 
$S_{\nu}$, is given by 
\begin{equation}
\begin{array}{ll}
S_{\nu}\left(\tau_{\nu},\mu,\phi\right) = & 
        \omega_{\nu} F^{\odot}_{\nu} e^{-\tau_{\nu}/\mu_{\odot}} \cdot
        p_{\nu}\left(\tau_{\nu},\mu,\phi,-\mu_{\odot},\phi_{\odot}\right)/4\pi  \\ & 
        + \left(1 - \omega_{\nu}\right) B_{\nu}\left(T\left(\tau_{\nu}\right)\right) \\ & 
        + \omega_{\nu} \int_{0}^{2\pi} d\phi^{\prime} \int_{-1}^{1} d\mu^{\prime} \cdot 
         I_{\nu}\left(\tau_{\nu},\mu^{\prime},\phi^{\prime}\right) 
         p_{\nu}\left(\tau_{\nu},\mu,\phi,\mu^{\prime},\phi^{\prime}\right)/4\pi \ ,
\end{array}
\label{eqn:srcfcn}
\end{equation}
where $ \omega_{\nu} = \omega_{\nu}\left( \tau_{\nu} \right) $ is the 
frequency-dependent single scattering albedo, $F^{\odot}_{\nu}$ is the 
top-of-atmosphere solar (or, more generally, stellar) irradiance, 
$\mu_{\odot}$ is the solar zenith angle, $\phi_{\odot}$ is the solar azimuth 
angle, $p_{\nu}$ is the scattering phase function, $B_{\nu}$ is the Planck 
function, and $T\left(\tau_{\nu}\right)$ is the atmospheric temperature 
profile.  The final term on the right-hand side of Equation~\ref{eqn:srcfcn}, 
which represents scattering from directions 
$\left(\mu^{\prime},\phi^{\prime}\right)$ into the beam at 
$\left(\mu,\phi\right)$, complicates radiative transfer calculations, as 
it turns the radiative transfer equation into an integro-differential 
equation.  Note that in the planetary literature it is common to split radiative 
transfer calculations 
into solar/stellar and thermal components, wherein the source function would then omit 
the second and first terms on the right-hand side of Equation~\ref{eqn:srcfcn}, 
respectively. In the case of hot Jupiters there can be substantial wavelength overlap
between the thermal emission and incident flux although they have very different
angular distributions.
The upwelling and downwelling fluxes are related to the 
angle-dependent radiances by
\begin{equation}
\begin{array}{c}
F^{+}_{\nu}\left(\tau_{\nu}\right) = \int_{0}^{2\pi} \int_{0}^{1} I_{\nu}\left(\tau_{\nu},\mu,\phi\right) \mu d\mu d\phi\\
F^{-}_{\nu}\left(\tau_{\nu}\right) = -\int_{0}^{2\pi} \int_{-1}^{0} I_{\nu}\left(\tau_{\nu},\mu,\phi\right) \mu d\mu d\phi\ .
\end{array}
\label{eqn:fluxint}
\end{equation}

A pair of boundary conditions are needed to solve the radiative transfer 
equation, and these typically specify the downwelling radiation field at 
the top of the model atmosphere and the upwelling radiation field at the 
bottom of the atmosphere.  For irradiated bodies, the  
top-of-atmosphere boundary condition for incident radiation
is simply a direct or diffuse stellar 
flux, and the bottom-of-atmosphere boundary condition is that either no 
flux returns from the deep atmosphere, or that the deep atmosphere has a 
specified albedo that reflects back some small radiative flux.  Thermal 
calculations assume zero downwelling flux at the top of the atmosphere.  
The thermal bottom-of-atmosphere boundary condition is less obvious, as 
the atmospheric column is assumed infinitely deep, but models usually 
take \citep[][ p. 165]{mihalas1970}, 
\begin{equation}
\begin{array}{lr}
I_{\nu}\left(\tau_{\nu}=\tau_{\nu}^{*},\mu,\phi\right) = B_{\nu}\left(T\left(\tau^{*}_{\nu}\right)\right) + \mu\frac{dB_{\nu}}{d\tau_{\nu}}\rvert_{\tau^{*}_{\nu}} & \forall~\mu > 0 \ ,
\end{array}
\end{equation}
where the gradient term allows some flux from deeper layers to contribute  
at the bottom model level (where $\tau_{\nu}=\tau^{*}_{\nu}$), and the factor of $\mu$ ensures that 
near-vertical streams `see' deeper into the atmosphere at the boundary.

\subsection{Approaches to Solving the RTE}

Solving the RTE requires a certain level of parameterization or 
simplification.  Most commonly, techniques are divided into either 
two-stream or multi-stream solutions, where a `stream' refers to a 
particular azimuth-zenith coordinate through the atmosphere.  Solutions 
in the two-stream category are more computationally efficient than 
multi-stream calculations, and offer analytic results that can help to 
provide insight into problems.  Multi-stream calculations provide more 
detailed information about the angular distribution of intensities and, 
thus, can provide more accurate solutions for radiant fluxes.

Two-stream solutions have roots in early theories of stellar 
atmospheres \citep{schwarzschild1906}, and provide techniques for rapidly 
computing $F^{+}_{\nu}$ and $F^{-}_{\nu}$ (which are the two streams).  
As discussed by \citet{meador&weaver1980}, by making assumptions about 
the scattering phase function and the distribution of intensity in azimuth 
and zenith angles, the RTE can be simplified to the two-stream equations:
\begin{equation}
\begin{array}{c}
\frac{dF^{+}_{\nu}}{d\tau_{\nu}} = \gamma_{1}F^{+}_{\nu} - \gamma_{2}F^{-}_{\nu} - S_{\nu}^{+} \\
\frac{dF^{-}_{\nu}}{d\tau_{\nu}} = \gamma_{2}F^{+}_{\nu} - \gamma_{1}F^{-}_{\nu} + S_{\nu}^{-} \ ,
\end{array}
\end{equation}
where the frequency-dependent $\gamma$-coefficients permit scattering from 
the upwelling stream into the downwelling stream and vice versa, and 
$S^{+/-}_{\nu}$ are level-dependent source terms.  A critical analysis of 
the accuracy of different two-stream implementations is also provided by 
\citet{meador&weaver1980} for cases without internal thermal sources.

\begin{textbox}
\section{GRAY RADIATIVE EQUILIBRIUM}
  The non-scattering, gray two-stream equations of thermal 
  radiative transfer, 
  \begin{equation*}
    \frac{dF^{+}_{t}}{d\tau} = D \left( F^{+}_{t} - \sigma T^4 \right)
  \end{equation*}
  \begin{equation*}
    \frac{dF^{-}_{t}}{d\tau} = -D \left( F^{-}_{t} - \sigma T^4 \right) \ ,
  \end{equation*}
  where $D$ is the diffusivity factor (i.e, a constant 
  that accounts for the integration of intensity over the 
  hemisphere, with preferred values ranging from 1.5--2) and 
  $\tau$ is the gray optical depth, which is related to a 
  gray opacity, $k$, via $d\tau = kdP/g$, can be combined 
  into a single differential equation, 
  \begin{equation*}
    \frac{d^2 F^{\rm net}_{t}}{d\tau^2} - D^2 F^{\rm net}_{t} = -2D\sigma\frac{dT^4}{d\tau} \ .
  \end{equation*}
  For a non-irradiated object with an atmosphere in 
  radiative equilibrium, the net thermal flux is constant 
  and equal to $\sigma T_{\rm eff}^4$, so that the previous 
  differential equation can be solved to yield the thermal 
  structure,
  \begin{equation*}
    T(\tau)^4 = \frac{1}{2} T_{\rm eff}^4 (1 + D\tau) \ .
  \end{equation*}
  Given either a database of gray opacities, or by making 
  assumptions about the pressure-dependence of $k$, the 
  temperature profile as a function of pressure can be 
  determined.  Simple, analytic expressions for the case 
  where convection is included are derived in 
  \citet{robinson&catling2012}.
\end{textbox}

A widely-used approach to solving the two-stream RTE with both solar 
and thermal sources is provided by \citet{toonetal1989} (which
regrettably contains numerous typographical errors). 
These authors describe a numerically stable matrix-based solution to the 
two-stream problem.  Additionally, to more accurately solve for the 
thermal radiation field in the presence of scattering, \citet{toonetal1989} 
develop an approach called the `two-stream source function' technique.  
Using the solution for the two-stream fluxes, 
direction-independent scattering source functions, $S^{+/-}_{\nu}$, are 
determined.  These source functions are then used in a multi-stream 
calculation, where the intensities from the different streams are 
integrated (i.e., via Equation \ref{eqn:fluxint}) to more accurately 
determine the upwelling and downwelling fluxes. The approach is accurate
in most cases excepting situations with very high albedo, very highly foward 
scattering particles. Thus, the two-stream 
source function approach represents an efficient middle ground for computing
the atmospheric energy balance and thermal profile between 
strict two-stream solutions and general multi-stream solutions, and is 
the approach currently used in the Marley/Saumon brown dwarf and giant 
planet models \citep[e.g.,][]{marleyetal2002,fortneyetal2005,
saumon&marley2008,morleyetal2012,marleyetal2012}.

True multi-stream solutions seek to solve the full RTE with self-consistent 
multiple scattering.  Such solutions commonly solve the azimuth-independent 
form of the RTE \citep[][p. 15]{chandrasekhar1960}, as aerosol scattering 
phase functions are typically expressed in terms of a single angle---the 
scattering angle.  More general techniques have been developed for solving 
the azimuth-dependent RTE, which are necessary when, for example, the 
solar/stellar source comes from a particular azimuth and zenith angle 
\citep{milkeyetal1975,stamnesetal1988}.

Numerical techniques for determining angle-dependent intensities 
along a set of discrete zenith angles (representing the streams) in 
scattering and/or emitting atmospheres/media have been developed in both 
the planetary and astrophysical literatures.  For the latter, the 
Accelerated Lambda Iteration (ALI) approach \citep{cannon1973,
olsonetal1986,hubeny&lanz1992,hauschildt1992} is most common 
\citep[see also review by][]{hubeny2003}.  In ALI, the level- 
and frequency-dependent source function ($S_{\nu}$) is iteratively 
adjusted, with each new iteration providing intensities that are 
increasingly accurate solutions to the RTE.  This technique is currently 
used in the PHOENIX atmospheric models \citep{allard&hauschildt1995,
barmanetal2001}, and in the TLUSTY radiative transfer model 
\citep{hubeny&lanz1995} implemented in models by Burrows~et~al. 
\citep{burrowsetal2002}.

Multi-stream techniques for solving the RTE developed in the Earth 
and planetary literature include adding-doubling and discrete ordinates.  
In the former, optically thin, homogeneous atmospheric layers---with 
given absorbing, emitting, and scattering properties---are combined to 
form an inhomogeneous atmospheric model \citep{vandehulst1963,
twomeyetal1966}.  A set of recursive relations, based on the linear 
interaction of the radiation with a thin layer, are used to compute the 
intensities within the inhomogeneous model \citep{hansen1969,
wiscombe1976,evans&stevens1991}.  The discrete ordinates method, 
originally developed by \citet[][p. 56]{chandrasekhar1960}, recasts the 
RTE as a system of ordinary differential equations and uses matrix 
techniques to find a solution to the system \citep{stamnes&swanson1981}.  
This method is used in the popular, and publicly available, radiative 
transfer model DISORT \citep{stamnesetal1988,stamnesetal2000}.

\section{Convection}
\label{sec:convection}

Absorption of stellar flux at a planetary surface or deep in the 
atmosphere of a gaseous world, or the presence of a large internal 
heat flux, can lead to thermal structures that are unstable to 
vertical convection.  Here, a parcel of gas that is displaced 
upwards would find itself in an environment whose density is 
greater than the parcel's internal density, so the parcel would 
continue to rise (or parcels displaced downwards continue to sink).  
This instability leads to a critical vertical density gradient in an 
atmosphere, with convection occurring when the gradient is too steep.  
Using an equation of state, this density gradient can be related to 
a temperature gradient (or ``lapse rate"), $-dT/dz$, that defines the 
limit between thermal structures that are convectively stable versus 
unstable.

Fortunately, modeling convection in brown dwarf and giant planet 
atmospheres is more straightforward than in the stellar structure 
literature \citep[e.g.,][Chapter 5]{hansenetal2004}, since the ideal 
gas law applies and the convection is, to a good approximation, 
adiabatic \citep{baraffeetal2002,ludwigetal2006,freytagetal2010}.  
Given these, the criterion for an unstable lapse rate is  
\begin{equation}
-\frac{dT}{dz} > \frac{g}{c_P} \ ,
\end{equation}
or, by including the equation of hydrostatic equilibrium,
\begin{equation}
\nabla = \frac{d\log T}{d\log P} > \frac{R_s}{c_P} = \nabla_{\text{ad}} \ ,
\end{equation}
where $R_{s}$ is the specific gas constant, which is equal to $k_{B}/m$.  
Thus, for lapse rates larger than the ``dry adiabatic lapse rate", 
$g/c_P$ (or $R_s/c_P$ in $d\log T/d\log P$), a parcel of gas that 
is perturbed upwards will continue to rise, meaning that convection would 
ensue.

\begin{textbox}
\section{DRY ADIABATIC LAPSE RATES AND TEMPERATURE GRADIENTS} 
  Combining the dry adiabatic lapse rate, $-dT/dz = g/c_P$, and the 
  equation of hydrostatic equilibrium (Equation~\ref{eqn:hydrostat}), 
  we find the equivalent of the dry adiabat in pressure-space,
  \begin{equation*}
    \frac{dT}{dP} = \frac{1}{c_P \rho} \ .
  \end{equation*}
  Then, using the ideal gas law to express density in terms of 
  pressure and temperature, we have
  \begin{equation*}
    \frac{dT}{dP} = \frac{k_{B}T}{mc_P P} \ ,
  \end{equation*}
  or
  \begin{equation*}
    \nabla_{\text{ad}} = \frac{d\log T}{d\log P} = \frac{R_s}{c_P} \ .
  \end{equation*}
  Recalling from kinetic theory that the specific gas constant 
  is equal to $c_P - c_v$, where $c_v$ is the specific heat at 
  constant volume, and the ratio of specific heats is, 
  \begin{equation*}
    \gamma = \frac{c_P}{c_v} \ ,
  \end{equation*}
  we can then write the dry adiabat as simply
  \begin{equation*}
    \nabla_{\text{ad}} = \frac{\gamma - 1}{\gamma}  \ .
  \end{equation*}
  Thus, over regions of the atmosphere where the internal degrees 
  of freedom of the gas are roughly constant, the 
  temperature-pressure relationship along a dry adiabat is given 
  by Poisson's adiabatic state equation,
  \begin{equation*}
    T(P) = T_{0}\left(\frac{P}{P_0}\right)^{(\gamma-1)/\gamma} \ ,
  \end{equation*}
  where $T_0$ and $P_0$ are a reference temperature and pressure, 
  respectively, along the adiabat.
\end{textbox}

It is important to remember that convection does not create or 
destroy energy---it simply redistributes heat.  Thus, convection 
schemes should be energy conserving.  Since lapse rates in a 
convectively unstable atmosphere are larger than the adiabatic 
lapse rate, convection will work to move heat from deeper atmospheric 
levels to levels at lower pressures (higher altitudes).  In this 
regard, convection serves to limit the thermal contrast between the 
deep atmosphere and the radiative-convective boundary.

Approaches to modeling convection (an inherently three-dimensional 
process) in one-dimensional thermal structure models are varied.  
Models rooted in the planetary science literature commonly employ 
an approach called `convective adjustment', while those derived from 
astrophysical sources will tend to use mixing-length theory.  These are 
discussed in turn below.

\subsection{Convective Adjustment}
\label{subsec:convadjust}
Convection in brown dwarf and giant planet atmospheres relaxes 
the thermal structure onto an adiabat \citep{ludwigetal2006}, so 
that the structure throughout the convective portion of the 
atmosphere can be  modeled using the dry 
adiabatic lapse rate.  In their simulations of Earth's atmospheric 
thermal structure, \citet{manabe&strickler1964} outlined a 
straightforward scheme for forcing convectively unstable atmospheric 
layers onto an adiabat, which these authors called `convective 
adjustment.'  In this work, radiative heating and cooling rates were 
used to timestep a model atmosphere towards radiative equilibrium.  
At each timestep, and proceeding upward from the surface, the lapse 
rate for each model layer, $|\Delta T/\Delta z|$, was compared to an 
adiabatic lapse rate.  If the layer was unstable, then the 
temperatures at the top and bottom of the layer were immediately 
changed, in an energy-conserving fashion, to give the adiabatic 
lapse rate.  An explicit description of this scheme is given in 
\citet{manabe&wetherald1967}.

For atmospheres experiencing a substantial amount of condensation, 
latent heat release causes the lapse rate in the convective portion 
of the atmosphere to be smaller than the dry adiabatic lapse rate.  
For example, on Earth the dry adiabatic lapse rate is 
$g/c_P = 9.8$~K~km$^{-1}$, whereas the actual average lapse 
rate\footnotemark in the troposphere is about 6.5~K~km$^{-1}$, 
which is smaller due to water vapor condensation.  So, in the 
original work by \citet{manabe&strickler1964}, the convective 
portion of the atmosphere was simply relaxed to the measured lapse 
rate instead of the dry adiabatic lapse rate.

\footnotetext{According to the U.~S. Standard Atmosphere (1976), 
NOAA-S/T76-1562.}

Of course, for a given background condensible gas, one expects that 
the influence of latent heating in the convective portion of an 
atmosphere to be less significant at lower temperatures, as cooler 
temperatures imply smaller amounts of the condensible.  The moist 
adiabatic lapse rate \citep[][p. 249]{satoh2004} captures these 
physics by accounting for the effects of latent heat release on a 
parcel or air that experiences condensation while being lifted 
adiabatically through the background atmosphere. While multiple
species indeed condense in ultracool dwarf atmospheres, in practice
the contribution of latent heating is only of first order importance
for water clouds, so we forego a deeper discussion of this issue here.


The strength of the convective adjustment approach is that it 
is computationally efficient---the thermal structure is simply 
taken to lie on a dry or moist adiabat.  Given this assumed structure, 
numerical models need only determine the location of the 
radiative-convective boundary, and then ensure that temperature and 
thermal flux are continuous across this boundary.  


Convective adjustment has several notable shortcomings beyond the 
difficulties associated with condensible gases discussed above.  First, 
convective adjustment may not realistically capture the dynamical 
response of an atmosphere to vertical motions.  Super-adiabatic layers 
are immediately adjusted onto an adiabat, without regard for any vertical 
mixing timescales.  This assumption is justified when only a steady-state 
solution is required, but could become problematic when studying 
time-dependent atmospheric variability (see 
Section~\ref{subsec:variability}).  Additionally, convective adjustment 
does not straightforwardly predict an eddy diffusion coefficient, which 
is a critical input to chemistry and cloud models.  One potential bypass 
for this shortcoming is to follow the scaling arguments of 
\citet[][Equation~16]{gierasch&conrath1985}, who use mixing length 
theory to derive an expression for the eddy diffusivity.


\subsection{Mixing Length Theory}
\label{subsec:mlt}

Mixing length theory \citep{prandtl1925,vitense1953,bohmvitense1958} 
uses dimensional and scaling arguments to model convective fluxes, 
and has seen successful applications and development in both the 
astrophysical and planetary science literatures \citep{henyeyetal1965,
gierasch&goody1968,spiegel1971,castellietal1997}.  In this approach, 
convection is modeled as the diffusion of heat through a turbulent 
medium.  The turbulent diffusivity is taken as
\begin{equation}
  K_{h} = w l \ ,
\label{eqn:mldiffus}
\end{equation}
where $l$ is a characteristic length over which turbulent mixing 
occurs (i.e., the mixing length), and $w$ is a characteristic upward 
transport velocity.  For a parcel lifted adiabatically 
through $l$, the temperature difference between the parcel and its 
surroundings is $\Delta T = -l\left(dT/dz + g/c_P\right)$, where 
$dT/dz$ is the lapse rate for the parcel, so that the convective 
flux is
\begin{equation}
  F_{c} = w \rho c_P \Delta T = - \rho c_P K_{h} \left(\frac{dT}{dz} + \frac{g}{c_P}\right) \ ,
\label{eqn:mlflux}
\end{equation}
where, of course, a convective heat flux is only present when the 
atmosphere is unstable to convection (i.e., $-dT/dz > g/c_P$).

Models usually assume that the mixing length is proportional to the 
pressure scale height, with $l = \alpha H$, where $\alpha$ is a free 
parameter, typically of order unity \citep[see][for exploration 
of this parameter]{burrowsetal2001,baraffeetal2002,
robinson&marley2014}.  The characteristic transport velocity can be 
derived from buoyancy force arguments 
\citep[][p. 62]{kippenhahnetal2012}, yielding 
$w = l \left[ -g/T \left(dT/dz + g/c_P\right) \right]^{1/2}$.  Thus, 
the turbulent diffusivity in a convectively unstable portion of 
the atmosphere is given by
\begin{equation}
  K_{h} = l^{2} \left[-\frac{g}{T} \left(\frac{dT}{dz} + \frac{g}{c_P}\right) \right]^{1/2} \ .
\label{eqn:mldiffusivity}
\end{equation}

\begin{textbox}
\section{CONVECTIVE EFFICIENCY} 
  Using the turbulent diffusivity in Equation~\ref{eqn:mldiffusivity}, 
  we can write the convective flux (Equation~\ref{eqn:mlflux}) as
  \begin{equation*}
    F_c = \rho c_P l^2 \left( \frac{g}{T} \right)^{1/2} \left[-\left(\frac{dT}{dz} + \frac{g}{c_P} \right) \right]^{3/2} \ .
  \end{equation*}
  If we then assume that the atmospheric lapse rate is some  
  fraction larger than the dry adiabatic lapse rate, with
  \begin{equation*}
    \frac{dT}{dz} = -(1+f)\frac{g}{c_P} \ ,
  \end{equation*}
  and that the mixing length is simply the pressure scale 
  height, then the convective flux is 
  \begin{equation*}
  \begin{array}{rl}
    F_c & = \rho c_P H^2 \left(\frac{g}{T}\right)^{1/2}\left(f\frac{g}{c_P}\right)^{3/2} \\
        & = f^{3/2} R_{s} P \left( \frac{T}{c_P} \right)^{1/2} \ ,
  \end{array}
  \end{equation*}
  where the second step used the ideal gas law and the definition 
  of the pressure scale height.  Notably, the gravitational 
  acceleration has cancelled out of this expression.
  
  We can now get a sense for how efficient convection is at 
  returning a super-adiabatic lapse rate to near the dry adiabatic 
  value.  In the convective region of a mid-L dwarf 
  ($T_{\rm eff}=1900$~K, $g=100~\rm{m~s^{-1}}$), with $P=10$~bar, 
  $T=3500$~K, and $c_P = 10^{9}~\rm{erg~g^{-1}~K^{-1}}$, and with 
  $R_s = 4\times10^{7}~\rm{erg~g^{-1}~K^{-1}}$, we find 
  $F_{c} = 7\times10^{11}f^{3/2} ~\rm{erg~cm^{-2}~s^{-1}}$.  This 
  flux is impressively large---even for a 1\% deviation from the 
  dry adiabat (i.e., $f=0.01$), the convective flux is large 
  enough to carry the entirety of the internal heat flux, or 
  $\sigma T_{\rm eff}^4 = 7.4 \times10^{8}~\rm{erg~cm^{-2}~s^{-1}}$.
  Thus, even very minor deviations of the lapse rate from the dry 
  adiabatic value will be promptly eliminated.
\end{textbox}

Mixing length approaches benefit from being simple and computationally 
efficient.  Additionally, these techniques directly compute the convective 
heat flux and turbulent diffusivity (Equations~\ref{eqn:mlflux} and 
\ref{eqn:mldiffusivity}, respectively).  The latter gives information 
about mixing processes in the convective portion of the atmosphere, and 
can be used in chemical and aerosol transport models.  Also, since 
convective heating and cooling rates can be computed from the convective 
flux profile, mixing length models can be used to study time-dependent 
atmospheric processes and phenomenon.

A common criticism of mixing length theory is that the mixing 
length, $l$, is ambiguous.  Additionally, the theory does not directly 
predict the thermal structure in the convective regime (unlike 
convective adjustment which imposes the adiabat),
but, instead, must solve for the thermal structure 
using the relevant fluxes and heating rates.  

\subsection{Beyond One-Dimensional Models of Convection}

Studies of stellar structure and work in the Earth and planetary 
sciences have benefited from two- and three-dimensional 
modeling of convection, and from comparing these models to 
one-dimensional methods for simulating convection.  For example, 
\citet{chanetal1982} performed two-dimensional simulations of 
turbulent convection in the deep atmospheres of stars and noted 
that the vertical velocities in their models were correlated 
over a characteristic length comparable to the pressure scale 
height, which lends some support to the mixing length picture 
outlined above.  \citet{cattaneoetal1991} used three-dimensional 
models to study convection in an astrophysical setting, and 
noted that local mixing length theories can adequately represent 
the turbulent transport of energy.


The brown dwarf and irradiated giant planet literature has seen 
very little development of convection-resolving models.  Numerous 
global circulation models (GCMs) for hot Jupiters 
\citep[see, e.g., review by][]{showmanetal2008} and brown dwarfs 
\citep{showman&kaspi2013,zhang&showman2014} exist.  However, only 
\citet{freytagetal2010} have applied convection-resolving models 
to brown dwarfs in their study of dust transport in cool and 
sub-stellar atmospheres.  Application of models and techniques 
from the Earth, planetary, and stellar literatures may, then, 
prove fruitful in advancing our understanding of convection in 
brown dwarf and giant exoplanet atmospheres.

\section{Chemistry}
\label{sec:chemistry}

In addition to convection and radiative transfer, of course,
a key characteristic of an atmosphere is its chemical makeup, which 
can in turn affect the computation of the thermal structure as gas  
abundances influence many processes (such as, e.g., opacities and, 
thus, radiative energy transport). In this section we briefly review 
influences on gas concentrations and discuss how the atmospheric 
composition is computed in the context of one-dimensional modeling. 
More extensive reviews of this topic can be found in 
\citet{lodders&fegley2006} and \citet{burrowsetal2001}.

\subsection{Abundances}

Of central importance to modeling the chemistry throughout a 
brown dwarf or giant planet atmosphere are the abundances of 
the underlying elements that make up the more complex molecules 
that form in these atmospheres.  These abundances are affected 
by two key processes.  First, the overall elemental abundances 
for the object determine the baseline distribution of elements.
Second, the formation and rainout of condensates will influence 
the availability of certain elements, possibly starving upper 
atmospheric layers (i.e., those at lower pressures and cooler 
temperatures) of certain elements.

\subsubsection{Elemental}

A key assumption in any atmosphere model is the underlying 
elemental abundances. The most important individual abundances 
being those of carbon and oxygen (C and O), as $\rm H_2O$, 
CO, and $\rm CH_4$ are key absorbers whose concentrations are 
controlled by the availability of C and O.  Abundances are 
typically referenced to those of the Sun, and an assumption 
of `solar abundances of the elements' is usually the starting 
point for atmospheric modeling.  \citet{asplundetal2009} 
reviews the challenges in defining such an abundance set.  
Unfortunately the solar C and O abundances are uncertain, and 
the generally accepted values have varied with time, even 
over the short history of ultracool dwarf modeling 
\citep[e.g.,][]{anders&grevesse1989,allendeprietoetal2001,
allendeprietoetal2002,asplundetal2009,caffauetal2011}.  For 
this reason any model comparison between different modeling 
groups must begin with a comparison of the assumed elemental 
abundances. \citet{barmanetal2011}, for example, utilize
the abundances of \citet{asplundetal2009} which are very similar 
to the \citet{lodders2003} abundances utilized by the Marley/Saumon
group.

Not all atmospheres will have solar abundances of course. 
Individual brown dwarfs or giant planets will sport a variety 
of compositions, and the defining characteristic of extrasolar 
giant planets may well be atmospheric compositions that depart 
from that of their primary stars. Thus any given atmosphere model
must make some choice for the initial elemental abundances.

An additional important aspect of abundance determination is 
the C/O ratio, which not only affects the relative abundance of 
the major carbon- and oxygen-bearing species, but other 
compounds as well.  As the C/O ratio increases towards unity, 
the condensation temperatures of oxides and silicates fall and 
C-bearing compounds become more prevalent 
\citep[see discussion in ][and references therein]{lodders2003}.  
Indeed the accepted `solar' value of C/O has ranged from 0.4 to 
0.6, and is known to vary amongst other stars 
\citep[for a recent discussion, see][]{fortney2012}.  Thus the 
assumed model C/O ratio should also always be noted when 
presenting modeling results.  \citet{fortney2012} has 
suggested that brown dwarf spectra are especially well-suited to 
measuring the C/O ratio in the solar neighborhood.  In the extreme 
case, if a brown dwarf were to be formed with atmospheric 
$\rm C/O > 1$, the entire atmospheric chemistry would be grossly 
altered, with essentially all available O going to form CO 
instead of $\rm H_2O$, with the excess C forming $\rm C_2$ and 
HCN.  However, no such brown dwarfs have yet been found and we 
neglect this possibility here.

\subsubsection{Rainout}
\label{subsubsec:rainout}

A crucial distinction between modeling approaches arises in 
the treatment of the chemistry of condensates. Once a condensate is 
formed, two limiting cases can be imagined.  In one case the solid 
grain or liquid drop will continue to chemically interact with the 
surrounding gas to arbitrarily low temperatures. This is a description 
of `true' chemical equilibrium where a given initial set of elements is 
always assumed to be in chemical equilibrium at a specified temperature 
and pressure.  This assumption was explored by \citet{burrows&sharp1999}, 
and is implicit in the COND and DUSTY models \citep{chabrieretal2000,
allardetal2001}. In the other extreme, the condensate can be imagined to 
rain out of the atmosphere, precluding further reactions with the 
neighboring gas, a limit sometimes called `rainout chemistry' (although 
`sedimentation chemistry' is probably a better phrase, `rainout' has 
become ingrained in the literature).

Discussions of rainout and the gas and cloud chemistry in sub-stellar 
atmosphere can be found in \citet{lodders1999}, \citet{burrows&sharp1999}, 
\citet{lodders2004}, and \citet{lodders&fegley2006}.  An important 
consequence of the rainout chemistry is that reactions that would be expected 
to take place between condensed species and the gas at temperatures cooler 
than the condensation temperature are excluded.  For example, without the 
rainout assumption, iron grains would react with $\rm H_2S$ gas to form 
FeS, thus removing $\rm H_2S$ from the atmosphere at around 500~K. In 
contrast, under the rainout assumption the iron grains are presumed to 
fall out of the atmosphere after they form at around $2000$~K, thereby 
precluding further reaction with $\rm H_2S$, which then stays in the gas 
phase. Since $\rm H_2S$ is indeed observed in Jupiter's atmosphere, this 
is presumably the correct limit \citep{barshay&lewis1978,
fegley&lodders1994, niemannetal1998}.

Another example of the importance of rainout arises in the alkali 
chemistry and the interaction of the alkali metals with silicate grains 
that initially form around 1800~K. Without the rainout assumption, 
gaseous sodium and potassium undergo a series of reactions with these 
grains that ultimately produce alkali feldspar ($\rm [Na,K]AlSi_3O_8$), 
thereby removing Na and K from the gas at around 1400~K.  With 
rainout the silicate grains that form near $1800$~K fall,  do not react with 
the atmosphere at cooler temperatures, and alkali feldspar does not form 
\citep{lodders1999}. In this limit sodium remains in the atmosphere 
until about 700~K where it forms $\rm Na_2S$ and KCl. 
\citet{marleyetal2002} argue that the far red optical spectra of 
T dwarfs support the rainout hypothesis.  More recently, 
\citet{morleyetal2012} convincingly demonstrate the presence of such 
alkali condensates in cool T dwarf spectra. Taken together these 
lines of evidence support the rainout limit.

\subsection{Equilibrium Chemistry}

Given an abundance of all relevant elements and appropriate 
thermodynamic data, the abundances of gaseous species at any 
pressure and temperature can be computed.  Different groups utilize 
one of primarily two different methods for solving for the gas 
abundances in chemical equilibrium. The quality of the results 
ultimately depends upon the number of species considered and the 
veracity of the underlying thermodynamic data, which are not always 
available for all reactions and species of interest.

The two methods for finding equilibrium are commonly termed 
`mass action' and `Gibbs minimization.'  The mass action approach 
utilizes the equilibrium constants for all relevant chemical 
reactions, as well as mass conservation, to find the abundance of 
each molecule and condensate of interest at a given pressure and temperature. 
Gibbs minimization solves for the mixture of species that has 
the lowest Gibbs free energy, given the pressure, temperature, and 
assumed elemental abundances. With identical thermodynamic 
data---and the same assumptions regarding rainout---the two 
approaches should give identical results. However, in practice, typically because
of computational limitations,
mass conservation is not always achieved by Gibbs minimization 
methods. Computational considerations also often limit the number 
of elements and compounds that can be included in this approach.
The best source for learning about the details of chemical 
equilibrium calculations is \citet{vanzeggeren&storey1970}.

For specific application to brown dwarf atmospheres, Section~4 
of \citet{sharp&burrows2007} provides ample discussion and 
examples of the free-energy minimization procedure.  Likewise, 
the mass action approach is described by 
\citet{fegley&lodders1994} in the context of Jupiter's atmosphere, 
and the method has been used to study the chemistry of the major 
elements in brown dwarf and exoplanet atmospheres by 
\citet{lodders&fegley2002} and \citet{visscheretal2006}.

Regardless of the method employed, the curation and vetting of 
relevant thermodynamic data is a time-consuming process. For 
example Lodders, Fegley, and collaborators follow approximately 
2,000 gaseous and 1,700 solid or liquid species in their code. 
Construction of the relevant thermodynamic database poses a 
substantial ``barrier to entry" for new scientists interested 
in computing their own atmospheric chemistry.

Examples of equilibrium gas abundances, or ratios of gas 
abundances, for several key species of interest 
\citep[from the rainout models of][]{lodders&fegley2002,
visscheretal2006} are shown in Figure~\ref{fig:abunds}, along 
with model temperature profiles for a few representative objects.
The effects of condensation removing Ti and Fe can clearly be 
seen.  Also, the ratios of the concentrations of CH$_4$ to CO 
and NH$_3$ to N$_2$ demonstrate the preferred chemical state 
of carbon and nitrogen at any given temperature and pressure.

\subsection{Disequilibrium Chemistry}

Although Jupiter's atmosphere is primarily in the expected state of 
chemical equilibrium, its atmospheric composition departs 
from equilibrium in some important respects, and these provide a 
guide for understanding similar excursions in brown dwarf and 
exoplanet atmospheres. First, convection is known to mix CO 
from the deep atmosphere, where it is the favored carbon-bearing 
species, up to the observable atmosphere, where $\rm CH_4$ should 
be overwhelmingly dominant \citep{prinn&barshay1977,bezardetal2002}.
This can happen in situations where the mixing time for species to 
be transported by convection is shorter than the timescale for a 
species to come into chemical equilibrium with its surroundings.

Because carbon is tied to oxygen by a strong triple bond in CO, it
can be difficult for atmospheric chemical processes to convert
the species to methane, even when this would be thermodynamically favored.
  Figure~\ref{fig:ch4}  illustrates the somewhat 
tortuous chemical pathway linking CO to ${\rm CH_4}$. For this reason 
\citet{fegley&lodders1996} predicted that CO would be discovered in 
the atmospheres of what were then termed the `methane dwarfs' (now 
the T dwarfs) and, indeed, several observational studies 
\citep{nolletal1997,oppenheimeretal1998,saumonetal2000,geballeetal2009} found 
excess CO in these objects.  Likewise, excess CO was found in the 
atmospheres of the cooler transiting planets and the repeated 
``discovery'' of chemical disequilibrium became, for a while, 
something of an industry within the exoplanet community.

The efficiency with which CO can be transported upwards depends 
both upon the vigor of eddy mixing and by the chemical equilibrium 
timescale. The former is usually parametrized in models with 
$K_{h}$, the eddy diffusivity for heat discussed earlier, while the 
latter depends upon the details of the chemical pathway. The usual 
approximation used in modeling this process is that the atmospheric 
abundance of CO is fixed at the point where the chemical equilibrium 
timescale is equal to the mixing timescale, commonly known as the 
`quench' approximation \citep[e.g.,][]{fegley&lodders1996,
saumonetal2000,lodders&fegley2002,hubeny&burrows2007}. An example is 
shown in Figure~\ref{fig:quench}.  For this model a pure equilibrium 
calculation would find that carbon in the atmosphere should be almost 
entirely in the form of methane above about 1~bar when the thermal 
profile has entered in the methane stability field. However, in the 
presence of vertical mixing, CO is transported upwards and can be 
the dominant species when mixing is vigorous. The most comprehensive discussions
of this process can be found in \citet{hubeny&burrows2007} and
\citet{zahnle&marley2014}

\begin{textbox}
\section{QUENCHING}
  Turbulent motions will tend to cause the vertical mixing 
  of chemical species throughout an atmosphere.  Chemical 
  disequilibrium can be established if mixing supplies a 
  species to an atmospheric level faster than the local 
  chemical reactions can remove it.  The concentration 
  of the species is then locked into its value from deeper in 
  the atmosphere, in a process called `quenching.'  The 
  quenching level is roughly where the mixing timescale 
  equals the chemical loss timescale, or 
  $t_{\rm mix} = t_{\rm chem}$.
  
  Given the characteristic turbulent velocity, $w$, mixing 
  over a pressure scale height, $H$, occurs at a timescale
  \begin{equation*}
    t_{\rm mix} = \frac{H}{w} = \frac{H^2}{K_h} \ ,
  \end{equation*}
  where the second step used Equation~\ref{eqn:mldiffus} 
  with $l=H$ (but see \citet{smith1998} for an alternative approach
  to the use of $H$ as well as the discussion in \citet{zahnle&marley2014}).
  The value of $K_h$ can either be evaluated 
  from a thermal structure model 
  using mixing length 
  theory (Section~\ref{subsec:mlt}), or simply specified as a 
  constant.  Above the convective region, the eddy diffusivity 
  from convective mixing is formally zero, but other processes 
  can still drive mixing (e.g., gravity waves), as have been 
  studied for the Solar System planets and brown dwarfs
  \citep[e.g.,][]{bishopetal1995, freytagetal2010}.

  Chemical loss timescales are more difficult to specify.  As 
  reviewed by \citet{zahnle&marley2014}, one option is to 
  determine the loss timescale by associating it with the 
  rate-limiting step in the loss reaction (although determining 
  this step can be difficult).  Alternatively, gridded results 
  from a large suite of photochemical models can be used to 
  determine a functional form of $t_{\rm chem}$.  This approach 
  was adopted by \citet{zahnle&marley2014}, who propose using 
  an Arrhenius-like rate, 
  \begin{equation*}
    t_{\rm chem} = A P^{-b} {\rm [M/H]}^{-c} e^{B/T} \ ,
  \end{equation*}
  where $\rm [M/H]$ is the metallicity, and $A$, $B$, $b$, 
  and $c$ are coefficients of the fit.  For the CO-CH$_4$ 
  system, these authors find
  \begin{equation*}
    t_{\rm chem} = 1.5\times10^{-6} P^{-1} {\rm [M/H]}^{-0.7} e^{42000/T}~{\rm sec} \ .
  \end{equation*}  
\end{textbox}
 
\citet{zahnle&marley2014} review the disequilibrium 
mixing literature, and argue that this process is more gravity-dependent 
than previously recognized. They suggest that, in the lowest mass 
extrasolar planets, such as those that are expected to be discovered 
by the Gemini Planet Imager (GPI) and the Spectro-Polarimetric 
High-contrast Exoplanet REsearch (SPHERE) surveys for young planets 
\citep{macintoshetal2008,beuzitetal2008}, CO may be the dominant 
carbon bearing species until temperatures as low as about 600 K.  
Furthermore, CO is not the only species that can be in disequilibrium 
because of mixing. Similar arguments apply to the equilibrium 
between $\rm N_2$ and $\rm NH_3$, as well as the chemistry of 
$\rm PH_3$ and $\rm GeH_4$.  Indeed signs of $\rm NH_3$ 
under-abundance compared to the expectations of equilibrium 
chemistry have already been reported in brown dwarfs \citep{saumonetal2006}, which has 
also been reviewed by \citet{zahnle&marley2014}.

One final important disequilibrium process should be noted---under 
the presence of incident ultraviolet radiation, methane, hydrogen 
sulfide, and other species may be photochemically destroyed, leading 
to the creation of more complex molecules. In cool Solar System 
atmospheres the resulting species often form haze particles which 
themselves become important opacity sources. A thorough 
post-{\em Voyager} review of methane photochemistry in the context 
of giant planets is \citet{bishopetal1995}, while a concise, 
informative overview of giant planet C, N, O, and S photochemistry 
can be found in \citet{moses2000}.  \citet{zahnleetal2009} consider 
photochemical processes that are unique to giants much warmer than 
Jupiter, specifically studying sulfur photochemistry in hot Jupiter 
atmospheres.  In general, photochemistry in any giant more heavily 
irradiated than Jupiter will be much more complex than in Solar System 
giants since species that are usually trapped in tropospheric clouds, 
such as $\rm NH_3$, $\rm H_2S$ and $\rm H_2O$, will instead be present 
in the upper atmosphere where the incident ultraviolet flux is far 
higher. 

\section{Gas Opacity}
\label{sec:opacity}

The transition from hot, atomic, continuum-opacity dominated solar-like
stars to the cool, mostly neutral, and molecular atmospheres of the
late M, L, T, and Y dwarfs is nowhere more apparent than in the
opacities which must be considered to model these atmospheres.
Properly accounting for the molecular opacity of key atmospheric
constituents is a major task of cool atmosphere modeling.  
This section reviews the important gaseous opacities in cool 
atmospheres, focusing on the treatment of molecular absorption lines. 
Techniques for solving the RTE given a complex spectrum of gas 
opacities are also discussed.

\subsection{Line Strengths and Shapes}

Each  molecular absorption band is formed from a collection of 
individual absorption lines, as shown in Figure~\ref{fig:abscoeff}, 
with each line marking a transition in the coupled rotational-vibrational 
energy state of the molecule.  The shapes and strengths of these lines 
depend on a number of processes, including structural properties of the 
molecule as well as the local atmospheric conditions where the line is 
formed. 

In general, an individual spectral line is described by three key 
parameters---the line position (i.e., the frequency or wavenumber at 
line center, $\nu_0$, typically in units of cm$^{-1}$), the line 
strength ($S=S(T)$, in units of cm$^{2}$ molecule$^{-1}$ cm$^{-1}$, 
or equivalent), and the line shape function ($f(\nu-\nu_0)$, in 
units of 1/cm$^{-1}$).  The frequency-dependent absorption 
coefficient for the line, $k_{\nu}$, is then expressed as 
\begin{equation}
  k_{\nu} = Sf(\nu-\nu_0) \ ,
\end{equation}
where the line shape function is normalized such that
\begin{equation}
  \int_{-\infty}^{+\infty} f(\nu-\nu_0) d\nu = 1 \ .
\label{eqn:lsnorm}
\end{equation}
Thus, the integrated area under an individual line is constant 
(equal to $S$), such that broader absorption lines increase opacity 
in line wings at the expense of opacity near line center.

Large databases, usually referred to as line lists, compile the 
necessary information for computing absorption line spectra.  These 
databases are based on either lab measurements or quantum chemistry 
simulations.  As a number of line parameters depend on temperature, 
line lists are typically referenced to a standard temperature (e.g., 
296~K).  Commonly used line lists include the HITRAN 
\citep{rothmanetal1987,rothmanetal2013}, HITEMP 
\citep{rothmanetal1995,rothmanetal2010}, and ExoMol 
\citep{tennyson&yurchenko2012} databases, where the latter 
is more appropriate for the range of temperatures encountered in 
sub-stellar atmospheres, but only includes data for H$_2$O, CO$_2$, 
CO, NO, and OH.  Sources of line lists for more exotic species, and 
discussion of how to implement these databases, are reviewed and/or 
tabulated in several recent papers 
\citep{sharp&burrows2007,freedmanetal2008,freedmanetal2014,lupuetal2014}.

Given a line strength $S_0$, defined at a standard temperature $T_0$, 
the temperature-dependent line strength can be computed from 
\citep{mcclatcheyetal1973}
\begin{equation}
  S(T) = S_0 \frac{Q(T_0)}{Q(T)}  
    \exp \left[ \frac{E^{\prime\prime}}{k_B} \left( \frac{1}{T_0} - \frac{1}{T} \right) \right] \frac{1 - \exp\left( -hc\nu_0/k_BT \right)}{1 - \exp\left( -hc\nu_0/k_BT_0 \right)} \ ,
\end{equation}
where $Q$ is the internal molecular partition function for vibrational 
and rotational states, $E^{\prime\prime}$ is the lower energy level for 
the rotational-vibrational transition (commonly supplied in line lists), 
$c$ is the speed of light ($c = 2.998 \times 10^{10}$ cm s$^{-1}$), and 
$h$ is Planck's constant ($h = 6.626 \times 10^{-27}~\rm{erg~s}$).  
The exponential term in the middle of this expression is the familiar 
Boltzmann distribution for energy states.  Standard methods exist for 
computing partition functions \citep[e.g.,][]{gamacheetal1990}, which 
describe the temperature-dependent partitioning of rotational-vibrational 
energy states.  The combination of the Boltzmann factor and the partition 
function weight the line strength by the probability of finding 
molecules of a given energy state within the ensemble of all states.  Finally, 
the ratio of terms at the end of this expression corrects for `stimulated 
emission,' where a molecule is de-excited from the higher energy state 
($E^{\prime\prime}+hc\nu_0$) via an interaction with a photon of energy 
$hc\nu_0$, thereby giving off two photons with the same energy as the 
initial photon.  Viewed as negative absorption, stimulated emission makes 
the overall strength of the absorption line slightly smaller.

The line shape depends critically on temperature, pressure, and 
atmospheric composition through so-called foreign broadening.  At 
lower pressures, where molecular collisions are relatively infrequent, 
line broadening is dominated by Doppler effects from thermal motions, 
which has a characteristic Gaussian line shape function, 
\begin{equation}
  f_{D}(\nu-\nu_0) = \frac{1}{\alpha_{D}} \sqrt{\frac{\ln 2}{\pi}}
    \exp \left[ - \frac{ \ln 2 \left(\nu - \nu_0 \right)^2}{\alpha_D^2} \right] \ ,
\end{equation}
where $\alpha_D$ is the Doppler line half-width at half-maximum 
(HWHM), given by
\begin{equation}
  \alpha_D = \frac{\nu_0}{c}\sqrt{2\ln 2 \frac{k_{B}T}{m}} \ ,
\end{equation}
where $m$ is th emolecular mass.  Thus, Doppler broadening is 
more effective at higher temperatures and for low mass molecules, 
which stem from an assumed Maxwellian distribution of molecular 
speeds.  Some atmospheric models, most notably the PHOENIX 
models \citep{allard&hauschildt1995,barmanetal2001}, add a 
`micro-turbulent velocity' to the thermal velocity, where the former 
is taken to scale with the characteristic convective velocity, 
$w$ (see Section~\ref{subsec:mlt}) \citep{husseretal2013}.  Physically, 
this scaling represents a turbulent cascade from the macro-scale to 
the micro-scale, where motions can broaden individual absorption 
lines.  These velocities are typically of order 1~km~s$^{-1}$ which, 
as an example, is comparable to the gas kinetic velocity of water vapor 
molecules at roughly 1,000 K.  In practice, the micro-turbulent velocity 
has been used as a parameter to better reproduce observed spectra and to 
account for incomplete linelists \citep{kurucz1996}.  

Deeper in the atmosphere, at pressure levels from which most thermal 
emission spectra emanate, line broadening is dominated by molecular 
collisional effects on emission and absorption.  The line shape for 
so-called `pressure broadened' lines is typically taken as a 
Lorentzian function, 
\begin{equation}
  f_{L}(\nu-\nu_0) = \frac{1}{\pi} \frac{\alpha_L}{\left(\nu - \nu_0 \right)^2 + \alpha_L^2} \ ,
\end{equation}
where $\alpha_L$ is the pressured-broadened HWHM.  Typically 
$\alpha_L$ is computed from a width parameter, $\gamma$, obtained from 
a line list, where 
\begin{equation}
  \alpha_L = \gamma P \left( \frac{T_0}{T} \right)^{n} \ ,
\end{equation}
where $n$ is a temperature-dependence exponent, commonly supplied in 
line lists.  The width parameter is different for self-broadening, 
where collisions are from molecules of the same species as the 
absorber, versus foreign-broadening, where other species dominate the 
molecular collisions.  In some cases, most notably for H$_2$O in 
Earth's atmosphere and CO$_2$ in Venus' atmosphere, the Lorentzian 
line shape has been found to either over- or under-estimate 
absorption in the far wings of pressure-broadened lines.  In these 
situations, it is common to apply a line shape correction, called a 
`$\chi$-factor', to best reproduce the observed data 
\citep[see, e.g.,][]{wintersetal1964,fukaborietal1986,pollacketal1993,
meadows&crisp1996,mlaweretal2012}.

Of course, there are large portions of the atmosphere where lines 
are influenced by both Doppler and pressure broadening.  Here, a 
convolution of the line shapes for these two processes is used, 
called the Voigt line shape, which is given by 
\begin{equation}
  f_{V}(\nu-\nu_0) = \frac{\alpha_L}{\pi^{3/2}} 
    \int_{-\infty}^{+\infty} \frac{\exp\left( -y^2 \right)}
      {\left( \nu-\nu_0 - \alpha_D \cdot y \right)^2 
      + \alpha_L^2} dy \ .
\end{equation}
\citet{schreier2011} discusses efficient methods for computing
this expression.
Figure~\ref{fig:lineshape} compares lineshapes for Doppler 
broadening (both with and without a micro-turbulent velocity), 
pressure broadening, and the combination of these into a Voigt 
lineshape.

An important source of uncertainty in opacity calculations is that 
the width parameter and the temperature-dependence exponent are 
usually measured for terrestrial air as the broadening gas, and are 
expected to have different values when other gases (e.g., H$_2$) are 
the primary source of pressure broadening. Pressure broadening 
parameters for either H$_2$ or H$_2+$He as the background gas are 
available for a limited set of molecules, and do not typically span a 
wide range of wavelengths \citep{bulaninetal1984,lemoal&severin1986,
margolis1993,margolis1996,brown&plymate1996,gamacheetal1996,gabard2013}. 
These data are commonly extrapolated to cover the full range of 
wavelengths of interest.  When broadening parameters are only available 
for terrestrial air, width parameters and temperature-dependence exponents 
are sometimes adjusted by a constant factor derived from insight and a 
limited set of available data.

Additionally, the power-law temperature dependence in the expression 
for the Lorentz HWHM likely does not hold over a wide range of 
temperatures.  This is especially problematic for astrophysical 
applications, where the physical environment is typically quite 
different from the laboratory conditions where line parameters are 
measured.  Thus theoretical modeling of line shape parameters is 
commonly employed, where a wide range of combinations of temperatures, 
pressures, and broadening gases can be investigated 
\citep{gabard2013,gamache&lamouroux2013}.

Even though most of the theory outlined above is straightforward, 
issues arise due to the scope of the problem.  Line lists can contain 
more than $10^{9}$ or even $10^{10}$ transitions, which makes assembling 
opacities computationally expensive.  Efficiency can be gained by omitting 
weak lines from opacity calculations, which is most effective when gas 
concentrations are known (or can be estimated) a priori.  Additionally, 
it isn't feasible to sum the contributions from such large numbers of 
lines out to arbitrarily large distances from line centers.  Thus, line 
shape profiles are commonly truncated at some distance from line center, 
which necessitates a re-scaling of the line shape to maintain the 
normalization of Equation~\ref{eqn:lsnorm}, thus preserving the 
integrated line strength \citep{sharp&burrows2007}.

Finally, while the basic procedure for applying molecular opacity 
databases to atmospheric modeling problems can be easily outlined, there 
are many subtleties (for example detailed choices about line broadening, 
or recognizing errors, which are not uncommon, in the databases).  Decision 
making in these cases benefits from deep experience in the construction and
use of line databases. However, relatively few young astronomers 
specialize specifically in the study of molecular opacity issues, and
this lack of new talent may represent a problem for the field in the 
future.

\subsection{Alkali Opacity}

At the time of the first T dwarf discoveries, a major shortcoming of 
the T dwarf models was that they predicted far too much flux in the 
far red ($\sim 900$ to $1000\,$nm) compared to observations. One early 
suggested solution even relied upon a high altitude, scattering haze 
to account for the discrepancy \citep{griffithetal1998}.  In fact the 
missing opacity source was soon found to be the highly pressure 
broadened wings of resonant Na and K lines \citep{burrowsetal2000}.

Correctly computing the opacity of these resonant lines is not 
straightforward and mismatches between models and data near 1000~nm 
are commonly attributed to shortcomings in the treatment of the 
resonant alkali lines. Recent work by \citet{allardetal2012} on 
these line shapes should improve the situation, but there are many 
subtleties in the use of the theoretical results and there is 
clearly room for more study of the alkali opacities.

\subsection{Continuum Gaseous Opacities}

In addition to line and molecular opacities, there are several 
continuum opacity sources that must be considered. Foremost among 
these is the collision-induced absorption (or CIA, sometimes 
termed pressure-induced absorption) opacity arising from transitions 
within supramolecules transiently created during collisions of 
$\rm H_2$ molecules with other other gas species, predominantly 
$\rm H_2$ and He. From the uncertainty principle, because collision 
timescales are very short, energy level transitions within the 
short-lived supramolecules are not sharply defined and the resulting 
opacity is generally smoothly varying with wavelength. A recent 
compendium can be found in \citet{richardetal2012}.

While electron densities are generally quite low in these 
atmospheres (Figure~\ref{fig:abunds}), continuum opacity sources 
associated with free electrons should be accounted for. These 
include bound-free absorption by H and $\rm H^-$ and free-free 
absorption by H, $\rm H_2$, $\rm H_2^-$, and $\rm H^-$ (see 
\citet{lenzunietal1991}), as well as electron scattering.

Finally, Rayleigh scattering in the gas is also important for 
problems involving incident starlight. Rigorous calculations also 
account for Raman scattering, an important process in the UV for 
giant planet atmospheres \citep{pollacketal1986}.

\subsection{Opacities and the RTE}
\label{subsec:opacsRTE}

As discussed by \citet[][p. 125]{goody&yung1989}, radiative 
calculations involve four distinct scales of wavelength-variation: 
the scale at which the Planck function varies, the scale of gas 
rotational-vibrational absorption bands, the scale of individual 
absorption lines, and, most finely, the `monochromatic' scale at 
which the absorption coefficient can be considered to be constant. 
Techniques exist for averaging wavelength-dependent atmospheric 
opacities across this full range of scales, with the purpose of 
increasing the speed at which net radiative heating rates can be 
determined.

\subsubsection{Gray Opacities}

Early studies of the structure and evolution of brown dwarfs 
\citep{lunineetal1986,lunineetal1989} and recent analytic models 
of the thermal structure of irradiated planetary atmospheres 
\citep{hansen2008,guillot2010,robinson&catling2012,
parmentier&guillot2014} have used opacities averaged over the 
entire spectral range, or `gray' opacities, combined with the 
gray two-stream equations \citep[e.g.][p.~84]{andrews2010}, to 
compute thermal fluxes in sub-stellar atmospheres.  Typically 
either the Planck mean opacity, $k_{P}$, or the Rosseland mean 
opacity, $k_{R}$, are used, whose averaging are defined, 
respectively, by
\begin{equation}
  k_{P}(P,T) = \frac{\int_{0}^{\infty} k_{\nu}(P,T) B_{\nu}(T) d\nu} 
                    {\int_{0}^{\infty} B_{\nu}(T) d\nu} \ ,
\end{equation}
\begin{equation}
  k_{R}^{-1}(P,T) = \frac{\int_{0}^{\infty} k_{\nu}^{-1}(P,T) dB_{\nu}(T)/dT d\nu} 
                         {\int_{0}^{\infty} dB_{\nu}(T)/dT d\nu} \ .
\end{equation}
The Planck mean is weighted towards spectral regions with large 
opacities, yields the correct flux emitted by an atmospheric layer 
of temperature $T$, and is best applied in the higher, more optically 
thin regions of an atmosphere.  Conversely, the Rosseland mean is 
weighted towards spectral regions of low opacity, and is designed to 
ensure the radiation diffusion limit in the opaque regions deep in 
an atmosphere \citep[][p. 38]{mihalas1970}.

Tabulating Planck and Rosseland mean opacities as a function of 
temperature and pressure using pre-computed, high resolution gas 
opacities is a difficult and tedious task.  This is especially 
true for the Rosseland mean, whose weighting towards low-opacity 
spectral regions places extra emphasis on the poorly-understood 
far wings of spectral lines.  Earlier models relied on 
\citet{tsuji1971} for gray opacities.  More recently, 
\citet{freedmanetal2008,freedmanetal2014} presented tables of 
Rosseland and Planck mean opacities appropriate for use in brown 
dwarf and gas giant thermal structure models.

\subsubsection{Band Models}

At the scale of molecular bands, numerous so-called `band models' 
have been historically applied to determine thermal radiative 
fluxes in planetary and stellar atmospheres.  While not widely 
used today, band models take advantage of the overall smooth nature of 
band-averaged transmission functions (particularly true for
the relatively dense and cool brown dwarf and giant planet
atmospheres), $\mathcal{T}_{\Delta \nu}$, 
with 
\begin{equation}
  \mathcal{T}_{\Delta \nu} = \frac{1}{\Delta \nu} \int_{\Delta \nu} \mathcal{T}_{\nu} d\nu 
              = \frac{1}{\Delta \nu} \int_{\Delta \nu} e^{-\left[ \tau_{\nu}(P_{2}) - \tau_{\nu}(P_{1}) \right]} d\nu \ ,
\label{eqn:meantrans}
\end{equation}
where $P_{1}$ and $P_{2}$ are atmospheric pressures (with 
$P_{2} > P_{1}$), and the final step uses the definition of 
transmissivity.  The band-average transmission functions for an 
individual gas can be tabulated using high-resolution, 
wavelength-dependent opacities, and are often parameterized by a 
fit to a particular type of band model, such as the Godson, 
Malkmus, or Elsasser models \citep[][Chap.~4]{goody&yung1989}.  
The treatment of band-averaged transmission functions is more 
complicated in spectral regions where more than one absorbing 
gas have overlapping features 
\citep[][ Sec.~4.4.5]{pierrehumbert2010}, but, nevertheless, 
combined transmission functions can be used to solve the 
two-stream thermal RTE.

A particular type of band-model, the Just Overlapping Line 
Approximation (JOLA), was used in early studies of brown 
dwarf atmospheres by Allard and collaborators 
\citep[e.g.,][]{allard&hauschildt1995} and by Tsuji and 
collaborators \citep{tsuji1984,tsuji1994,tsuji2002}. 
\citet{carbon1979} reviews this and other types of band 
models.


\subsubsection{Correlated-$k$}

Spectral integrals of functions that have complicated variation 
in wavelength (e.g., Equation~\ref{eqn:meantrans}) are 
computationally expensive.  However, over a wavelength range 
where the source function and the scattering properties of the 
atmosphere can be taken to be roughly constant, a great deal of 
computational efficiency can be gained by using the 
{\it distribution} of opacities to replace the integral over 
frequency with a more well-behaved integral over a new dependent 
variable.  This, effectively, re-orders the opacities in 
wavelength space, creating a smooth, monotonic description of 
the absorption coefficients, and is the root of the 
$k$-distribution approach \citep{ambartzumian1936,kondratyev1965,
arking&grossman1972,domoto1974}.

Given the distribution function of gas opacities, $f(k)$ (where, 
then, the fraction of absorption coefficients between $k$ and 
$k+dk$ is $f(k)dk$), the mean transmission through an atmospheric 
layer (Equation~\ref{eqn:meantrans}) can be expressed as 
\citep[][Sec.~4.8]{goody&yung1989},
\begin{equation}
  \mathcal{T}_{\Delta \nu} = \int_{0}^{\infty} f(k) e^{-k\mathcal{M}} dk \ ,
\label{eqn:kdistr}  
\end{equation}
where the distribution function is over the interval $\Delta \nu$, 
and $\mathcal{M}$ is the column mass of the layer.  Since the 
distribution function is a relatively smooth function of $k$, 
the integral in Equation~\ref{eqn:kdistr} can be computed with 
much less computational effort than that in 
Equation~\ref{eqn:meantrans}.  The integral in 
Equation~\ref{eqn:kdistr} can be made even more straightforward 
if we replace the distribution function with the cumulative 
distribution function, $g(k)$, with
\begin{equation}
  g(k) = \int_{0}^{k} f(k')dk' \ ,
\end{equation}
so that the mean transmission is then
\begin{equation}
  \mathcal{T}_{\Delta \nu} = \int_{0}^{1} e^{-k_{g}\mathcal{M}} dg \ ,
\label{eqn:corrk}
\end{equation}
where $k_{g}$ indicates the mapping between $k$ and $g$ (i.e., 
$k_{g}$ is the value of $k$ that corresponds to the independent 
variable $g$).  Given the smooth, monotonic nature of $g(k)$, 
this integral can be accurately evaluated using only 10--20 
intervals in $g$ or less \citep{goodyetal1989}.  

Instead of determining the transmission over $\Delta \nu$ 
using Equation~\ref{eqn:corrk} and then solving the RTE, which 
would not allow for the treatment of scattering, many 
$k$-distribution models will, instead, use the $k$--$g$ mapping 
to determine characteristic absorption coefficients for 
ranges of $g$ \citep{yamamotoetal1970,lacis&hansen1974,liou1974,
ackermanetal1976,mlaweretal1997}, called $k$-coefficients.  
Along with the relevant thermal and scattering source terms 
(assumed constant over $\Delta \nu$), the $k$-coefficients 
are then used to solve the RTE.  The fluxes from these 
calculations are then combined based on the width of the 
range of $g$-values used, effectively swapping the order of 
integration over frequency and with solving the RTE.  As 
opacities depend on temperature, pressure, and gas composition, 
the $k$-coefficients must be computed for each model atmospheric 
layer, or, most commonly, obtained from pre-tabulated results. 

In practice, two key complications arise when using 
$k$-distributions to perform radiative transfer calculations in 
inhomogeneous atmospheres.  First, the particular ordering of 
opacities in wavelength space that maps to $g(k)$ for a particular 
atmospheric level need not be the same ordering as at any other 
level in the model.  If the absorption coefficients are indeed 
spectrally uncorrelated with those at another level, then different 
wavelengths are being mixed when solving the RTE for the entire 
atmosphere.  Thus, it is assumed, as is often the case, that 
absorption coefficients throughout the atmosphere are spectrally 
correlated, leading to the so-called correlated-$k$ method 
\citep{lacis&oinas1991}. In practice this limitation can be 
overcome by using narrow spectral intervals  and choosing interval 
boundaries so that one main absorber dominates each interval.

Second, the blending of $k$-coefficients, computed for an 
individual gaseous species, with those for other species to 
represent a realistic mixture of gases in an atmosphere remains 
controversial.  Most correctly, the $k$-coefficients should be 
computed for a mixture of gases, but it is often the case in 
atmospheric modeling that gas mixing ratios are not known a 
priori (which happens when, e.g., the concentration of a gas 
depends on other atmospheric state variables).  Thus, a more 
flexible approach is to compute the $k$-coefficients for each 
individual gas, and then blend these `on the fly' in model runs.  
However, as discussed by \citet{goodyetal1989}, this approach 
is only valid when the absorption coefficients for the individual 
gases are spectrally uncorrelated.  Furthermore, blending 
$k$-coefficients for gas mixtures can be computationally 
expensive, as a spectral interval with $n$ different absorbers 
requires calculation $n^{n}$ permutations of $k$-coefficients.  
Because of the uncertain magnitude of the error introduced
by blending single-gas k-coefficients, this approach
should only be used with some care.

Complications aside, correlated-$k$ techniques have proven to be 
efficient and accurate.  Extensive testing in the Earth science 
literature has shown that correlated-$k$ methods can achieve flux 
and heating rate accuracies to within  1--5\%, when compared to 
more precise techniques \citep{lacis&oinas1991,fu&liou1992,
mlaweretal1997}.  \citet{goodyetal1989} note occasional cases 
where errors are much larger than several percent, and caution 
that expanding the use of correlated-$k$ techniques into new 
applications (i.e., beyond the Earth sciences) requires testing 
against more reliable methods to ensure the validity of the 
approach.  Internal tests by the Marley/Saumon group have found 
that, for cloud-free models, the correlated-$k$ approach is 
accurate to within about 1\% for brown dwarfs when compared to 
rigorous line-by-line calculations---a level of accuracy that 
exceeds most other uncertainties in the problem.

\subsubsection{Line-by-Line}

At spectral scales where the radiative source terms and all 
opacity sources can be considered to be constant (i.e., at 
resolutions typically less than about 0.01~cm$^{-1}$), radiative 
transfer calculations are monochromatic, and computing 
spectrally-averaged opacities is no longer an issue.  Models 
that solve the monochromatic RTE over a fine grid of wavelengths 
are referred to as `line-by-line' radiative transfer models.  This 
technique is considered to be the most accurate method for dealing 
with complex opacities, but the `brute force' nature of the 
approach makes the technique computationally expensive.  
Line-by-line methods can be made more efficient through the use 
of spectral mapping techniques 
\citep{westetal1990,meadows&crisp1996}, where monochromatic 
elements with similar optical properties at all atmospheric 
levels are binned together, with the RTE being solved only once 
for the bin instead of for every monochromatic element.

A sample comparison between a line-by-line model spectrum 
($T_{\rm eff} = 1900~\rm{K}$) and data for an L2 dwarf at a 
spectral resolution of $\lambda/\Delta\lambda=50,000$ is shown 
in Figure~\ref{fig:high_rez}. This spectral resolution, which 
is much higher than typical brown dwarf spectra, is sufficient 
to resolve individual lines.  While the model 
\citep[from][]{marleyetal2002} has been convolved with a rotational 
kernel, the overall agreement in this particular case demonstrates 
that the model line broadening treatment generally reproduces the data.  Note that 
this particular model employed assumes no micro-turbulent 
broadening.

Finally, `direct opacity sampling' (dOS) \citep{hauschildtetal2001}, 
like line-by-line calculations, solves the RTE monochromatically, and 
is the method for handling opacities in the RTE currently used in the 
PHOENIX group of models \citep{allard&hauschildt1995,barmanetal2001}.  
Note that dOS is distinguished from standard opacity sampling methods, 
which use statistical techniques (e.g., weighting by a Planck 
function) to sample the wavelength-dependent opacity distribution, in 
that dOS uses a pre-specified spectral grid, which can be set to 
very fine resolutions.  In practice, computational cost limits dOS 
calculations to $\sim 1$~cm$^{-1}$, which is sufficient to accurately 
compute the radiative fluxes, but is not as fine as true line-by-line 
calculations.

\section{Cloud Opacity}
\label{sec:condense}

The problem of modeling, to some degree of fidelity, the clouds in 
ultracool dwarfs and giant planets is a central challenge to understanding
the atmospheric structure, reflected and emitted spectra, and even evolution
of these objects.  The subject is complex and there are many different
approaches.  In this section we give an overview of the problem and some
perspective on the various modeling approaches. A more detailed review can 
be found in \citet{marleyetal2013}, and detailed comparisons of cloud 
treatments can be found in \citet{hellingetal2008c} (which is now slightly 
dated).

\subsection{Condensates Overview}

As in  the cool, dense molecular atmospheres found within the Solar 
System, a variety of species condense within the atmospheres of brown 
dwarfs.  The resulting clouds present the single greatest obstacle 
impeding our understanding of these objects. To correctly account for 
cloud opacity, it is necessary to model each condensate-forming species, 
estimating particle sizes and vertical distributions. Predicting
whether condensates are sub-micron in size and are distributed 
vertically throughout the atmosphere, to millibar pressures for example, 
or whether particles are large and mostly confined to thin cloud decks, 
or are something in between, is the central task of condensate modeling. 
The difficulty in accomplishing this, and of reconciling diverse modeling 
approaches with data, limits progress and understanding.  While a number 
of innovative and insightful cloud models have been developed to 
facilitate modeling of brown dwarf atmospheres and the interpretation of 
data, there has been remarkably little progress in the past decade and 
the problem of cloud opacity provides the greatest opportunity to 
realize improvements in model atmosphere fidelity.

At the relatively cool atmospheric temperatures of brown dwarfs and 
giant planets, important atmospheric constituents are expected to be 
found in condensed phases, particularly Fe, Si, and Mg, but also the 
more refractory components, including Al, Ca, Ti, and V.  Table 1 lists 
many of the important condensates predicted by rainout chemical 
equilibrium.  

For the case of homogeneous condensation, where the gas phase species
condenses to form a solid or liquid of the same species 
(e.g., Fe or $\rm H_2O$), condensation
first occurs in a rising parcel of gas when the local partial pressure of the 
condensing gas first
exceeds the saturation vapor pressure. This defines the cloud base.
Supersaturation is a measure
of how far in excess of the saturation vapor pressure the gas must
be in order to condense.  In more complex 
cases (e.g., Ca and TiO forming $\rm CaTiO_3$)
the cloud base is determined in principle through chemical equilibrium
calculations. Figure~\ref{fig:cond} shows the condensation boundaries, 
again for rainout chemistry, for many key species along with a 
collection of model pressure-temperature profiles for reference.  The 
marked boundaries in the figure are the locations where the labeled 
solid or liquid species will form as its progenitor gaseous species are  
carried upwards in a rising air parcel.

As Figure~\ref{fig:cond} attests, there are numerous atmospheric 
condensates within the brown dwarf pressure-temperature regime. However 
not all of them are equally important. Some species, such as TiO, play a 
leading role in controlling gaseous opacity in the M and early L dwarfs, 
but because of their low abundance, are relatively unimportant cloud 
opacity sources.  The more abundant Fe-, Si-, and Mg-bearing species have 
a greater contribution to column grain optical depths. The role of 
elemental abundances and particle sizes can best be appreciated by 
constructing a highly simplistic cloud model \citep{marley2000}, a task 
that also illustrates the challenge faced by cloud modelers. 

If we assume that all molecules of a species with a fractional number 
mixing ratio, $f$, resident above a given pressure level, $P_{\rm c}$, 
condense, then, for a hydrostatic atmosphere, the column mass of the 
condensate, $\cal M_{\rm c}$, is given by
\begin{equation}
{\cal M}_{\rm c} = f \left( \frac{m_{\rm c}}{m} \right) \left( \frac{P_{\rm c}}{g} \right) \ ,
\end{equation}
where $m_{\rm c}$ is the molecular mass of the condensed species and 
$m$ is the mean molecular mass of the atmosphere. Table 2 gives 
$\varphi = fm_{\rm c}/{  m}$ and the condensation temperature at 3~bar 
in a solar composition gas for several important species.  Given this 
column mass, the total condensate optical depth as a function of wavelength, 
$\lambda$, can be estimated given the particle size, $r_{\rm c}$, the 
extinction efficiency, $Q^{\rm ext}_{\lambda}\left(r_{\rm c}\right)$ (which 
can be derived from Mie theory), and the condensate mass density, 
$\rho_{\rm c}$,
\begin{equation}
\tau_\lambda = 75 \epsilon Q^{\rm ext}_{\lambda}(r_{\rm c}) \varphi \biggl({P_{\rm c}\over {1~\rm bar}}\biggr) \biggl({{10^5~\rm cm\,s^{-2}}\over g}\biggr)\biggl({{1~\rm \mu m}\over r_{\rm c}}\biggr)\biggl({{1~\rm g\,cm^{-3}}\over \rho_{\rm c}}\biggr) \ .
\label{eqn:simpcondmdl}
\end{equation}
In a real atmosphere, not all of the condensible species is found in 
the condensed phase, so this parametrization introduces a term, 
$\epsilon<1$, to account for such effects.

Equation~\ref{eqn:simpcondmdl} exemplifies many of the challenges 
presented by atmospheric condensates. While we can estimate the column 
mass of a given material that may be found in the condensed phase, the 
actual opacity depends sensitively upon the particle size, both through 
the mass partition term and the Mie extinction.  Figure~\ref{fig:miegrid} 
shows spectra of Mie absorption and scattering efficiencies, where 
$Q^{\rm ext}_{\lambda} = Q^{\rm abs}_{\lambda} + Q^{\rm scat}_{\lambda}$, 
for a collection of condensates and for several different particle sizes.
Furthermore, the mass balance in Equation~\ref{eqn:simpcondmdl} tells us 
nothing about the vertical distribution of the grains above a condensation 
level.  An additional limitation of course is that basic Mie theory 
\citep[for a modern review of the history of the method and summary of 
available codes, see][]{wriedt2012} assumes that the condensates are 
ideal, spherical particles.

All else being equal, however, reference to Table 2 demonstrates that 
knowing the fractional abundance of a condensate is crucial to understanding 
its potential contribution to cloud opacity. The single most important 
condensate over the range of temperatures expected for brown dwarf 
atmospheres is $\rm H_2O$. Iron and silicate phases are the second most
important species, while the refractory oxides, such as $\rm CaTiO_3$, and 
the alkali species, such as KCl, are of tertiary importance.

\subsection{Modeling Approaches}

One could begin to compute the vertical distribution of cloud opacity 
by considering several possible fates for atmospheric grains after they 
have formed.  In one extreme, the grains might immediately fall out of
the atmosphere, leaving behind a clean gas.  Fortuitously, the first 
indisputable brown dwarf, Gliese~229B, had such an atmosphere, and 
models that entirely neglected grain opacity did a reasonable job 
of reproducing the observed spectra.  Many groups today continue to 
produce cloud-free models, although they differ in the details of how 
the chemistry is treated. Examples include the COND models of the 
PHOENIX group, and various model formulations from other groups 
including the cloudless models of \citet{saumon&marley2008}.  At the other 
extreme, grains might stay in chemical equilibrium with the local gas 
and not precipitate at all. This is the domain of the DUSTY models of 
\citet{allardetal2001}. Recall that these two extremes also describe 
treatments of rainout chemistry discussed in 
Section~\ref{subsubsec:rainout}.  All other cloud models attempt to 
define some intermediate case where grains grow and fall from the 
upper reaches of the atmosphere, but do not completely disperse, thus 
forming cloud decks of some finite vertical extent. 

One approach attempts to empirically describe clouds by varying only 
a few parameters. Typically this is a cloud bottom, at the condensation level, and top pressure 
(or, equivalently, a bottom pressure and fractional vertical scale 
height) and a particle size.  Examples of such `defined cloud' models 
include the `unified' models of Tsuji~et~al. \citep{tsuji2002} and 
various cloud models employed by Burrows and collaborators  
\citep{burrowsetal2002,currieetal2011}. The strength of this approach is that no 
cloud physics need be crafted, but rather a set of empirical cloud 
descriptions can be accumulated as progressively more objects are 
compared to models. In principle, more sophisticated models could be 
constructed that attempt to explain trends in particle sizes or cloud 
thicknesses once a sufficient number of objects have been characterized. 
In practice, however, this has not yet happened and we are left with 
individualized descriptions of particular objects. This approach 
also has little predictive value, as it is not obvious what 
parameters might be appropriate to describe new objects. 

Nevertheless, this empirical method has contributed to the 
definitive finding that neither the cloudless nor fully dusty concepts are 
correct for the L dwarfs---brown dwarf clouds are found in 
discrete layers, as are clouds throughout the Solar System.  Indeed, 
this limitation of the DUSTY modeling approach is an important one. 
Since the DUSTY models assume that condensates do not form cloud layers, 
for cooler models the atmospheric column becomes optically thick very 
rapidly, even when just accounting for the grain opacity. By 
$T_{\rm eff}\sim 1300\,\rm K$, DUSTY models are far redder, with far 
shallower molecular bands, than even the cloudiest known L dwarfs. For 
this reason, fits of DUSTY models to data seldom find effective 
temperatures below 1400 or even 1500~K. 

One example of a model that attempts to capture some aspects of
cloud physics to predict particle sizes and vertical distributions
is the {\tt Eddysed} model developed by \citet{ackerman&marley2001}.
This model balances upwards and downwards transport of condensible 
gases and condensates by way of a tunable `sedimentation efficiency'
parameter, $f_{\rm sed}$. As $f_{\rm sed}$ increases, downwards
mass transport by falling particles becomes more efficient and the
clouds become thinner. The code computes the particle sizes that are
required to produce the implied mass balance rather than attempting
to model grain growth explicitly. 

A unique cloud modeling paradigm has been developed by Helling~et~al. 
in a series of papers \citep[e.g.,][and references
therein]{hellingetal2001,helling&woitke2006,hellingetal2008a,
hellingetal2008b}. The key underlying assumption of this approach 
is that---not unlike the conversion of CO to $\rm CH_4$---condensation 
is difficult; the formation of many of the condensates predicted by 
chemical equilibrium requires the collisions and reactions of multiple 
gaseous components (e.g., Ca and TiO molecules, both of which are 
relatively rare in the predominantly hydrogen-helium gas). This 
approach thus posits that an equilibrium progression of distinct 
cloud layers, as predicted by equilibrium chemistry, is unlikely.  
Instead this paradigm follows the growth of condensate `seed 
particles' that fall down from the top of the atmosphere, facilitating 
the nucleation of a sequence of compounds that are formed during the 
downward drift of particles from above.  Instead of the canonical 
equilibrium cloud layers, this approach predicts a blend of `dirty' 
grains, with a complex, mixed composition that varies continuously 
with height. 

Helling and collaborators have explored the various ramifications of this 
approach in greater detail than can be summarized here (recently in
\citet{witteetal2011}). The main 
takeaway, however, is that the canonical condensation sequence of 
Figure~\ref{fig:cond} is an overly idealized conception of reality.  
However, to reach this conclusion, the Helling~et~al. model posits what 
may be an equally unlikely scenario.  Here, the highly refractory `seed 
particles' are taken to be carried by strong updrafts from the deep 
atmosphere to the top of the atmosphere, where they begin their downward 
journey. These $\rm TiO_2$ particles are presumed to be transported 
upwards rapidly enough such that no condensation takes place onto these 
nuclei and that none of the other condensible species that are 
also entrained in the updraft condense during the upward journey. In 
the limit of cold Solar System giant planets, we know that this is not 
the case as there is no such high altitude refractory haze observed. 
Another practical difficulty with this model is that the rich brew of 
condensates imprints a complex signature on models of emergent spectra, 
such that it is challenging to appreciate the effect of any single 
component of the model.

Despite the aforementioned concerns, there is much to recommend the  
Helling~et~al viewpoint, and only detailed comparisons of model 
predictions to data will ascertain the validity of this model 
conception. Indeed the `Drift-PHOENIX' set of forward models has 
adapted a version of the Helling~et~al models, although as of the 
time of preparation of this review, the models have not been 
described in the literature.



\section{Deriving the Thermal Profile}

In practice, a one-dimensional radiative-convective thermal structure 
profile must be derived iteratively, starting from a first guess. There 
are several approaches that can be followed to determine the equilibrium 
profile, which satisfies Equation~\ref{eqn:equilibrium}, and where the 
temperature and flux profiles are all continuous.  A straightforward 
technique is to simply timestep the atmosphere to equilibrium, which 
dates back to some of the earliest radiative-convective models 
\citep{manabe&strickler1964}.  However, more efficient computational 
approaches to determining the equilibrium thermal structure have been 
developed, and, here, simply as a means of bringing together the previous 
sections, we sketch out one technique that has been successfully 
applied to Solar System, extrasolar planet, and brown dwarf 
atmospheres \citep{mckayetal1989,marleyetal1996,marley&mckay1999,
fortneyetal2005}.
 
We begin with an atmosphere grid that has some number, $N$, of discrete 
levels, denoted by $i$ (Figure 1). The topmost level is taken as $i=1$, and each 
level is associated with a fixed pressure. If the radiative-convective 
boundary falls at level $i_{\rm RC}$, then, as was discussed in 
Section~\ref{subsec:convadjust}, the temperature at each point in the 
convection zone (i.e., the points where $N \ge i \ge i_{\rm RC}$) can be 
found by following an adiabatic gradient downwards from the temperature 
at the radiative-convective boundary, $T_{\rm RC}$. Thus, our task is 
to solve for the temperature at each level where $i < i_{\rm RC}$. Here, 
the atmosphere is in radiative equilibrium, and, for non-irradiated 
bodies, the net thermal flux through each level must be 
(from Equation~\ref{eqn:equilibrium}) 
\begin{equation}
  F^{net}_{i,\rm target} = \sigma T_{\rm eff}^4 \ ,
\label{eqn:target}
\end{equation}
where $T_{\rm eff}$ is the desired effective temperature for the model, 
and the thermal subscript has been dropped for clarity. For irradiated 
planets, the net thermal flux must be increased by the net stellar flux 
at each level.

For a first guess atmosphere profile $T(P)$, chemical abundances must 
be determined (Section~\ref{sec:chemistry}) and distributions of any 
relevant aerosols must be found (Section~\ref{sec:condense}).  From 
these, the wavelength-dependent opacities can be computed (or obtained 
from a look-up table; Section~\ref{sec:opacity}), thereby permitting 
the calculation of the net thermal flux profile 
(Section~\ref{sec:radiation}).  In general, though, the net thermal 
fluxes at the model levels throughout the radiative zone will not sum 
to the target value, implying a net radiative heating or cooling 
(Equation~\ref{eqn:heatrate}).  From the mismatch with the target, a 
flux error can be computed, and is given by 
$\delta F^{\rm net}_i = F^{\rm net}_i - F^{\rm net}_{i, \rm target}$. 
The task of the model is then to correct the first guess $T(P)$ profile to 
find an improved guess. This is a complex problem, as the flux at each 
level in the atmosphere depends on the temperature structure of all other 
layers. So $F^{\rm net}_i=F^{\rm net}_i(T_1,T_2,T_3,\ldots,T_{\rm RC})$. 
Note that, since $T_{\rm RC}$ controls the temperature profile down 
the adiabatic gradient through the convection zone, $F^{\rm net}_i$ does 
not explicitly depend on the temperature at levels with $i>i_{\rm RC}$, 
as these temperatures are uniquely specified by $T_{\rm RC}$ and 
the adiabat.  A Jacobian matrix, $\overleftrightarrow{\boldsymbol{A}}$, 
of partial derivatives describes how the flux at each level in the 
atmosphere depends upon the temperature at each level, 
\begin{equation}
\overleftrightarrow{\boldsymbol{A}} = \begin{pmatrix}{\partial F_{1} \over \partial T_1}&{\partial F_1 \over \partial T_2}&\ldots&{\partial F_1 \over \partial T_{\rm RC}}\cr 
                {\partial F_{2} \over \partial T_1}&{\partial F_2 \over \partial T_2}&\ldots&{\partial F_2 \over \partial T_{\rm RC}}\cr
                \vdots&\vdots&\ddots&\vdots \cr
                {\partial F_{\rm RC} \over \partial T_1}&{\partial F_{\rm RC} \over \partial T_2}&\ldots&{\partial F_{\rm RC} \over \partial T_{\rm RC}}\cr
       \end{pmatrix} \ .
\end{equation}
The individual terms in the Jacobian matrix are computed by 
iteratively perturbing the temperature at each level in the model 
and re-computing the net fluxes throughout the atmosphere. This is 
a time-consuming step as the entire radiative transfer must be 
solved $i_{\rm RC}$ times, each occasion with a temperature 
perturbation introduced at a single level. Once 
$\overleftrightarrow{\boldsymbol{A}}$ is in hand it must be 
inverted to find the vector,
\begin{equation}
\delta \boldsymbol{T} = \boldsymbol{A}^{-1} \cdot \delta \boldsymbol{F} \ ,
\end{equation}
where $\delta \boldsymbol{F}$ is the vector of flux errors. Then 
$T^{\prime}_i = T_i + \delta T_i$ would ideally bring all of the 
net thermal fluxes to the desired value 
(i.e., Equation~\ref{eqn:target}).

Once a temperature correction is applied and a new thermal structure  
is in hand, the process will repeat. In practice, the new model will 
not be precisely at $F^{\rm net}_{\rm target}$, because the problem 
is not linear and molecular abundances, opacities, and cloud 
structure will all respond to the new profile, further perturbing the 
computed fluxes. Thus, many iterations are required until the net 
thermal flux change from one iteration to the next is smaller than 
some small target value throughout the atmosphere, typically taken 
as $10^{-5}$--$10^{-6}$. In addition, if the temperature gradient in 
the bottom-most radiative layer exceeds the adiabatic lapse rate, then 
the lapse rate in that layer must be reset to the adiabatic value, and 
the radiative-convective boundary moved up to level $i_{\rm RC}-1$. In 
practice there are many more subtleties that arise, but, nevertheless,
approaches such as this are robust and eventually converge on a desired 
solution. 

The computational intensity of the process outlined above, particularly 
the repeated calculation of $\overleftrightarrow{\boldsymbol{A}}$ and 
the attendant radiative transfer calculations, is the reason
why rapid methods for computing the radiative transfer are required for 
atmospheric modeling (e.g., Section~\ref{subsec:opacsRTE}).  
Additionally, various methods for streamlining this process and 
minimizing the number of matrix inversions that must be performed
have been devised.

The greatest impediment to model convergence is, of course, clouds. If 
an iterative step cools the overall profile, then the cloud base for a 
photospheric condensate will move down to thicker atmospheric levels and, 
as a result, the cloud optical depth will increase. However, a thicker 
cloud will trap more thermal radiation in the atmosphere (i.e, provide a 
stronger greenhouse effect), which will, in turn, tend to heat the 
atmosphere and warm the temperature profile. This, then, leads to a 
higher cloud base, a thinner cloud, and greater cooling. Without some 
numerical approach to smooth convergence, it is not unusual to
see model profiles vacillate between different cloud states while never
converging to a mean solution.  In modern terrestrial cloud modeling, 
in fact, clouds are never treated in one dimension because of this 
problem.  Approaches that have been tried in brown dwarf atmospheric 
modeling include using trailing averages of past cloud states to smooth 
out jumps in the cloud state, and using two atmospheric columns, one 
cloudy and one clear, to allow for horizontal patchiness and reduced 
sensitivity to opacity changes.  Details of such numerical `tricks' are 
not always well described in the literature.

\section{Results}

Several research groups actively construct radiative-convective 
equilibrium atmosphere models, compute emergent spectra, and compare 
the results to data. Table 3 lists a few of the most active 
collaborations, and selected theory and model-data comparison papers.
These papers serve as jumping off points to further explore the 
approaches employed by and the science results of each group.

Figure~\ref{fig:datmod_comp} present some example comparisons 
between models computed by the Marley/Saumon group and spectral 
data for a broad range of L and T dwarfs. Details of the models 
and data are presented in \citet{stephensetal2009}. The individual
best fitting models were selected from a large grid of forward models 
computed for this purpose. As the figure attests, the quality of 
the matches varies from object to object.  Overall the fits for 
the T dwarfs are generally quite good.  Notably for the early 
(T0 to T2) and the one T5.5 dwarfs, thin clouds ($f_{\rm sed}\sim 3$ 
to 4) are favored while for the T4 dwarfs cloudless models fit best. 

In contrast, for the L dwarfs, generally thicker clouds ($f_{\rm sed}$ 
mostly in the range of 1 to 2) are required. Cloudless models (not 
shown) fit far more poorly, thus demonstrating that even though there 
are no `smoking-gun' spectral features indicating that clouds are present, 
their overall impact on the spectra are undeniable (the absence of TiO and
other spectral features in the spectra of L dwarfs that are found in 
M dwarfs indicates that these species have condensed, but not necessarily 
into discrete cloud decks). 
Figure~\ref{fig:datmod_comp} also demonstrates that while the cloudy 
models generally reproduce the spectral shapes of each object, there 
are important mismatches between models and data. Since we know from 
the comparisons with the cloudless T dwarfs that the atmospheric
chemistry is reasonably well understood, these mismatches point to 
shortcomings in the cloud model.  Understanding the sources of these 
mismatches and how the cloud description should be modified in each 
case represents an important task for future research. 

\citet{witteetal2011} fit many of the same dwarfs as 
\citet{stephensetal2009}, and this provides an opportunity to compare 
derived parameters between two groups. For 2MASS J1507, an L5.5 dwarf 
for example, the best fitting model of \citet{stephensetal2009} has 
$g=3\times10^{3}~\rm{m~s^{-2}}$ and $T_{\rm eff} = 1600\,\rm K$. Witte et al., 
using the cloud model approach of Helling's group (but fitting only 
to the 1.0 to $3.0\,\rm \mu m$ data), find $g=1\times10^{3}~\rm{m~s^{-2}}$ and 
$T_{\rm eff} = 1800\,\rm K$. This discrepancy is almost certainly 
attributable to the differing cloud models and again points to the 
need for higher fidelity models or a new approach.

\section{Current Issues}

\subsection{Variability and Patchy Clouds}
\label{subsec:variability}

The emergent spectra of many L and T dwarfs are known to vary with time 
\citep[e.g.,][]{tinney&trolley1999,bailerjones&mundt1999,gelinoetal2002,
artigauetal2009,heinzeetal2013}.  Broadband observations have revealed 
periodic and non-periodic flux variations as large as 10--30\% 
\citep{radiganetal2012,gillonetal2013}, occurring on timescales from 
1--100~hr \citep{bailerjones&mundt2001}, and spectroscopic observations 
have shown that variability can be strongly wavelength-dependent 
\citep{buenzlietal2012}.  While a variety of dynamical processes can 
influence brown dwarf spectra \citep{robinson&marley2014,
zhang&showman2014}, it is generally expected that clouds play an 
important, if not central, role in these brightness variations, as cloud 
structures provide a continuum opacity source that sculpts the emergent 
spectra of nearly all spectral classes of brown dwarfs.  Indeed, a variety 
of patchy cloud models have demonstrated that, in many cases, observed 
variability can be explained by changes in cloud distribution and 
thickness \citep{marleyetal2010,apaietal2013,burgasseretal2014,
morleyetal2014b}.  Furthermore, a recent report of a spatially-resolved 
map of a nearby L--T transition dwarf \citep{crossfieldetal2014} revealed 
a patchy photosphere consistent with complex cloud structures.

Clearly observations of brown dwarf variability present an important 
opportunity for constraining cloud models and dynamical simulations.
For example, \citet{zhang&showman2014} used a simple, cloud-free 
dynamical model to study how different circulation regimes in brown 
dwarf atmospheres could influence broadband lightcurves.  However, 
realistically simulating the influence of patchy, time-evolving clouds 
on the emergent spectra of brown dwarfs presents a great modeling 
challenge.  The study of brown dwarf atmospheric circulation is 
only just beginning \citep[e.g.,][]{showman&kaspi2013}, and, as a 
result, these models have not yet incorporated chemistry, aerosols, 
or wavelength-dependent radiative transfer.  Meanwhile, one-dimensional 
brown dwarf atmospheric models, which do incorporate chemistry, clouds, 
and realistic radiative transfer, are typically only used to study an 
atmosphere in its steady-state.  A combination of these two approaches 
will be needed if we are to understand the true nature and complexity 
of brown dwarf atmospheres.


\subsection{Atmospheric Retrieval}

The actual atmospheric state of a world, including the thermal structure and 
gas concentrations, can be constrained by a group of techniques called 
`retrieval.'  Atmospheric retrieval techniques, somtimes referred to as 
'inverse methods', have their origins in the Earth remote sensing 
literature \citep{rodgers1976,rodgers2000}, and operate by extracting 
information about atmospheric conditions from emitted-, reflected-, or 
transmitted-light spectra.  From the perspective of radiative-convective 
modeling, the utility of retrieval is obvious---these techniques provide a 
direct and independent means of constraining the parameters and physics of 
atmospheric thermal structure models.

Hot Jupiters have seen extensive applications of retrieval techniques.  
\citet{madhusudhan&seager2009}, by comparing observations of HD~189733b 
and HD~209458b to a multi-dimensional grid of models, reported constraints on 
the atmospheric thermal structure and concentrations of several key gases for 
these worlds.  More sophisticated retrieval methods have also been applied 
to observations of hot Jupiters, including optimal estimation techniques 
\citep{leeetal2012,lineetal2012} and Markov chain Monte Carlo methods 
\citep{madhusudhanetal2011,benneke&seager2012,lineetal2013}. Care must
be taken when interpreting retrieval results, especially when constrained only
by photometric data, as the results hinge on the assumptions incorporated
into the forward model, including the particular set of gases included.

Historically, evolutionary and atmospheric processes of brown dwarfs 
have been constrained by comparing observed spectra of individual 
objects to a large grid of model spectra \citep[e.g., ][]{burrowsetal1993,
burgasseretal2007}.  Uncertainties in key parameters (e.g., effective 
temperature) are sometimes estimated within the grid-based model 
comparison approach \citep{cushingetal2008,riceetal2010}.  A drawback of 
this approach is that it imposes model assumptions onto the resulting 
fits, and the fitting approach amounts to hunting for the model parameters 
that best reproduce an observed spectrum

Very recently, optimal estimation retrieval techniques have been applied to 
hot, young directly-imaged gas giant planets \citep{leeetal2013} and brown 
dwarfs \citep{lineetal2014}.  Figure~\ref{fig:retrieval} compares thermal 
profiles from a grid-based approach \citep{saumonetal2006} and an optimal 
estimation retrieval approach \citep{lineetal2014}, as applied to Gl~570D 
\citep{burgasseretal2000}.  The agreement between the two methods  
is quite good, except in the upper atmosphere, where the optimal 
estimation approach indicates a warmer thermal structure.  This result 
hints that important physics may be missing from the models used in the 
grid-based approach, such as non-local thermodynamic equilibrium 
processes \citep{sorahanaetal2014}.


\section{Conclusion}

Since the discovery of the first unmistakable brown dwarf in 1995 
there have been well over 3,000 papers published on this topic alone. 
Almost without exception, these papers, which generally aim to 
understand the spectra, formation, and evolution of brown dwarfs, 
relate in one way or another to the atmospheres of these objects. While 
the numbers of papers about the directly-imaged planets are--for 
now--smaller, there is little doubt that this field is likewise 
on the verge of a rapid expansion.

The science yield from these new exoplanet discoveries, as well as 
from ongoing studies of brown dwarfs, hinges in large part in our 
ability to model and understand the atmospheres of these worlds.  While 
the first two decades of brown dwarf science has seen remarkable 
advances in the fidelity of atmosphere modeling, there is still much 
room for improvement.  Better cloud models, greater exploration of
the effect of varying elemental abundances, particularly atmospheric 
C/O ratios, and greater studies of departures from equilibrium chemistry 
are all important areas for improvement. Ultimately, retrieval methods for 
determining atmospheric thermal, cloud, and chemical profiles likely offer 
the best avenue for truly constraining the properties of these objects. 
Retrieval methods, however, require high fidelity data---ideally taken over 
a large wavelength range---and this may be difficult for directly-imaged
exoplanets in 
the foreseeable future.

Nevertheless, the convergence of modeling approaches derived 
from the studies of Earth, Solar System planets, exoplanets, and, of 
course, stellar atmospheres will continue to enrich this field. Brown 
dwarfs are often explained as being `failed stars', but there is little 
doubt that the study of their complex, fascinating atmospheres has been 
one of the great successes of theoretical astrophysics over the past two 
decades.

\section*{Acknowledgements}

MSM acknowledges support of the NASA Planetary Atmospheres, Astrophysics 
Theory, and Origins programs.  TDR gratefully acknowledges support from 
an appointment to the NASA Postdoctoral Program at NASA Ames Research 
Center, administered by Oak Ridge Affiliated Universities.  Helpful feedback 
on earlier versions of this manuscript was provided by Jonathan Fortney, 
Didier Saumon, and David Catling.  We thank Roxana Lupu for sharing results 
from high-resolution opacity calculations, Richard Freedman for interesting 
discussions regarding line shapes and strengths, Caroline Morley for sharing 
thermal structure models and condensation curves, Michael Line for providing 
output from retrieval models, Sandy Leggett, Cullen Blake, and Mike Cushing for 
sharing data on short notice, Kevin Zahnle and Bruce Fegley for discussions 
about chemistry, and many other colleagues for generously offering their 
insights and comments.

\bibliographystyle{apj_edit}
\bibliography{biblio.bib}

\newpage


\begin{table}[ht]
  \centering
  {\bf Table 1. Selected Elemental Abundances and Species of Interest} \\
  \vspace{2mm}
  \begin{tabular}{c c | l}
    \hline
    \hline
    Element &   Relative Abundance$^1$                 &   Important Species$^2$      \\
            &   $\log [n({\rm El})/n({\rm H})] + 12 $  &    (rainout chemistry)       \\
    \hline
      H     &   12       &   $\rm H_2$, H    \\
     He     &   10.9     &   He     \\
      O     &   8.7      &  CO, $\rm H_2O\,(v,s,l)$, silicates    \\
      C     &   8.4      &  CO, $\rm CH_4$, $\rm CO_2$    \\
      N     &   7.8      &  $\rm N_2$, $\rm NH_3\,(v,s)$, $\rm NH_4SH\,(s)$     \\
     Mg     &   7.6      &  $\rm MgSiO_3\,(s)$, $\rm Mg_2SiO_4\,(s)$, MgH  \\
     Fe     &   7.5      &  FeH, Fe\,(l,s)    \\
      S     &   7.2      &  $\rm H_2S$, $\rm NH_4SH\,(s)$, MnS\,(s), $\rm Na_2S\,(s)$, ZnS\,(s)    \\
     Al     &   6.5      &  $\rm Al_2O_3\,(s)$,  \\
     Na     &   6.3      &  Na, $\rm Na_2S\,(s)$    \\
      P     &   5.5      &  $\rm PH_3$, $\rm P_4O_6$    \\
      K     &   5.1      &  K, $\rm KCl\,(s)$    \\
     Ti     &   4.9      &  TiO, $\rm CaTiO_3\,(s)$  \\
      V     &   4.0      &  VO, V-oxides  \\    
    \hline
\end{tabular}
\noindent{\\For condensed phases, (v) = vapor, (l) = liquid, (s) = solid.\\
          $^1$Abundances from \citet{lodders2003}.\\
          $^2$See \citet{lodders2010} for a review.}
\end{table}
\begin{table}[ht]
  \centering
  {\bf Table 2. Parameters for Elementary Cloud Model of Selected Condensates} \\
  \vspace{2mm}
  \begin{tabular}{l c c c}
    \hline
    \hline
    Species &  $f$  &   $\varphi=f\frac{m_c}{m}$  & $T_{\rm cond}$ at 3 bar    \\
    \hline
     $\rm H_2O$      &   $1.2\times 10^{-3}$ & 9.7 & 265    \\
     KCl             &   $2.2\times 10^{-7}$ & $7.1\times10^{-3}$ & 820    \\
     $\rm Na_2S$     &   $1.7\times 10^{-6}$ & $5.8\times10^{-2}$ & 1025    \\
     $\rm MgSiO_3$   &   $5.9\times 10^{-5}$ & 2.6 & 1685    \\
     $\rm Mg_2SiO_4$ &   $3.0\times 10^{-5}$ & 1.8 & 1760    \\
     Fe              &   $5.3\times 10^{-5}$ & 1.3 & 1930    \\
     $\rm CaTiO_3$   &   $1.4\times 10^{-7}$ & $8.4\times10^{-3}$ & 2010    \\
    \hline
  \end{tabular}
\end{table}
\begin{landscape}
\begin{table}[ht]
  \centering
  {\bf Table 3. Ultracool Modeling Schools } \\
  \vspace{2mm}
  \begin{tabular}{c c c}
    \hline
    \hline
    School     &   Key Characteristics              & Selected Papers     \\
               &   (chemistry; cloud; opacity)      &      \\
    \hline
     Barman        &   true chem.~eq.; defined clouds; sampling     & {\citet{barmanetal2011}} \\
     Burrows       &   true chem.~eq.; defined cloud; sampling      & {\citet{burrowsetal2002}}  \\
                        &                                           & {\citet{currieetal2011}} \\
     Marley/Saumon &   rainout eq.; eddysed$^1$; correlated-k       & {\citet{saumon&marley2008}}    \\
                   &                                                & {\citet{stephensetal2009}} \\
     PHOENIX       &   true chem.~eq.; various clouds$^2$; sampling & {\citet{witteetal2011}}  \\
     Tokyo         &   true chem.~eq.; UCM$^3$; band model          & {\citet{sorahana&yamamura2012}} \\
                   &                                                & {\citet{tsuji2005}}  \\
    \hline
  \end{tabular}
\end{table}
\end{landscape}
\noindent {$^1$Eddy-sedimentation, a cloud physics model          
                \citep{ackerman&marley2001}.\\
           $^2$Various cloud physics models, including DUSTY 
                \citep{allardetal2001} and DRIFT \citep{witteetal2011}.\\
           $^3$The `Unified Cloud Model', a defined cloud model 
               \citep{tsuji2002}.}


\newpage
\begin{figure}
  \centering
  \includegraphics[scale=0.65]{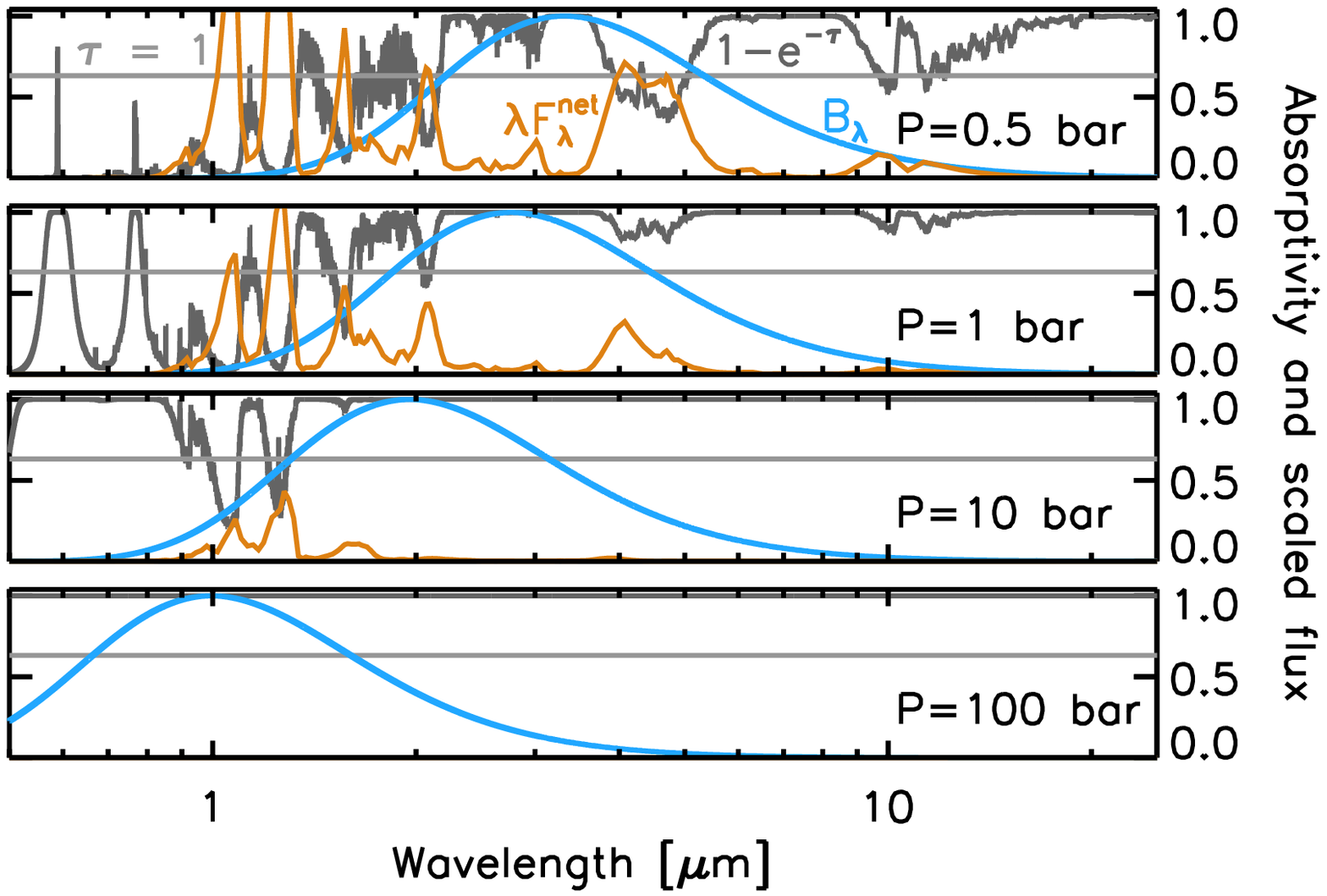}
\end{figure}
\begin{figure}
  \centering
  \includegraphics[width=2.45in]{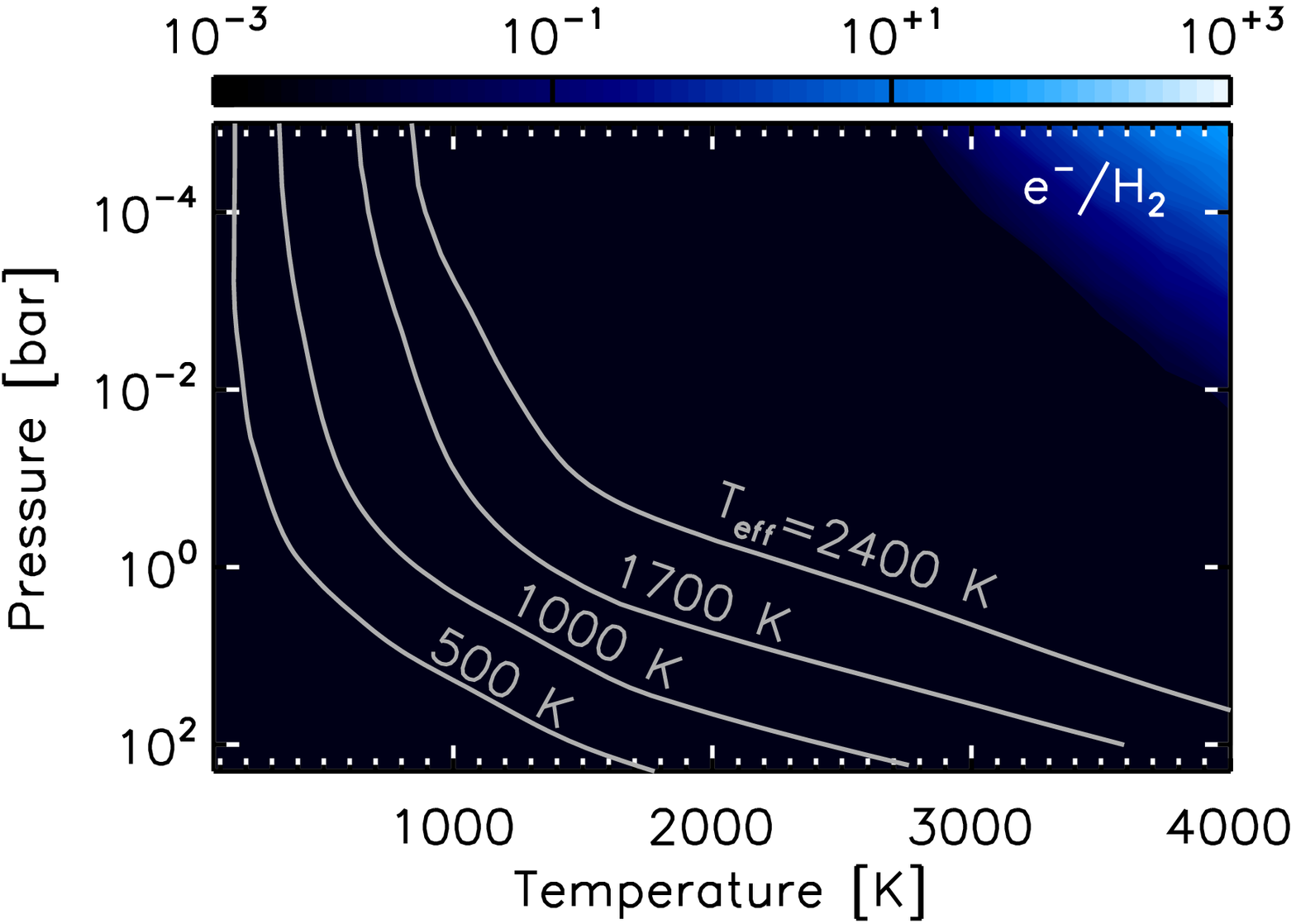}
  \includegraphics[width=2.45in]{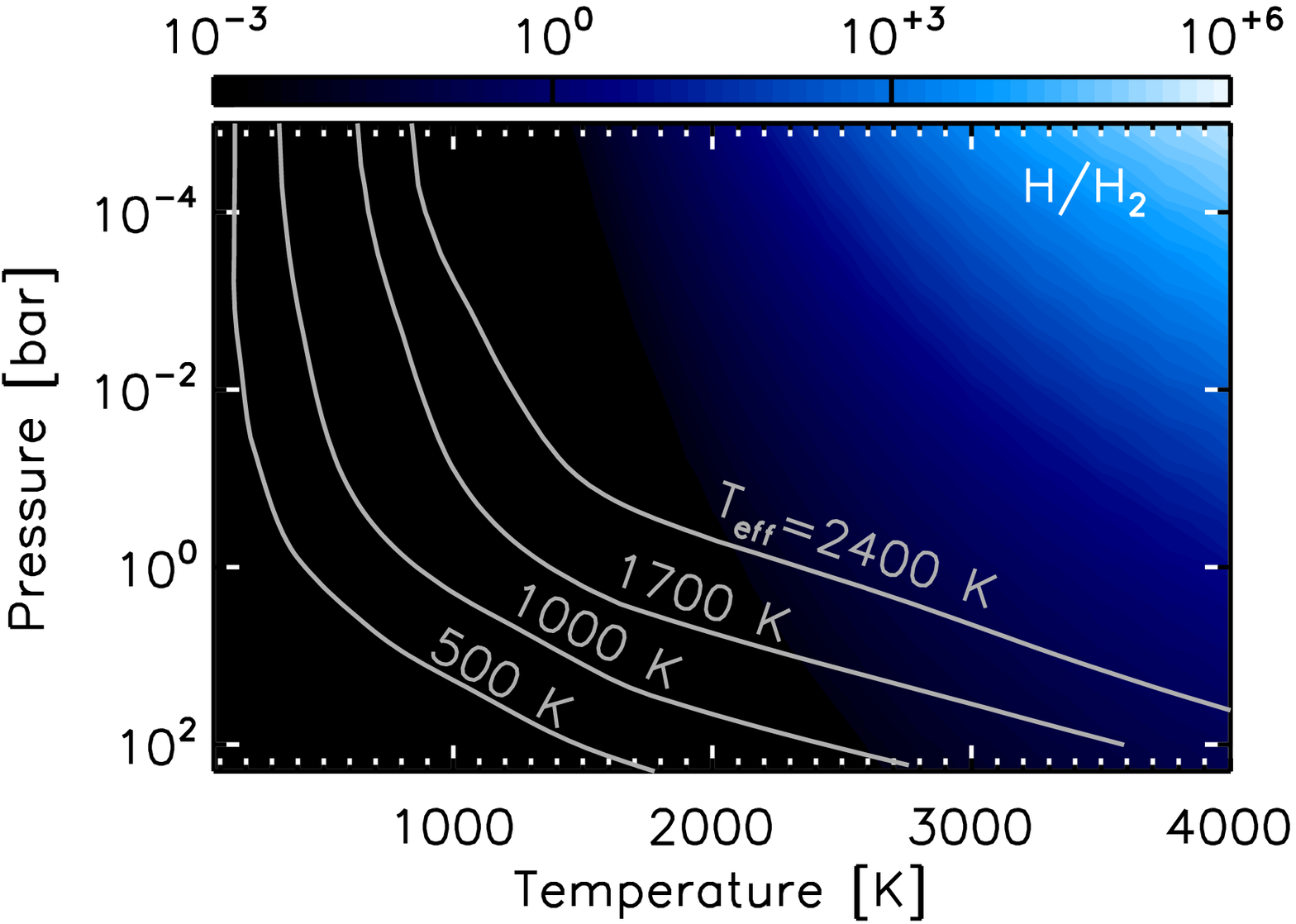}
  \includegraphics[width=2.45in]{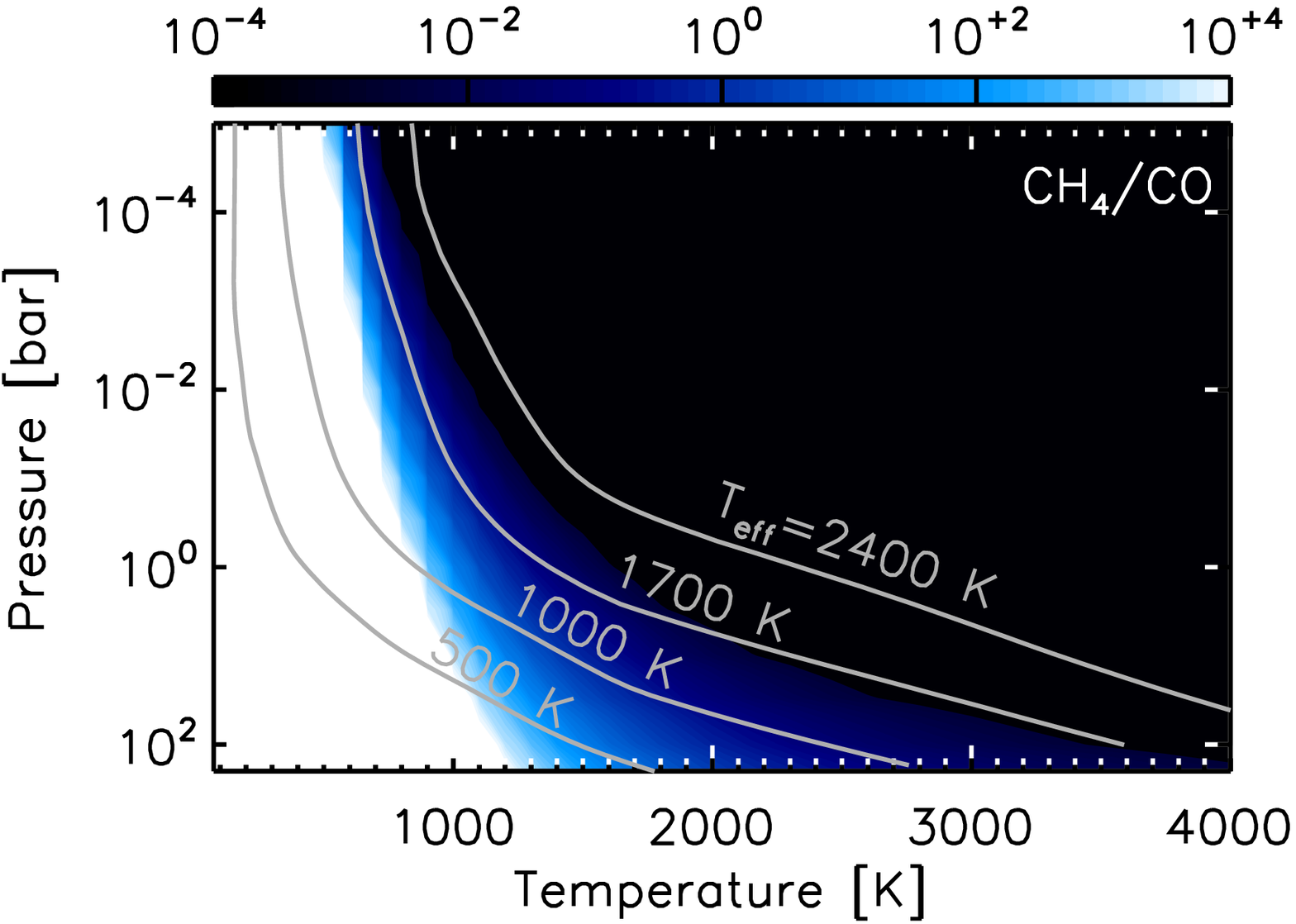}
  \caption{ {\normalsize (\textbf{Two pages.}) Abundances or ratios of 
  abundances for various species of interest, as a function of pressure 
  and temperature, from \citet{lodders&fegley2002} and 
  \citet{visscheretal2006}.  Species are listed in the upper-right of 
  each sub-plot, and the abundance, or abundance ratio, is according to 
  the color bar at the top of each sub-plot.  Thermal profiles from 
  cloud-free models of various effective temperatures, and with 
  $g=10^3~\rm{m~s^{-2}}$, are over-plot in gray.  Photospheres for 
  these models typically extend to depths of 3--30 bar.  Note the transition 
  from CO to CH$_4$ at cooler temperatures, and a similar transition for 
  N$_2$ to NH$_3$.  Both TiO and FeH are strongly depleted by rainout at 
  lower temperatures.}}
  \label{fig:abunds}
\end{figure}
\newpage
\begin{figure}
  \centering
  \includegraphics[width=2.45in]{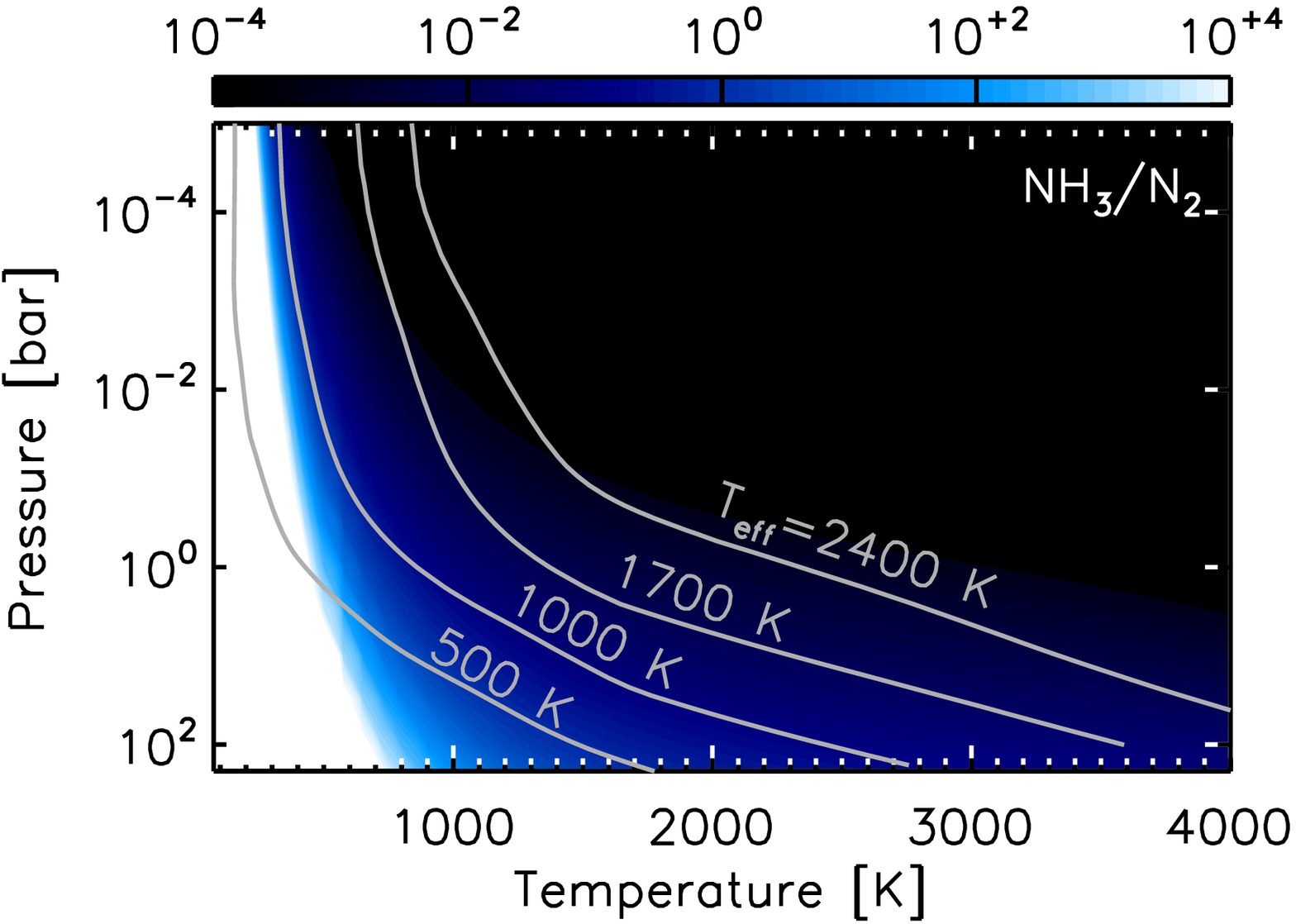}
  \includegraphics[width=2.45in]{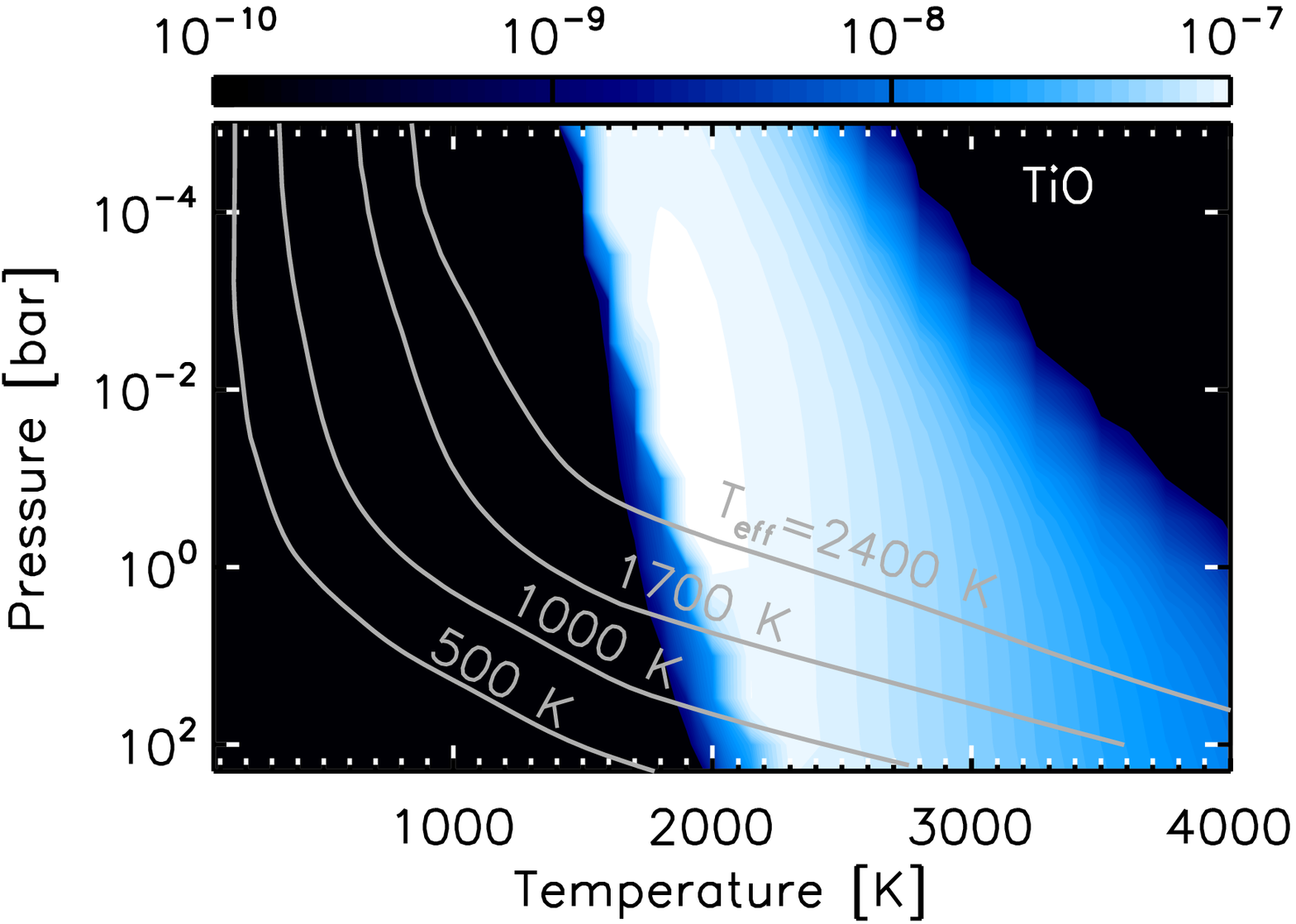}
  \includegraphics[width=2.45in]{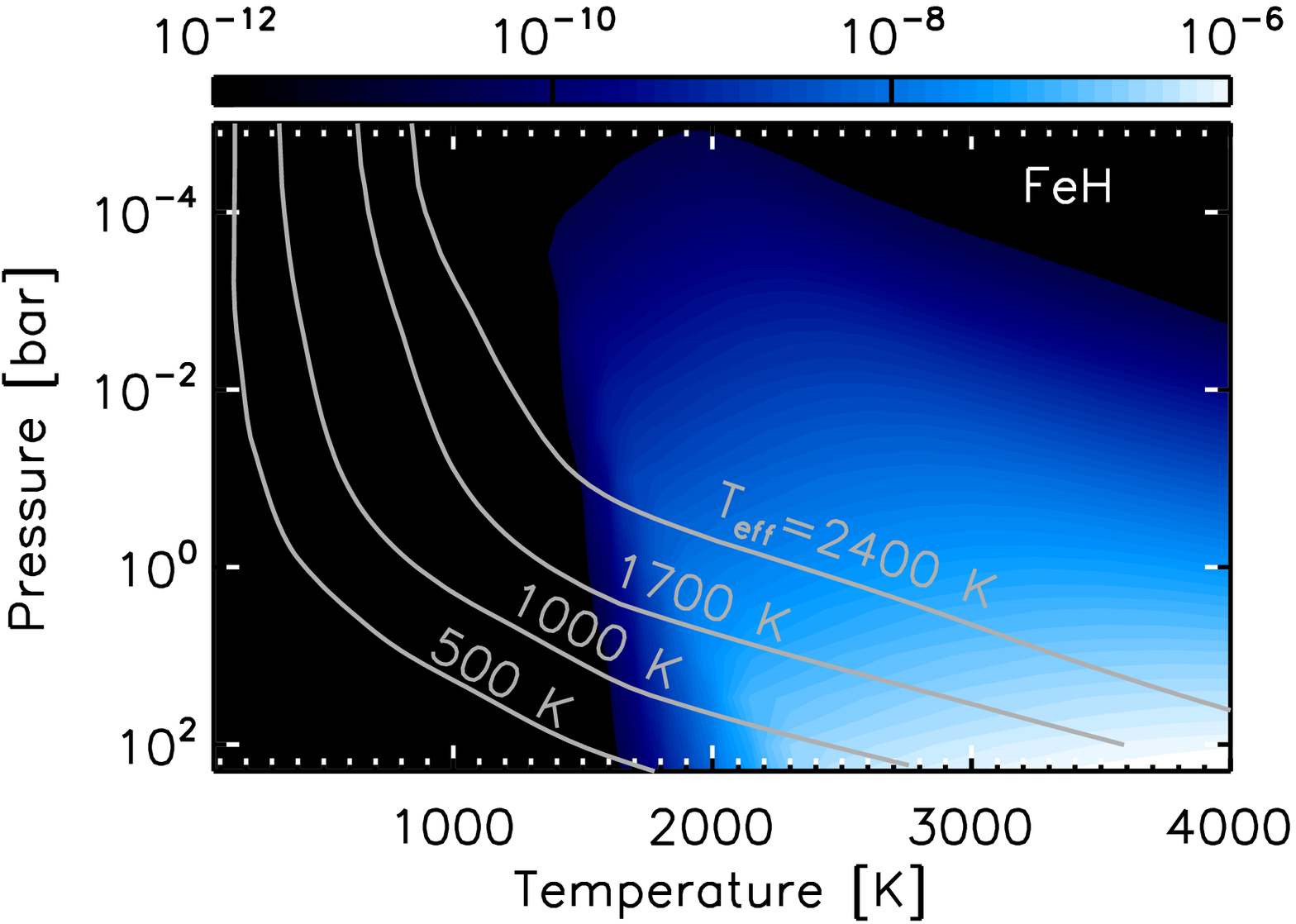}
\end{figure}
\begin{figure}
  \centering
  \includegraphics[scale=0.5]{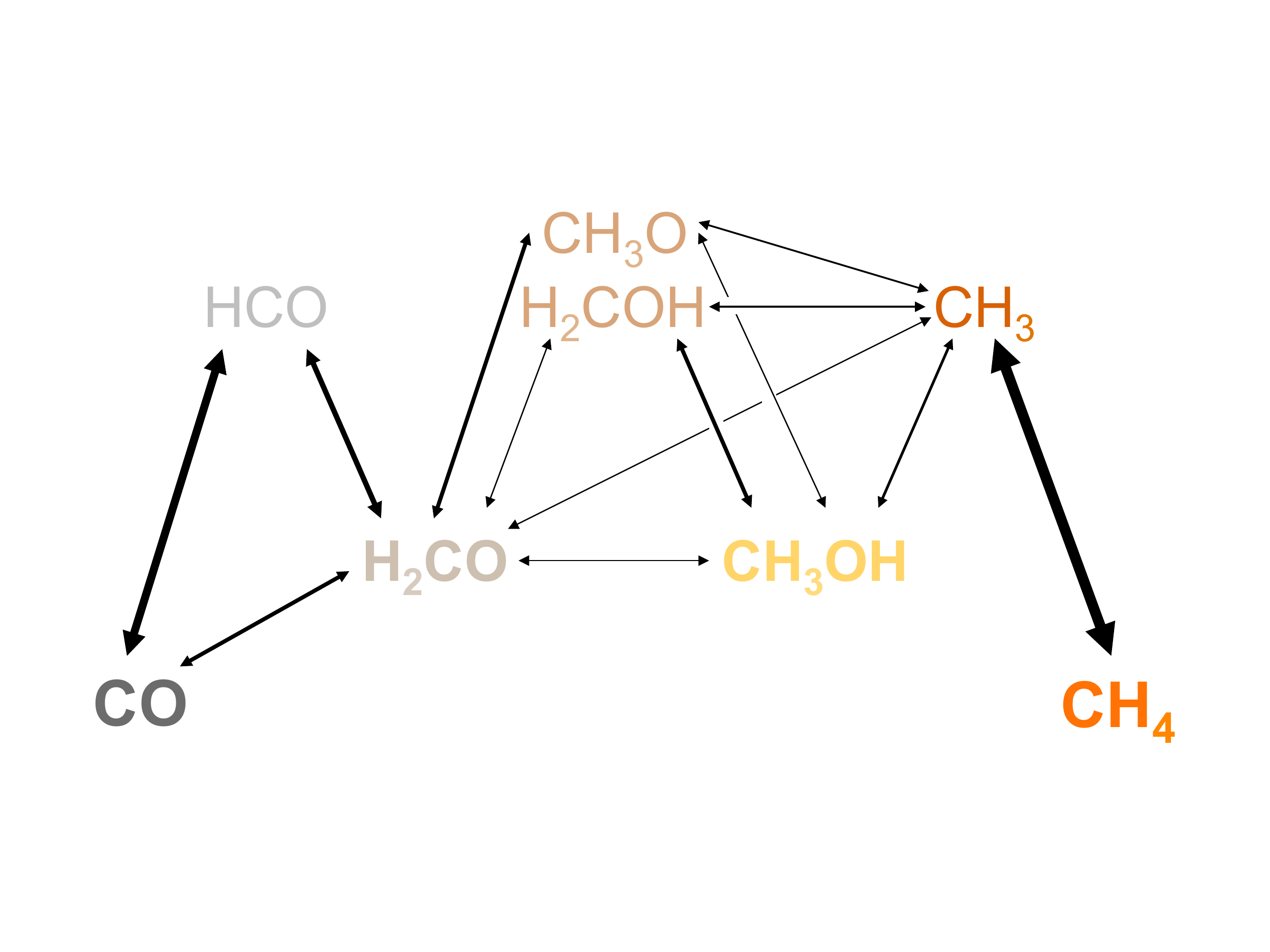}
  \caption{Major chemical pathways linking CO and CH$_4$ in an 
  H$_2$-rich atmosphere \citep[from][]{zahnle&marley2014}. Reactions 
  from left to right are with H$_2$ or H.  Key intermediate molecules 
  are formaldehyde (H$_2$CO) and methanol (CH$_3$OH) while other 
  intermediates (HCO, H$_2$COH, CH$_3$O, CH$_3$) are short-lived free 
  radicals.  The vertical position of individual species gives a rough 
  indication of the energetics.  Energy barriers correspond to breaking 
  C-O bonds---from triple to double, from double to single, and from 
  single to freedom.  Relative magnitudes of reaction rates are indicated 
  by arrow thickness.  Conceptually, higher temperatures and lower 
  pressures tilts the plot as a whole to the left, with carbon pooling in 
  CO. Lower temperatures and higher pressures tilts the plot as a 
  whole to the right, and carbon pools in CH$_4$. }
  \label{fig:ch4}
\end{figure}
\begin{figure}
  \centering
  \includegraphics[scale=0.65]{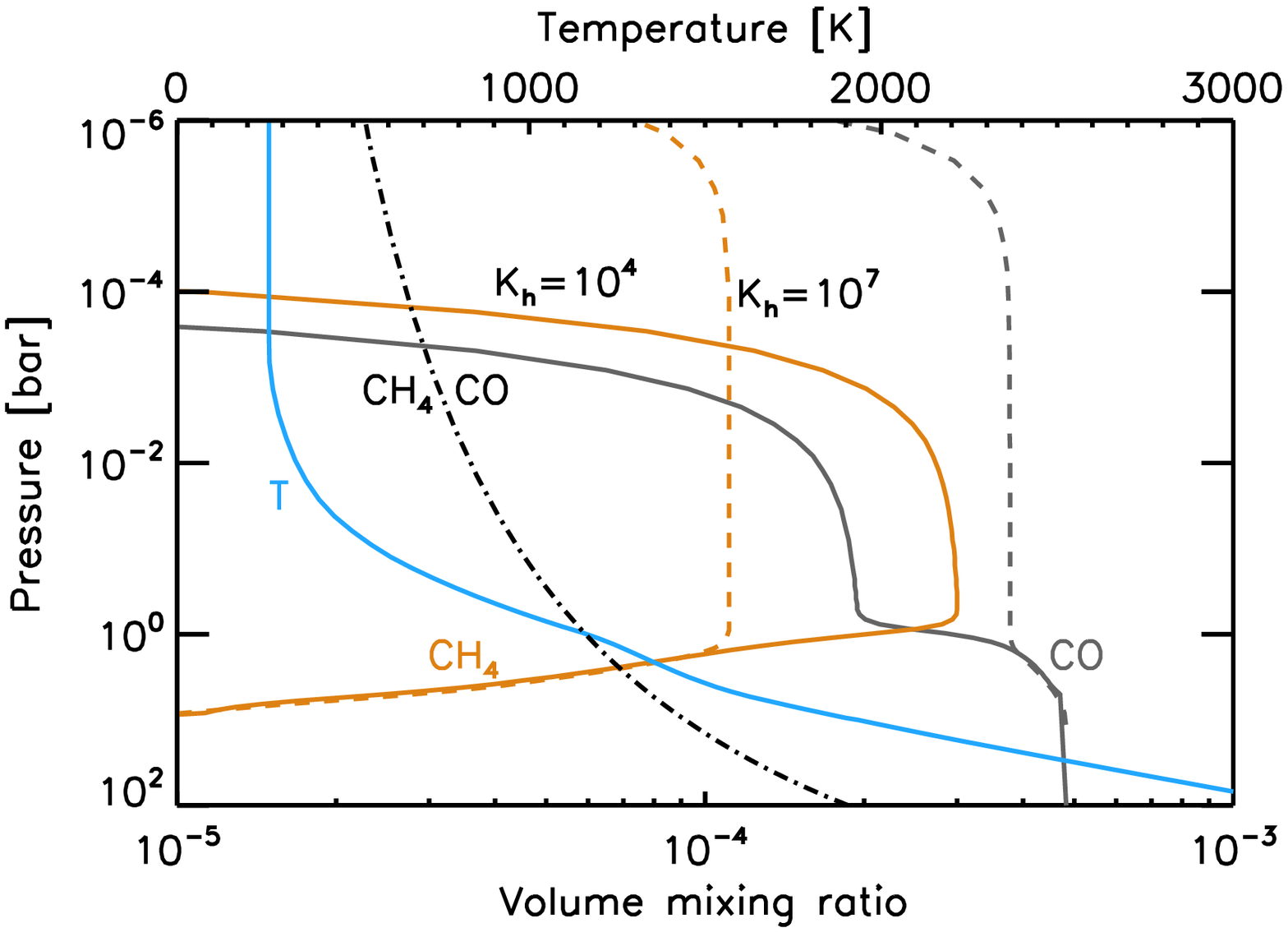}
  \vspace{2mm}
  \caption{An example of $\rm CH_4$ and CO disequilibrium. This 
   $\sim 2\,\rm M_J$ giant planet model is cloud-free, has $T=600~\rm{K}$, 
   $g=10~\rm{m/s^2}$, and no insolation.  The $P$-$T$ profile of 
   the atmosphere (blue) is compared to the $P$-$T$ curve (black 
   dot-dash) where the mixing ratios of methane and carbon monoxide 
   are equal under equilibrium conditions.  Methane is 
   thermodynamically favored when $T$ is to the left of this 
   curve.  Disequilibrium CO (gray) and CH$_4$ (orange) mixing ratios, 
   computed by \citet{zahnle&marley2014}, with
   $K_{h}=10^{4}$~cm$^2$/s (solid) and $K_{h}=10^{7}$~cm$^2$/s (dashed) 
   are shown.  Below the quench points, CO and CH$_4$ are in 
   equilibrium.  Above the quench points, the CO and CH$_4$ mixing 
   ratios are are constant with altitude until molecular diffusion 
   leads to the separation of CO and CH$_4$ from the background $\rm H_2$ gas at lower pressures.}
  \label{fig:quench}
\end{figure}
\begin{figure}
  \centering
  \includegraphics[scale=0.65]{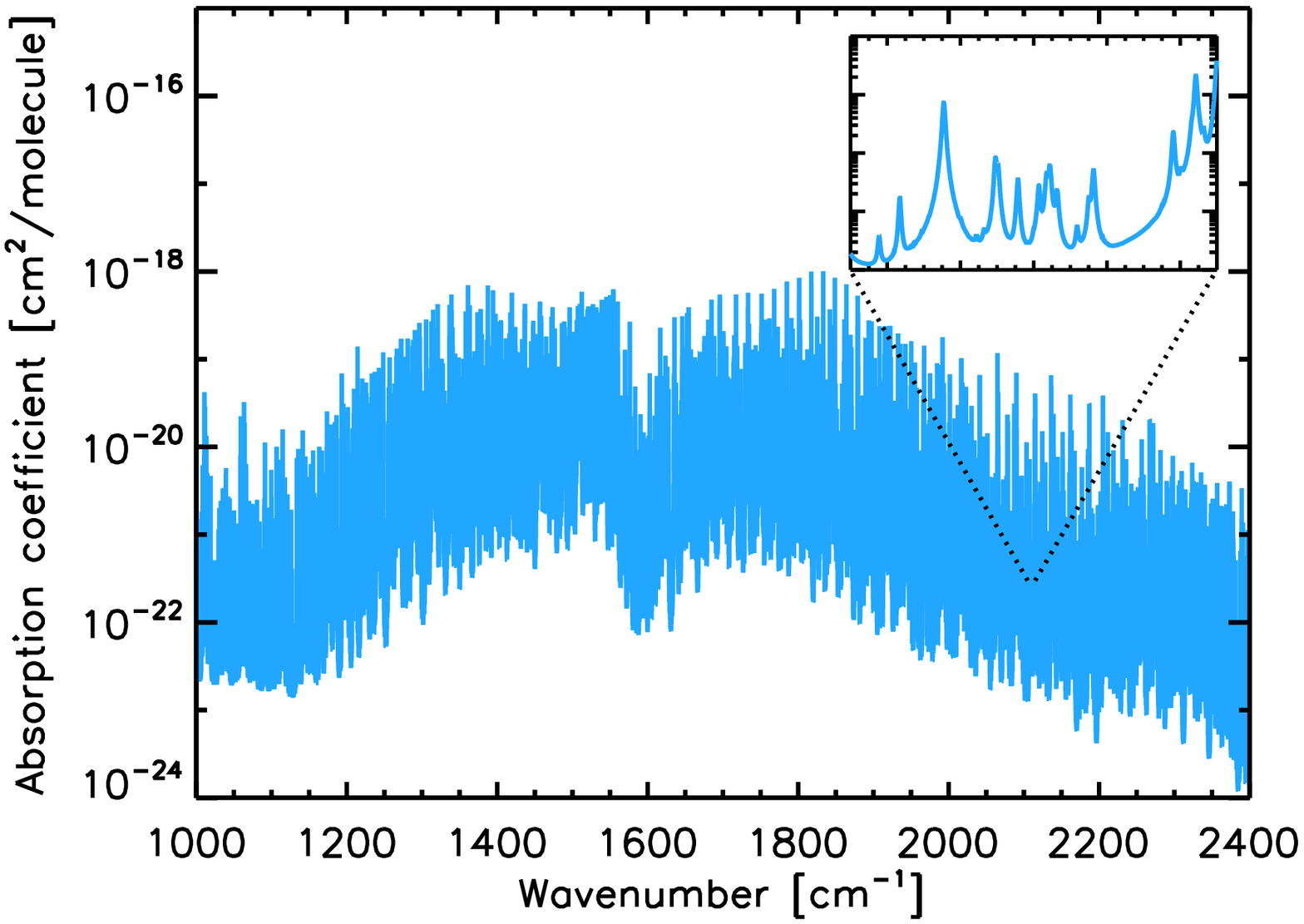}
  \vspace{2mm}
  \caption{Demonstration of molecular line absorption opacity.  Absorption 
  coefficients for the 1600~cm$^{-1}$ (6.3~$\mu$m) water vapor absorption 
  band are shown for a temperature of 1,000~K and a pressure of 1~bar.  Inset 
  shows a smaller range, centered at 2100~cm$^{-1}$ and spanning 10~cm$^{-1}$, 
  where individual absorption lines can clearly be distinguished.}
  \label{fig:abscoeff}
\end{figure}
\begin{figure}
  \centering
  \includegraphics[scale=0.65]{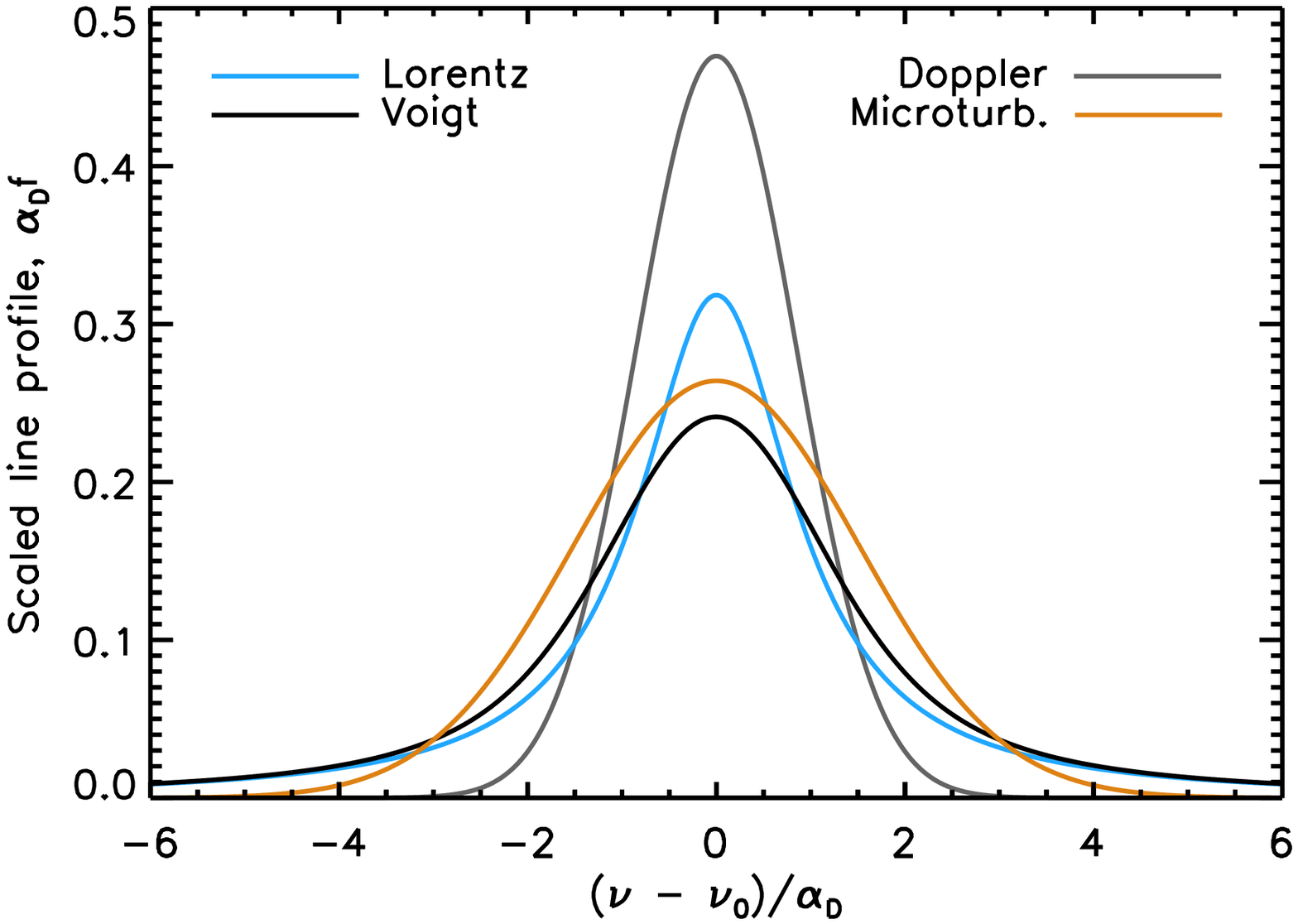}
  \vspace{2mm}
  \caption{Comparison of different lineshapes.  The horizontal axis is 
  in units of the Doppler half-width ($\alpha_{D}$), so that a half-width
  of one occurs at unity along the x-axis.  The lineshape, $f$, is scaled 
  by the Doppler half-width, and the Lorentz and Voigt lineshapes use 
  $\alpha_{L} = \alpha_{D}$, so that the area under the curves are all 
  equal \citep[see][]{huang&yung2004}.  Also shown is a Doppler 
  lineshape where microturbulent broadening is included, where the 
  turbulent velocity is taken to be 1~km~s$^{-1}$.}
  \label{fig:lineshape}
\end{figure}
\begin{figure}
  \centering
  \includegraphics[scale=0.65]{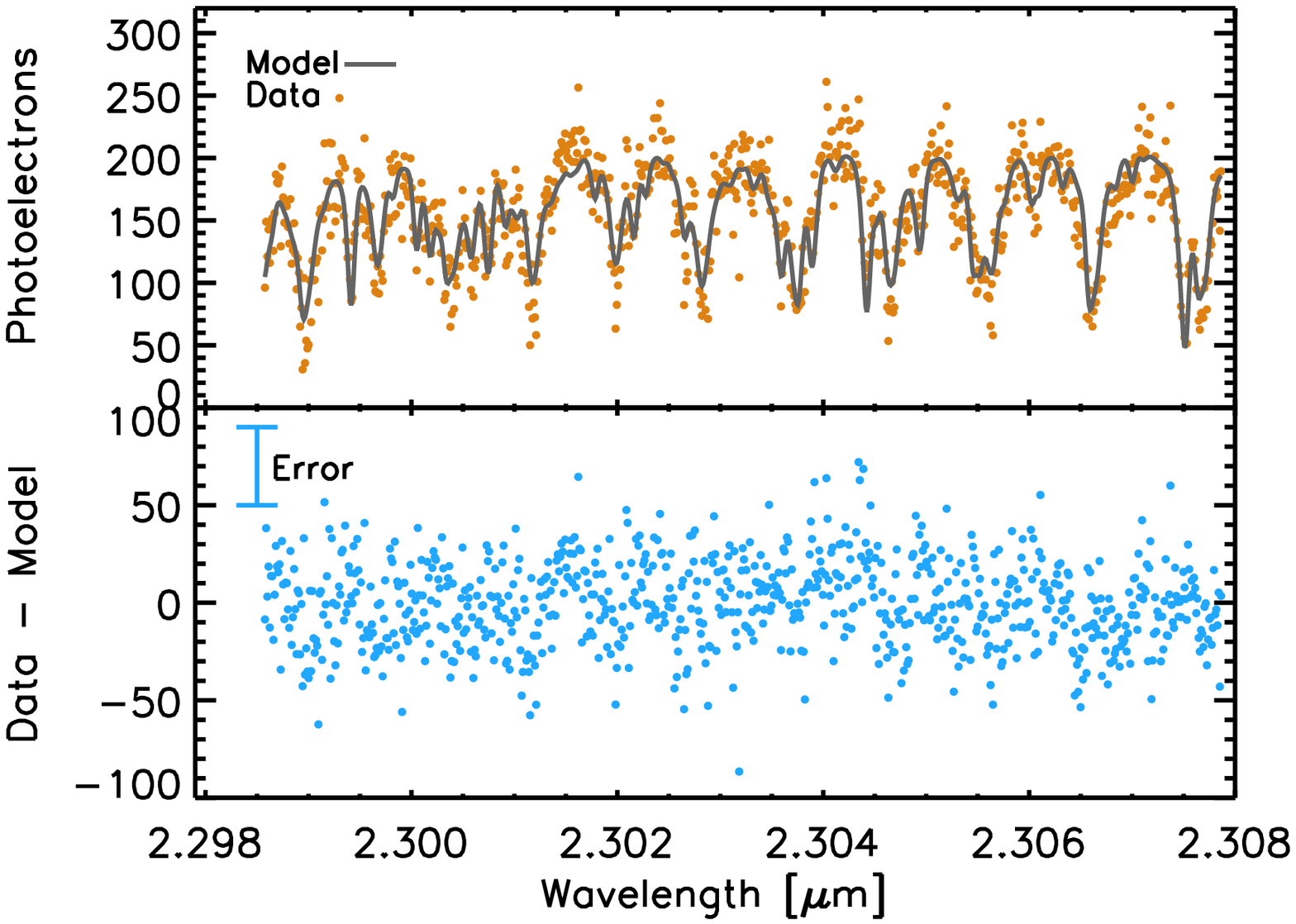}
  \vspace{2mm}
  \caption{Comparison of a model spectrum to observations of the 
  L2 dwarf 2MASS J01090150-5100494
  \citep[from][]{blakeetal2007}, obtained at a 
  spectral resolution of about $\lambda/\Delta \lambda=$50,000. The 
  top panel shows the observed data as small points (orange) with a 
  best-fit model overplotted as a line (gray, with 
  $T_{\rm{eff}}=2200$~K and $g=10^3$~m~s$^{-2}$). The bottom panel 
  shows the residuals of this fit (blue), and the error bar in the 
  upper left corner approximates the expected noise (photon and read 
  noise) of the spectrum. The agreement indicates that, in this case, 
  the model line shapes (primarily due to $\rm H_2O$ and CO lines) do 
  a good, although not perfect, job of reproducing the observations even at this high resolution.}
\label{fig:high_rez}
\end{figure}
\begin{figure}
  \centering
  \includegraphics[scale=0.65]{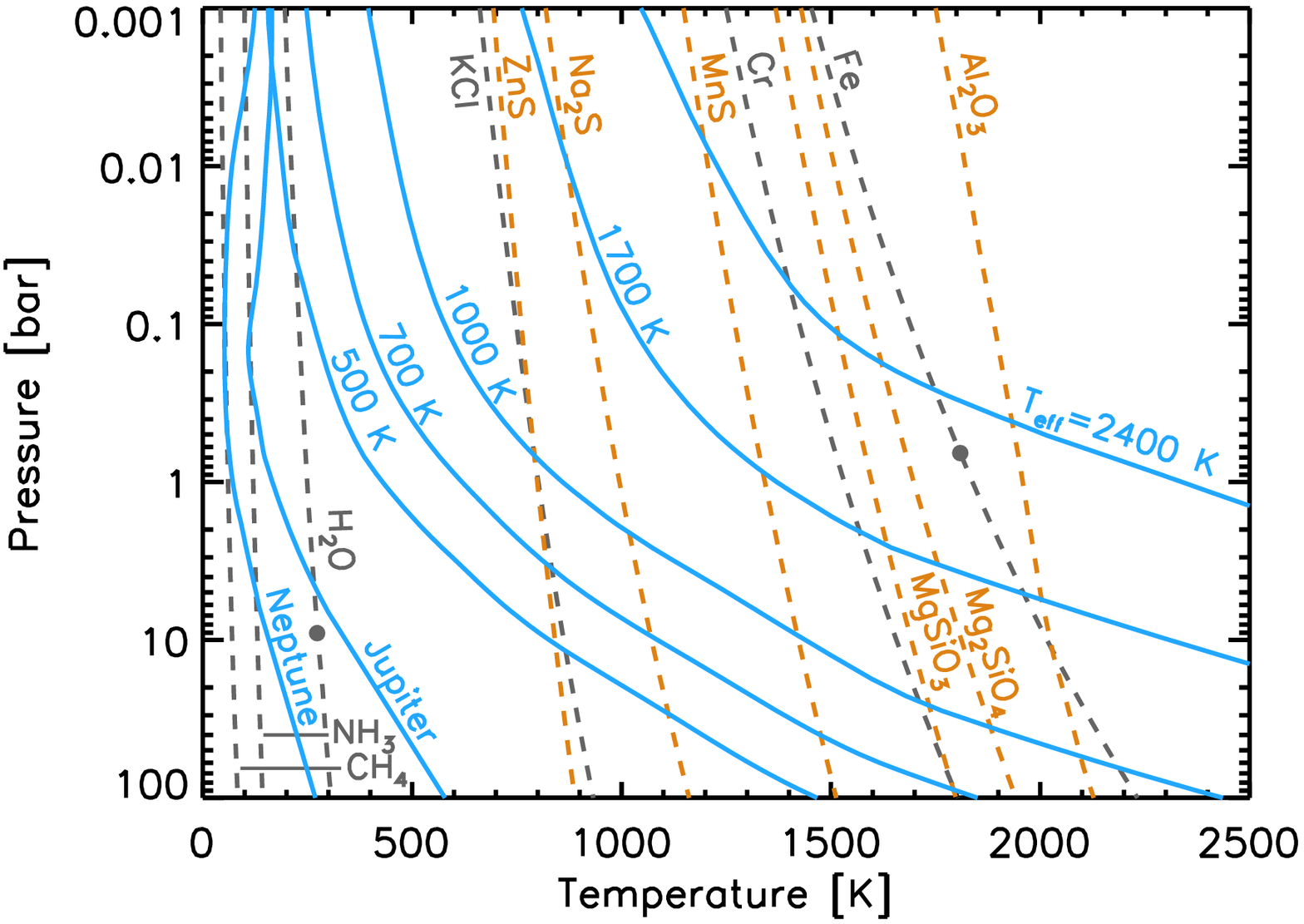}
  \vspace{2mm}
  \caption{Condensation curves for a variety of species (dashed), 
  assuming solar abundances from \citet{lodders2003}.  Gray curves 
  are for direct condensation, while orange curves are for condensates 
  that form due to chemical reactions.  Filled circles indicate a 
  liquid-solid transition.  Several cloud-free model thermal profiles 
  are provided for comparison, as well as empirically-derived profiles 
  for Jupiter and Neptune.}
  \label{fig:cond}
\end{figure}
\begin{figure}
  \centering
  \includegraphics[width=2.41in]{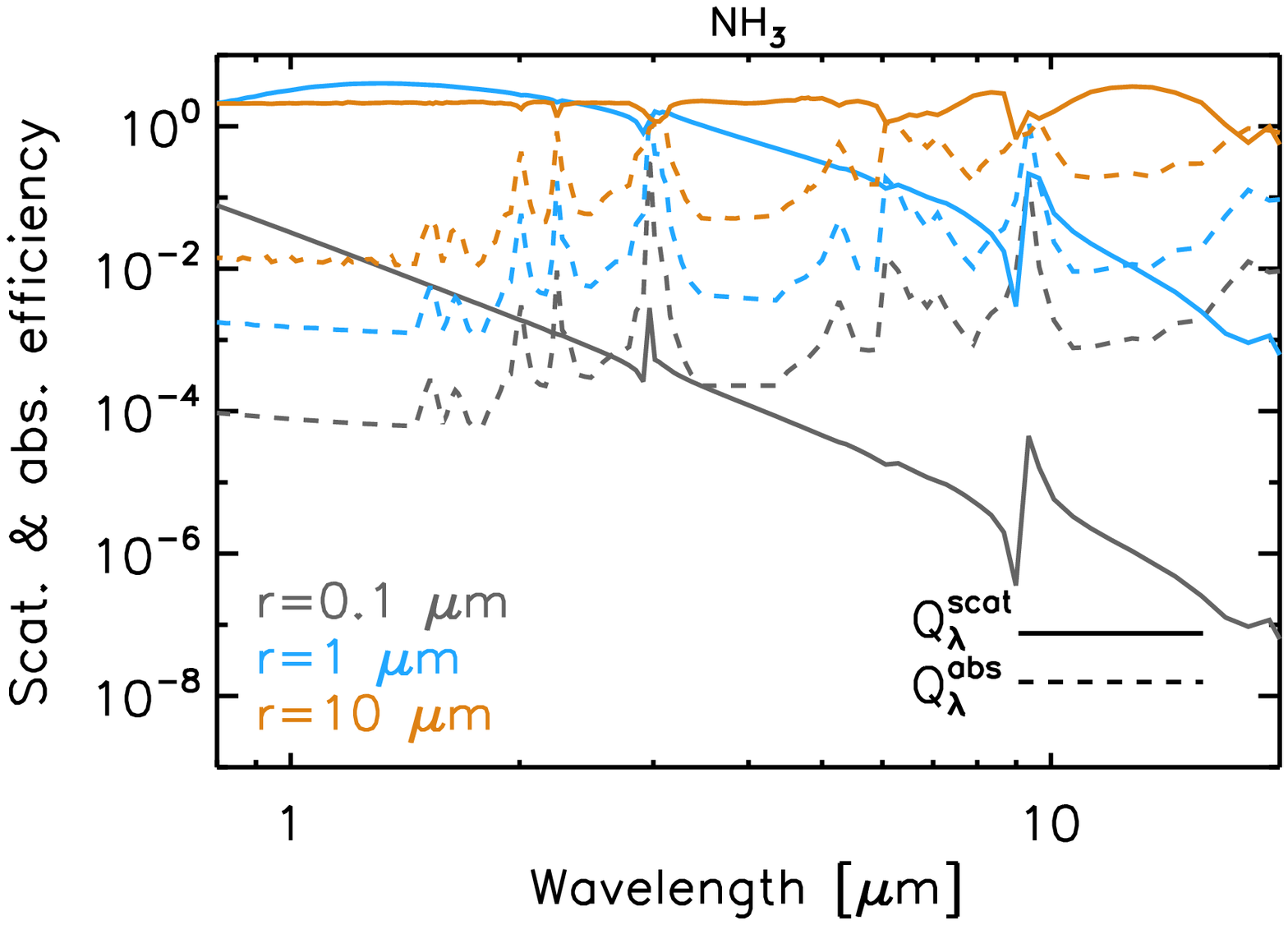}
  \includegraphics[width=2.41in]{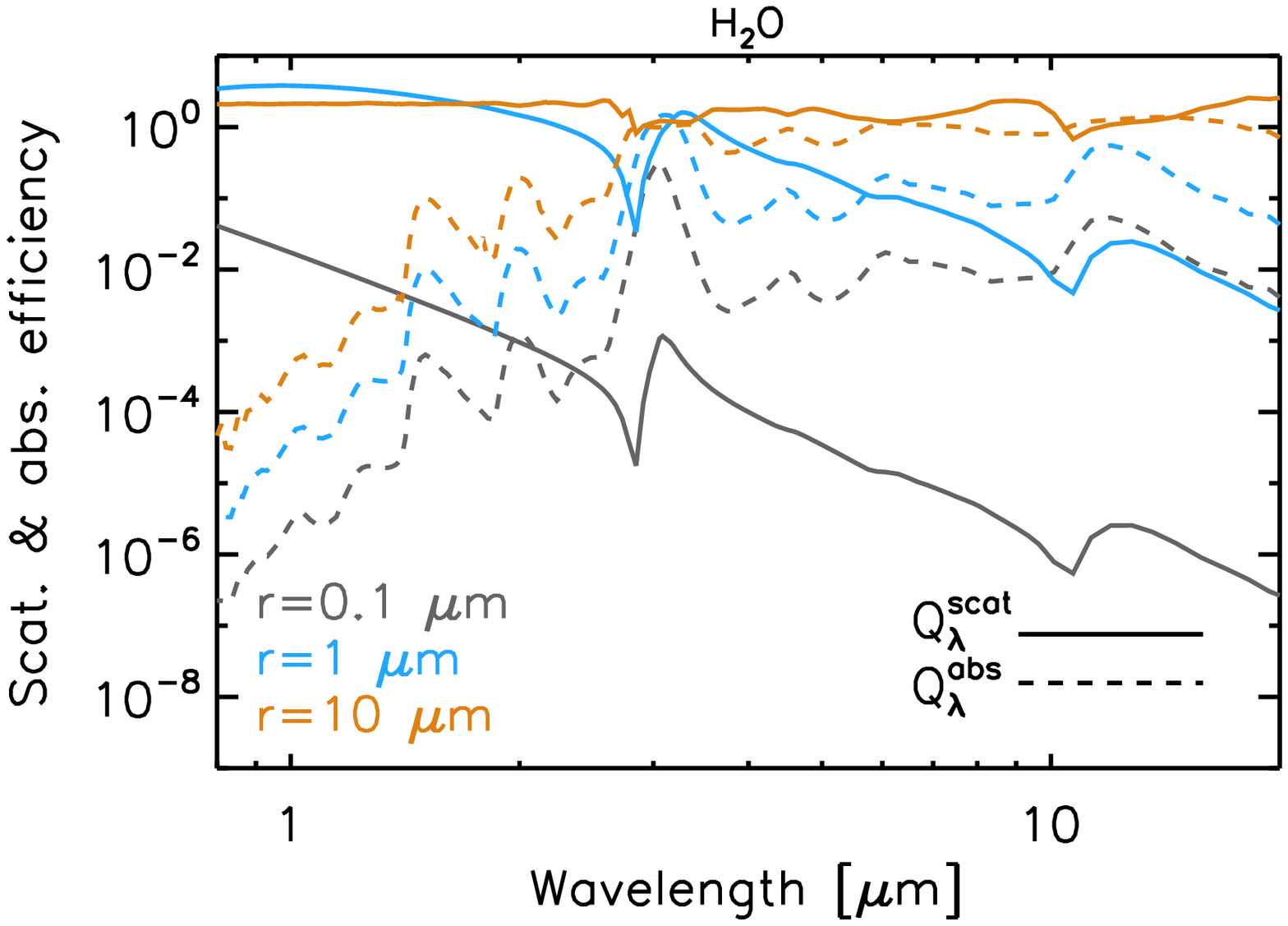}
  \includegraphics[width=2.41in]{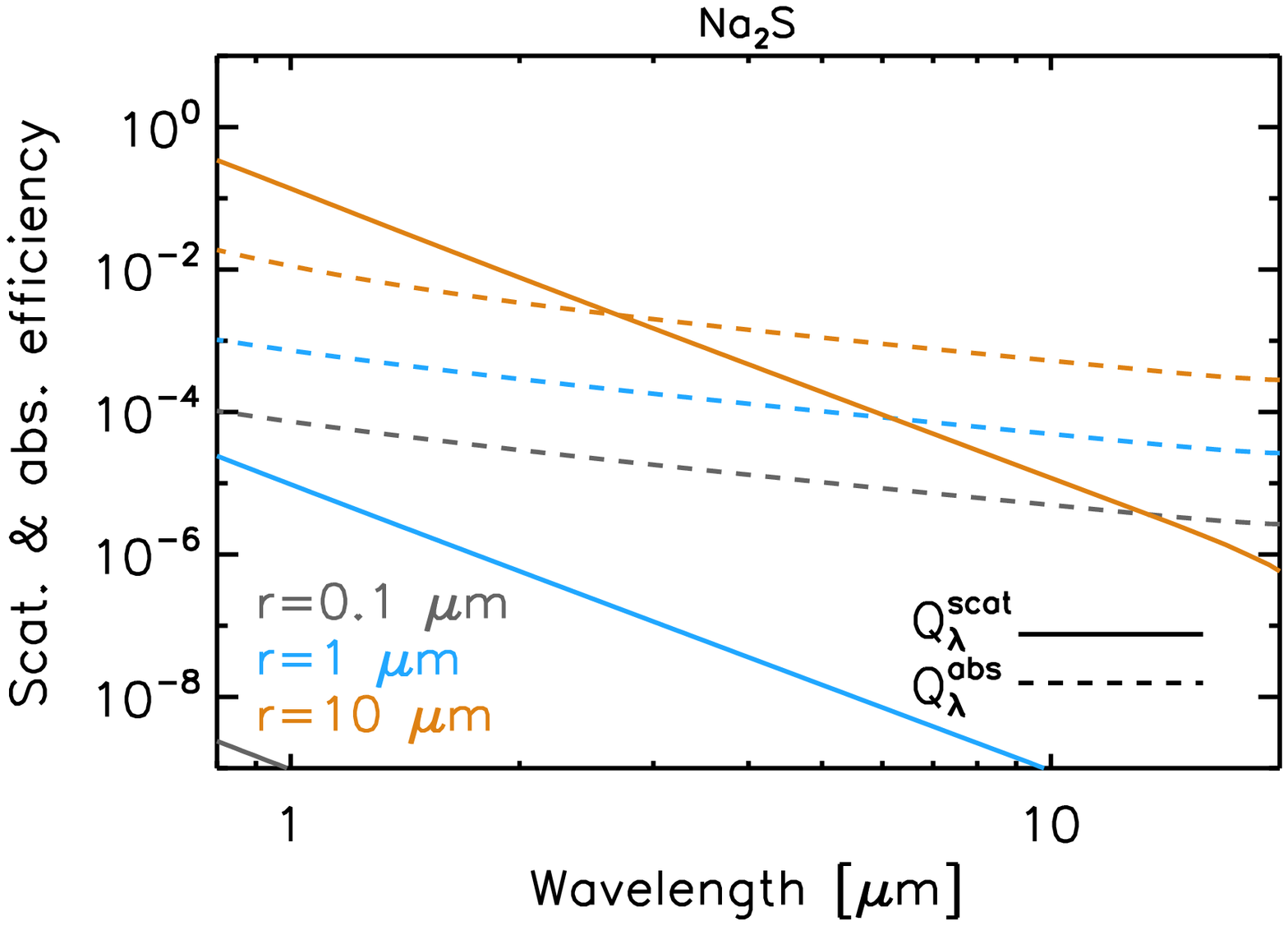}
  \caption{(\textbf{Two pages.})  Absorption and scattering efficiencies 
  for six different condensible species computed using Mie theory. In each 
  panel, the Mie scattering and absorption efficiencies, 
  $Q^{\rm scat}_{\lambda}$ (solid) and $Q^{\rm abs}_{\lambda}$ (dashed), 
  are shown for three particle sizes---$0.1~\mu$m (black), $1~\mu$m (blue), 
  and $10~\mu$m (orange).  Larger particles are more efficient at both 
  absorbing and scattering for most wavelengths. Figure generalized 
  from \citet{morleyetal2014a}.}
  \label{fig:miegrid}
\end{figure}
\newpage
\begin{figure}
  \centering
  \includegraphics[width=2.41in]{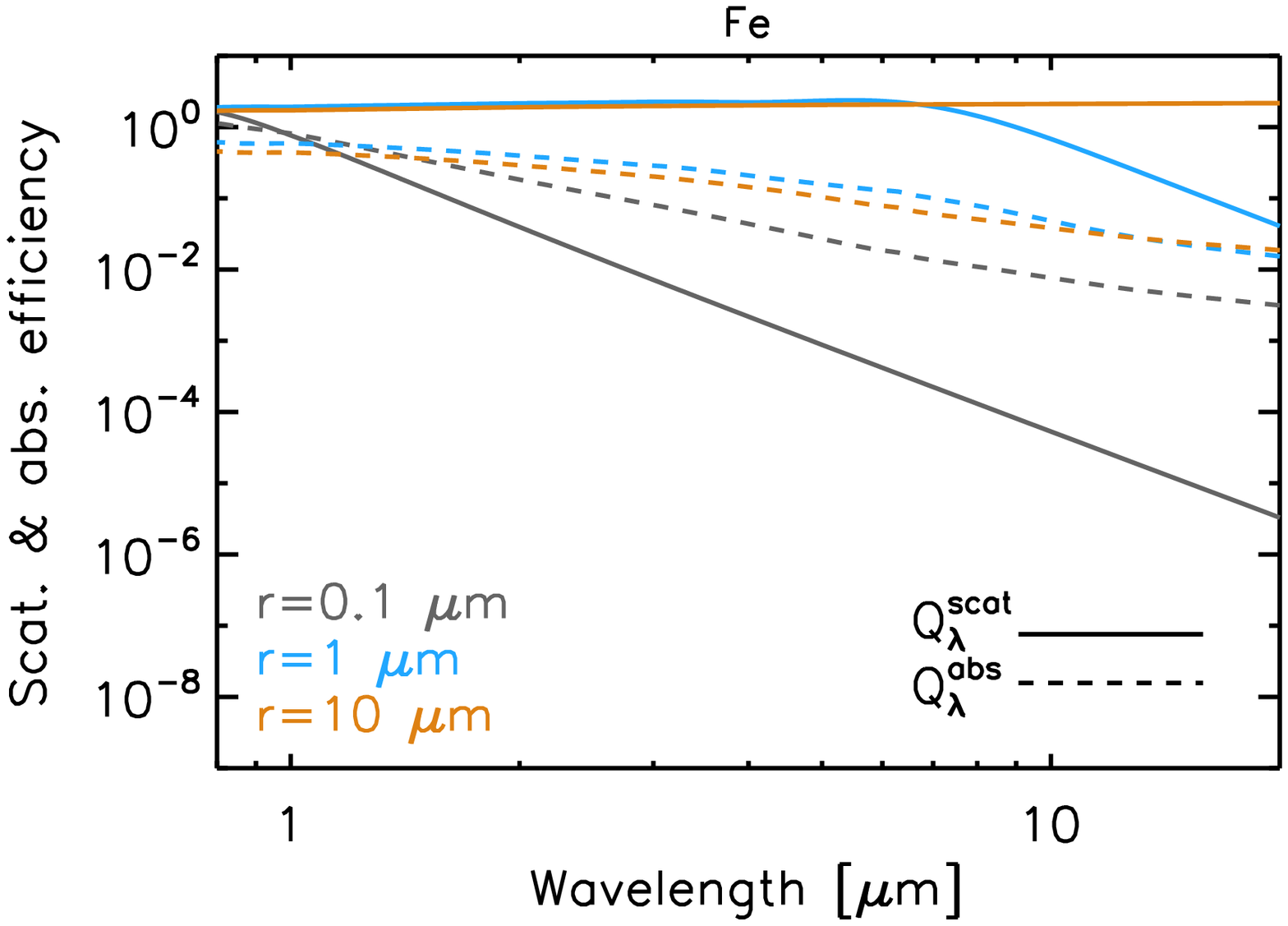}
  \includegraphics[width=2.41in]{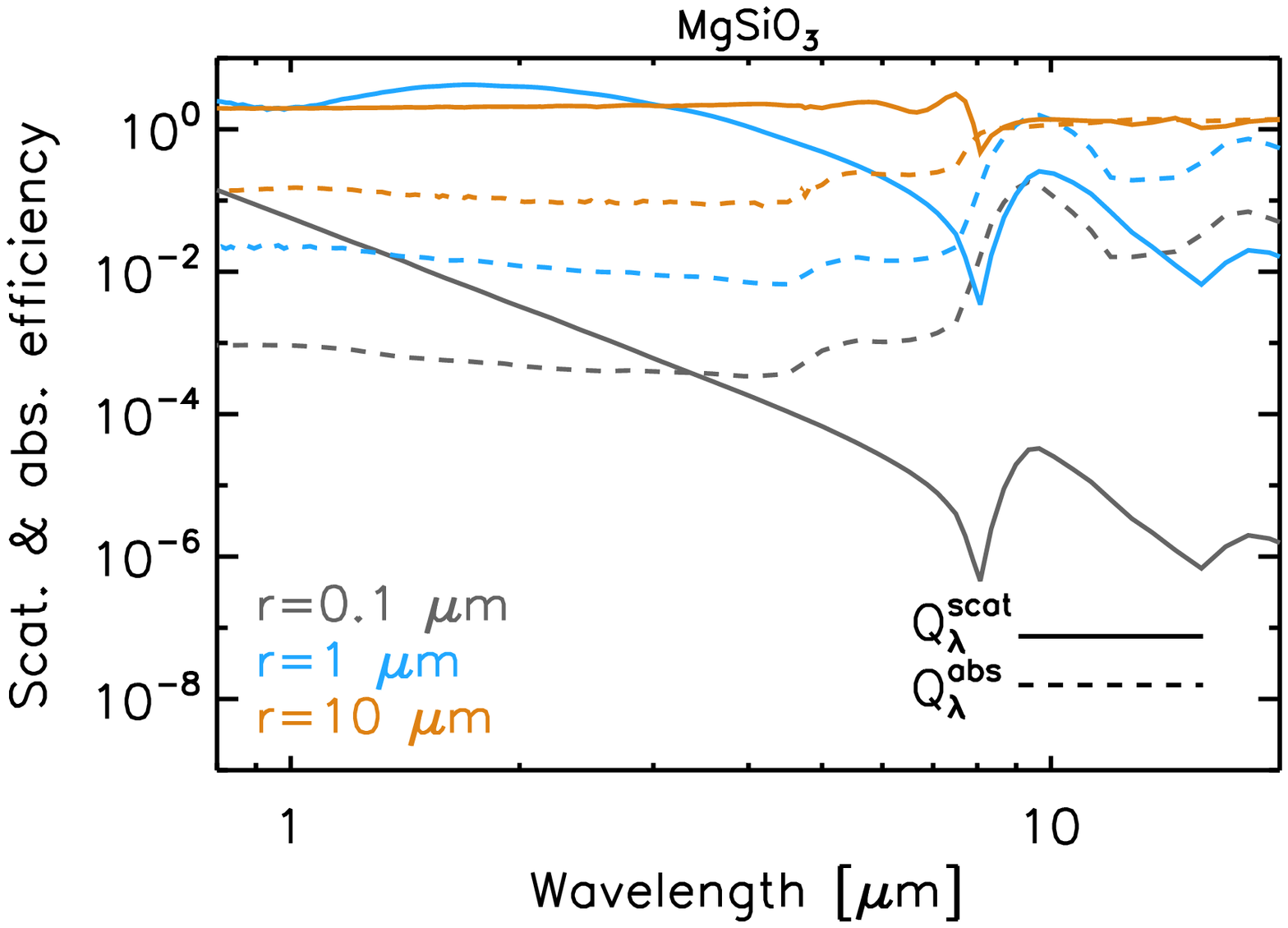}
  \includegraphics[width=2.41in]{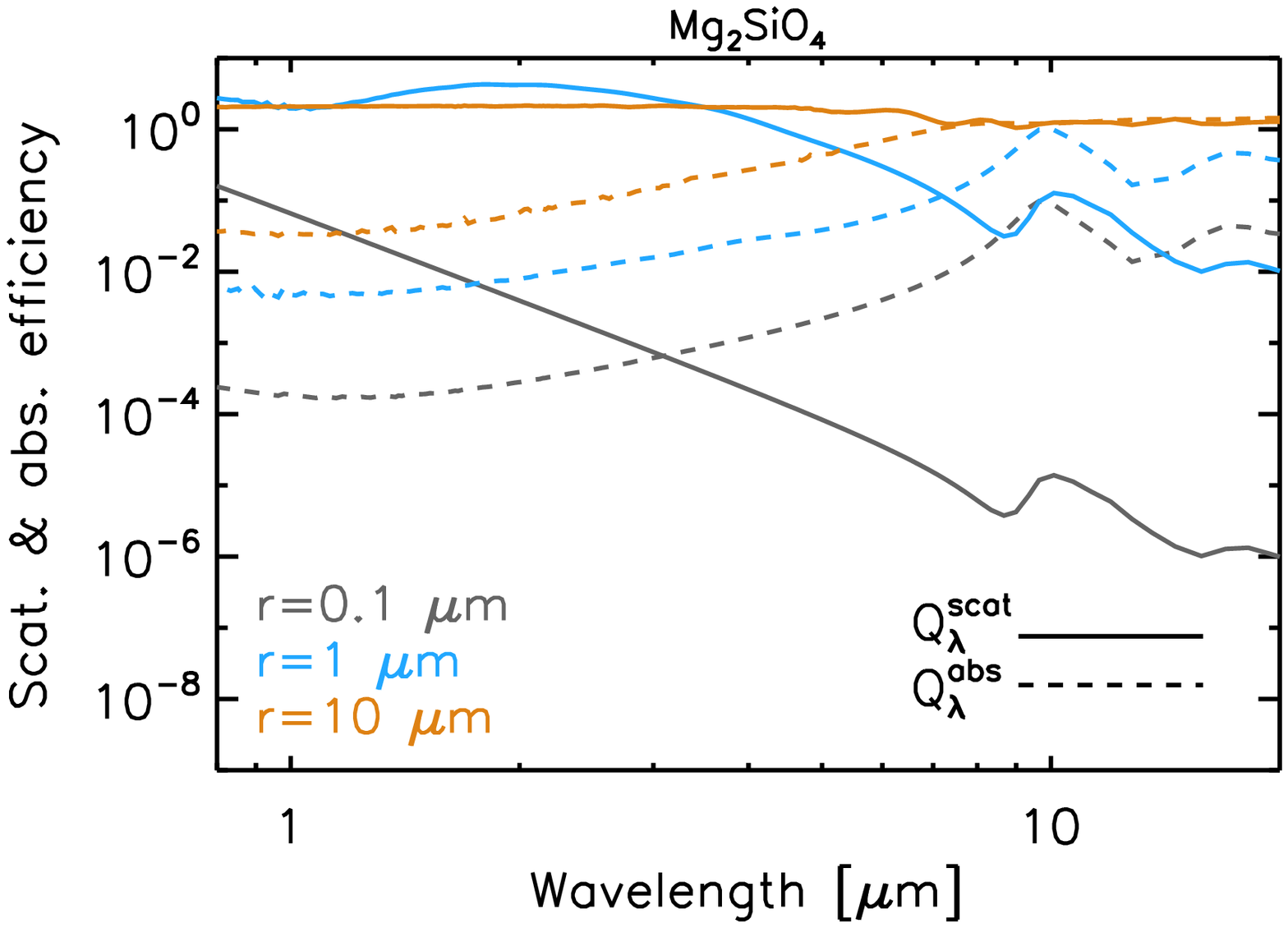}
\end{figure}
\begin{figure}
  \centering
  \includegraphics[scale=0.65]{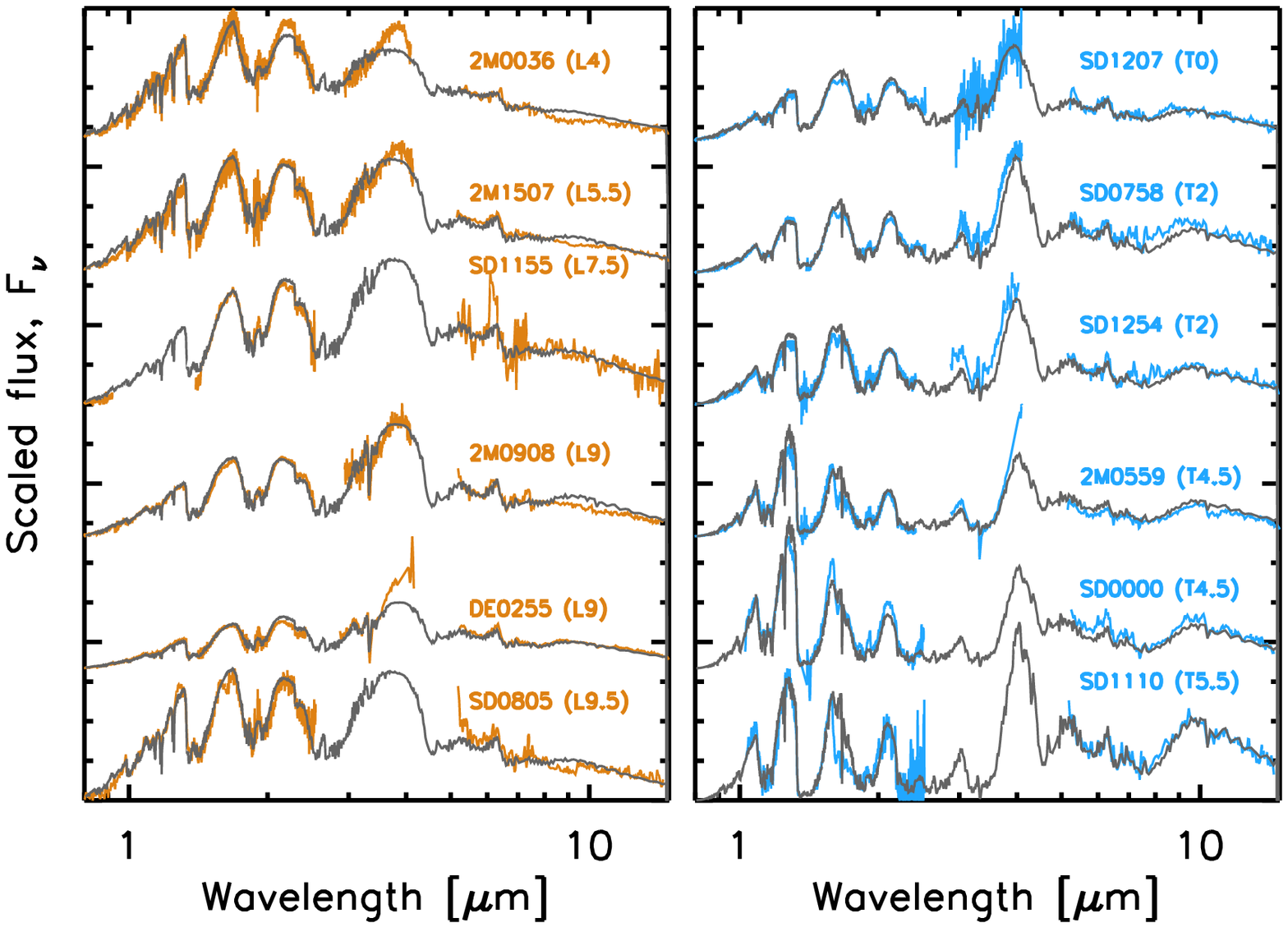}
  \vspace{2mm}
  \caption{Spectral comparison between observations and best-fit models 
  (gray), after \citet{stephensetal2009}.  The left sub-plot shows 
  L dwarfs (orange), and the right shows T dwarfs (blue).  Spectral 
  types are indicated in parentheses.  Details regarding data and model 
  properties can be found in \citet{stephensetal2009}.}
  \label{fig:datmod_comp}
\end{figure}
\begin{figure}
  \centering
  \includegraphics[scale=0.65]{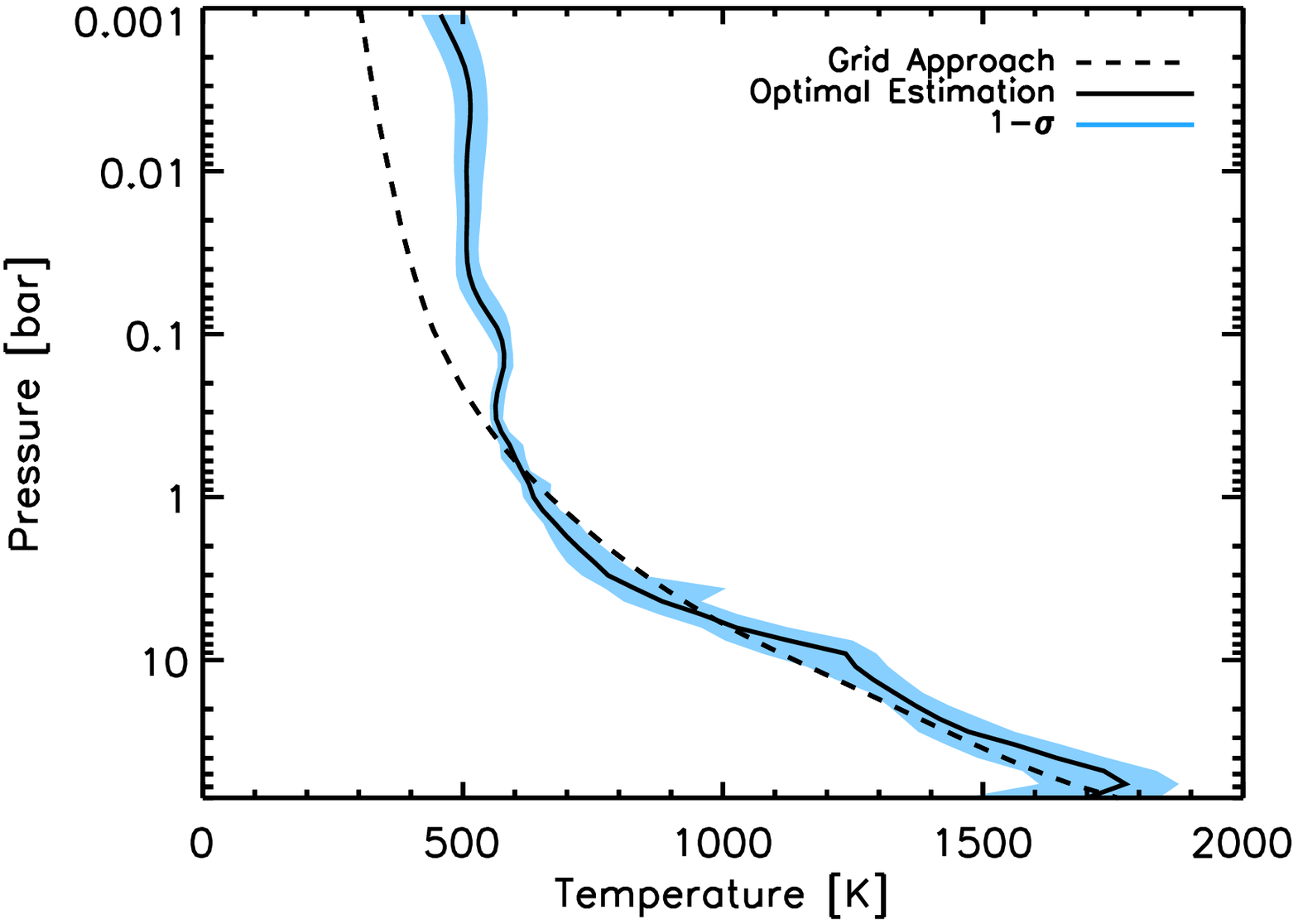}
  \vspace{2mm}
  \caption{Comparison of two approaches for determining the thermal 
  structure of a brown dwarf, as applied to GL~570D 
  \citep{burgasseretal2000}.  A grid comparison approach from 
  \citet{saumonetal2006} is shown (dashed), and a retrieved $P$-$T$ 
  profile (solid) with 1-$\sigma$ error bars (blue shaded), determined 
  using optimal estimation techniques, are also shown.  Figure adapted 
  from \citet{lineetal2014}.}
  \label{fig:retrieval}
\end{figure}

\end{document}